\documentclass[trackchanges,twocolumn,twocolappendix]{aastex701}

\usepackage{amsmath}
\usepackage{gensymb}
\renewcommand{\arraystretch}{1.5}
\usepackage{graphicx}
\usepackage{float}
\usepackage{array}
\usepackage{physics}
\usepackage{placeins}
\usepackage{booktabs}
\usepackage{multirow}
\usepackage{longtable}
\usepackage{rotating}
\usepackage{soul}

\begin{document}

\title{Satellite pairs in the Local Group: \\ from spatial clustering to dynamical associations in observations and $\Lambda$CDM simulations}
\shorttitle{Satellite pairs in the Local Group}

\shortauthors{J\'ulio et al.}
\received{July 23, 2026}
\revised{September 9, 2026}
\accepted{October 17, 2026}
\submitjournal{ApJ}

\author[orcid=0000-0002-2954-7312,gname=Mariana,sname='P. J\'ulio']{Mariana P. J\'ulio}
\affiliation{Leibniz-Institut für Astrophysik Potsdam (AIP), An der Sternwarte 16, D-14482 Potsdam, Germany}
\affiliation{Institut für Physik und Astronomie, Universität Potsdam, Karl-Liebknecht-Straße 24/25, D-14476 Potsdam, Germany}
\email{mpouseirojulio@aip.de}  

\author[orcid=0000-0002-9197-9300,gname=Marcel,sname='S. Pawlowski']{Marcel S. Pawlowski} 
\affiliation{Leibniz-Institut für Astrophysik Potsdam (AIP), An der Sternwarte 16, D-14482 Potsdam, Germany}
\affiliation{University of Southern Denmark (SDU), Department of Physics, Chemistry and Pharmacy, CP3-Origins, Campusvej 55, 5230 Odense, Denmark}
\affiliation{Danish Institute for Advanced Study (DIAS), Fioniavej 34, 5230 Odense, Denmark}
\email{mpawlowski@aip.de}

\author[orcid=0000-0002-6406-0016,gname=Noam,sname= I. Libeskind]{Noam I. Libeskind}
\affiliation{Leibniz-Institut für Astrophysik Potsdam (AIP), An der Sternwarte 16, D-14482 Potsdam, Germany}
\email{nlibeskind@aip.de}


\begin{abstract}
We investigate the incidence and physical nature of satellite galaxy pairs in the Local Group using updated observational data for the Milky Way (MW) and Andromeda (M31), together with analogues from the \textsc{hestia} constrained cosmological simulations. Our comparison uses three distinct simulated Local Group analogues drawn from the \textsc{hestia} suite, allowing us to assess the observed pair statistics against multiple cosmological realisations of the Local Group environment. We identify close nearest-neighbour associations by comparing the observed satellite distributions with isotropic populations that reproduce the radial distributions of each host, and investigate their physical nature using line-of-sight velocities, proper motions, and gravitational binding estimates. We find that the MW exhibits a significant excess of close satellite associations relative to isotropic expectations, with a pairing efficiency of $\sim53\%$, nearly twice that found in the simulated analogues ($\sim17$--$27\%$). This excess persists after applying a stellar mass cut ($M_\star > 10^4\,\mathrm{M}_\odot$) to match the resolution limits of the simulations. In contrast, the M31 satellite system is broadly consistent with isotropy, with a pairing efficiency of $\sim28\%$. However, only a minority of pair candidates exhibit coherent kinematics or are likely to be gravitationally bound, indicating that most represent transient alignments or remnants of common group infall rather than long-lived binary satellites. These results suggest that the MW hosts an unusually rich population of phase-space-correlated satellites, likely reflecting its recent accretion history, whereas M31 is more consistent with typical $\Lambda$CDM expectations.
\end{abstract}



\section{Introduction}
Satellite galaxies are the low-mass companions that orbit more massive host galaxies such as the Milky Way (MW) and Andromeda (M31). In the hierarchical framework of $\Lambda$ Cold Dark Matter ($\Lambda$CDM) cosmology, these systems are understood as the surviving remnants of smaller dark matter haloes accreted during the assembly of their hosts \citep{white_core_1978, blumenthal_formation_1984}. In this scenario, satellite galaxies are not necessarily accreted individually. Cosmological simulations predict that they can instead enter host haloes as members of larger satellite associations \citep{li_infall_2008, wetzel_satellite_2015}. Although tidal interactions with the host and dynamical processes such as phase mixing gradually disperse these associations into the general satellite population \citep{deason_satellite_2014}, remnants of their original configuration may persist as spatially or kinematically correlated pairs \citep{lynden-bell_ghostly_1995}.

The presence of satellite pairs may also be connected to the larger-scale organisation of satellite systems in the Local Group (LG). Observations of both the MW and M31 have revealed that their satellite populations are not distributed isotropically; instead, many satellites appear to reside in thin, spatially extended, and kinematically coherent structures known as planes of satellites \citep[see][for a review]{pawlowski_planes_2018,pawlowski_phase-space_2021}. The origin of the Vast Polar Structure (VPOS; \citealp{pawlowski_vpos_2012, pawlowski_vast_2014}) around the MW and the Giant Plane of Andromeda \citep[GPoA;][]{ibata_vast_2013, conn_three-dimensional_2013} remains a subject of active debate, as such strong spatial and orbital alignments are rare in standard $\Lambda$CDM simulations \citep{pawlowski_co-orbiting_2014}. Group infall provides a potential mechanism for producing these correlated structures. If a significant fraction of satellites are accreted together as part of a pre-existing group, they may naturally share similar orbital planes and angular momentum vectors \citep{lynden-bell_ghostly_1995, li_infall_2008, wang_spatial_2013}. Recent kinematic studies of the MW's satellite system further support this scenario, suggesting that a substantial fraction of the VPOS may have originated from a few major accretion events of dwarf galaxies \citep{taibi_portrait_2024}. A striking example is the first-passage of the Large Magellanic Cloud (LMC), as dynamical modelling suggests that the LMC likely brought with it a significant number of smaller satellites, many of which remain spatially and kinematically associated today \citep{vasiliev_effect_2023}. In this context, satellite pairs may represent the most compact surviving remnants of these larger accretion events. 

While galaxy pairs have long been used as a diagnostic of hierarchical growth for massive galaxies \citep[$M_\star \gtrsim 10^{10}\,\mathrm{M_\odot}$; e.g.][]{patton_dynamically_2002, duncan_observational_2019}, the low-mass regime of satellite galaxy pairs has been less studied. In the literature, definitions of a ``satellite pair'' vary, ranging from simple distance-velocity thresholds \citep[e.g.][]{stierwalt_direct_2017, besla_frequency_2018} to strict gravitational binding criteria \citep{sales_satellites_2013}. Regardless of the specific definition adopted, $\Lambda$CDM simulations predict that subhaloes are frequently accreted as groups \citep{wetzel_satellite_2015}, and cosmological simulations show non-zero fractions of surviving satellite pairs at the present day \citep{fattahi_galaxy_2013, callingham_group_2025}.

In the LG, several candidate satellite associations have been proposed over the past decades, mostly based on spatial proximity and line-of-sight velocities. Early examples include the Fornax--Leo--Sculptor system \citep{lynden-bell_fornax-leo-sculptor_1982}, which was later expanded to include Sextans and Phoenix \citep{majewski_fornax-leo-sculptor_1994}. Perhaps the most prominent example of an interacting satellite pair in the LG is the Large and Small Magellanic Clouds, which are widely believed to have been accreted together as part of a common system \citep{besla_are_2007, donghia_small_2008, patel_orbital_2020, vasiliev_effect_2023}. Beyond the Magellanic Clouds, several other possible associations have been discussed in the literature. More recent candidates include Pisces II and Pegasus III \citep{kim_heros_2015, garofalo_born_2021, richstein_structural_2022}, as well as the trio NGC 147, NGC 185, and Cassiopeia II around M31 \citep{van_den_bergh_binary_1998, arias_ngc_2016}. 
A particularly intriguing association in the MW halo involves the Crater-Leo system. Initially proposed to include Leo IV and Leo V, whose proximity in space and velocity was noted by \citet{jong_enigmatic_2010}, this group was later expanded to include Crater 1, Crater II, and possibly Leo II \citep{torrealba_feeble_2016}. Recent work has used \textit{Gaia} and Hubble Space Telescope (HST) proper motions (PMs) to investigate this configuration, finding dynamical support for the group infall scenario for a subset of its members \citep{julio_satellite_2024}.

Outside of the LG, another striking candidate for a recent group accretion is the asymmetric satellite distribution of the M81 system, which \citet{bell_ultrafaint_2022} interpret as evidence for the infall of a satellite group. Satellite pairs and small groups of dwarf galaxies have also been identified in the Local Volume through dedicated surveys such as TiNy Titans \citep{stierwalt_tiny_2015,pearson_local_2016}, as well as in individual systems such as NGC 4490 \citep{pawlowski_satellite_2024} and other nearby dwarf groups \citep{paudel_discovery_2024,paudel_groups_2026}. Statistical studies have further investigated correlations between satellite pairs around external galaxies, finding evidence for correlated satellite positions and velocities in both low-redshift galaxy samples \citep{ibata_velocity_2014} and massive galaxy groups \citep{gu_younger_2024}. Related spatial and kinematic associations of dwarf galaxies have also been identified in the Local Volume \citep{tully_squelched_2002,tully_associations_2006}, although these systems are generally more extended than the close satellite associations considered here.

However, establishing whether such systems represent real physical associations remains challenging. Many early suggestions were based primarily on projected spatial proximity and line-of-sight velocities, which provide only partial information about the true orbital configuration of satellites. As more precise proper motion measurements have become available, some proposed systems have been re-evaluated. One example is the recent re-examination of the long-suspected binary nature of the M31 satellites NGC 147 and NGC 185. Although their close projected proximity once suggested a bound state \citep{geha_local_2010}, recent HST PMs and high-precision distance measurements revealed that they only recently reached their minimum separation and likely entered the M31 halo independently \citep{sohn_hst_2020, savino_hubble_2022}. These developments illustrate the difficulty of distinguishing true physical pairs from chance alignments within the satellite population.

Despite these uncertainties, previous studies have suggested that satellite pairs may be relatively common in the LG. \citet{fattahi_galaxy_2013}, for instance, found that a substantial fraction ($30\%$) of satellites brighter than $M_V \leq -8$ appear to have unusually close neighbours compared to expectations from isotropic distributions. However, by applying the same method to a semi-analytic model of the satellite population in the six simulated haloes of the Aquarius Project \citep{springel_aquarius_2008}, they found that fewer than $4\%$ of such satellites were paired. This difference of one order of magnitude highlights a significant tension between observed satellite distributions and theoretical expectations and suggests that the standard cosmological model fails to capture the frequency of group-infall events or the survival rate of satellite associations.

In this work, we revisit the incidence of satellite pairs in the LG by combining updated observational data with full phase-space information (when available). We characterise satellite associations using both their spatial proximity and kinematic coherence, and determine their significance by comparing them to mock satellite populations drawn from cosmological simulations. This allows us to place the observed pair population in a statistically robust cosmological context and to evaluate whether the apparent excess of close associations from previous works can be reconciled with expectations from hierarchical galaxy formation.

In Section~\ref{sec:data} we describe the data of the dwarf galaxies that we used for this study, both for the observations (Section~\ref{data:observations}) and for the simulations (Section~\ref{data:simulations}). This is followed in Section~\ref{sec:methods} by the methods used to identify and quantify satellite associations. In Section~\ref{sec:results}, we present the statistical analysis of satellite pairing (Section~\ref{sec:statistical}) and investigate the properties of the identified pair candidates (Section~\ref{sec:paircandidates}). We summarise our conclusions in Section~\ref{sec:conclusion}. In Appendices~\ref{app:references},~\ref{app:cdfs_sims},~\ref{app:cdfs_opposite_hosts} and~\ref{app:kinematic_coherence}, we provide supplementary material in support of the results that we discuss in the main text.


\section{Data}\label{sec:data}
\subsection{Observations}\label{data:observations}
For our analysis, we rely on the up-to-date Local Volume Database \citep{pace_local_2025}\footnote{Last accessed on 12 March 2026.} as the source of the positions, distances, proper motions, line-of-sight velocities, and stellar masses of LG dwarfs. The references for the individual measurements used in our MW and M31 samples are provided in Appendix~\ref{app:references}. We exclude unconfirmed dwarf candidates and satellites with no distance measurements. In addition, we exclude dwarfs that have $d_\mathrm{host} > 350 \,\mathrm{kpc}$. This approximately corresponds to the outer regions of the virial volumes of both the MW and M31 \citep[e.g.][]{watkins_masses_2010}, beyond which satellites are likely to be weakly bound or part of the surrounding LG environment rather than the host halo itself. After this, our sample consists of 51 satellites of the MW and 39 satellites of M31. 

To ensure a fair comparison with the finite resolution of the \textsc{hestia} simulations (described below), we also consider a subsample of the observed satellites selected by imposing a lower limit on their stellar mass ($M_\star > 10^4\,\mathrm{M}_\odot$). This cut approximately matches the stellar mass completeness of the simulated satellite populations, in an attempt to alleviate possible biases introduced by the inclusion of ultra-faint systems that are unresolved in the simulations. We perform our analysis on both the full observational sample and this mass-limited subsample, allowing us to understand how the results are affected when limited to the mass range resolved by the simulations. On the observational side, we do not have galaxies with masses as low as this among the known M31 satellites, so we retain all 39 satellites. For the MW satellites, 26 satellites have masses lower than this number. In this case, we end up with 25 MW satellites. These numbers are summarised in Table~\ref{tab:number_satellites}.

\subsection{Simulations}\label{data:simulations}
The `High-resolution Environmental Simulations of The Immediate Area' \textsc{(hestia)} project is described in detail in \cite{libeskind_hestia_2020}. We briefly summarise the key features of the simulations here for the reader's convenience.

\textsc{hestia} \citep{libeskind_hestia_2020} is a suite of high-resolution cosmological magneto-hydrodynamical simulations of the LG in a $\Lambda$CDM cosmology ($\Omega_{\mathrm{m}}=0.270$, $\Omega_{\Lambda}=0.682$, $\Omega_{\mathrm{b}}=0.048$, and $\mathrm{H}_{0}=100\,h\mathrm{km\,s}^{-1}\mathrm{Mpc}^{-1}$, where $h=0.677$; \citealt{planck_collaboration_planck_2014}). It uses the code \textsc{arepo} \citep{springel_e_2010} and the galaxy formation model from \textsc{auriga} \citep{grand_auriga_2017}. These simulations are constructed from constrained initial conditions based on measurements of the local peculiar velocity field, allowing them to reproduce the large-scale environment of the LG, including the locations of major structures such as the Virgo cluster, the Local Void, and the filament connecting the LG to Virgo. Within this environment, the simulations also reproduce the observed masses, separation, and relative velocity of the MW--M31 pair \citep{libeskind_hestia_2020}.

In this work, we use the resulting three high-resolution LGs, corresponding to $8192^3$ particles within the region composed of two overlapping spheres around the two main $z=0$ LG members with a radius of $3.7\,\mathrm{Mpc}$ $(2.5\,\mathrm{Mpc}\,h^{-1})$ each. The mass and spatial resolution achieved is $\mathrm{m}_{\mathrm{dm}}=1.5\times10^5$\,$\mathrm{M}_{\odot}$, $\mathrm{m}_{\mathrm{gas}}=2.2\times10^4$\,$\mathrm{M}_{\odot}$ and $\epsilon = 220\,\mathrm{pc}$. In this set of simulations, dwarf galaxies are resolved and have similar stellar mass distributions and radial concentration as observed \citep{libeskind_hestia_2020}. The use of these constrained simulations is particularly motivated by their ability to reproduce the LG's unique dynamical environment. By reproducing the large-scale structure and density field of the local Universe, \textsc{hestia} provides a realistic cosmographic environment in which the assembly and accretion of satellite systems can be studied, including processes such as group infall that are expected in $\Lambda$CDM.

To construct our satellite sample, we keep only the luminous subhaloes. After this first selection, we then remove leftover baryon-dominated substructures associated with the central galactic discs. These objects are identified as subhaloes for which the baryonic mass fraction, defined as $f_b =(M_\star + M_{\mathrm{gas}})/M_{\mathrm{vir}}$, exceeds 0.5, indicating that they are not dark matter dominated and are therefore unlikely to represent satellite galaxies. 

\begin{figure}[ht]
    \centering
    \includegraphics[width=0.9\columnwidth]{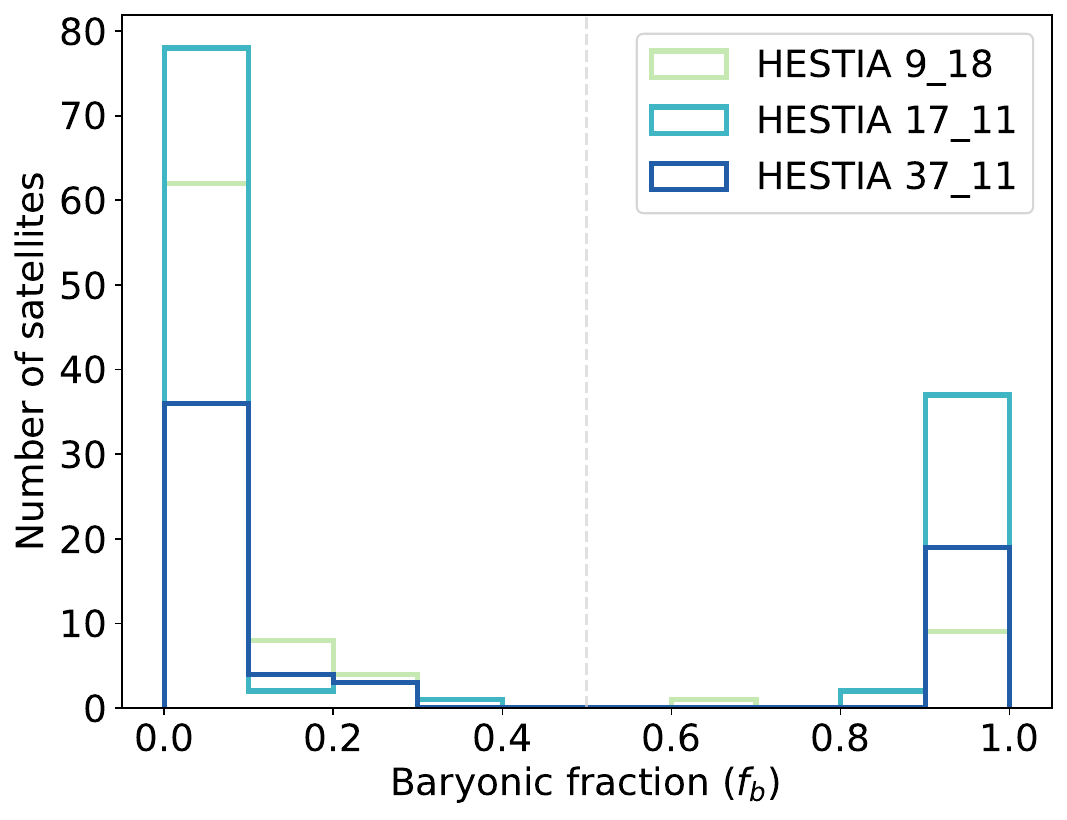}
    \caption{Distribution of baryonic fractions $f_b$ for all luminous satellites (with number of star particles $N_\star > 0$) in the \textsc{hestia} simulations. The satellites of MW- and M31-like hosts are combined for each of the three simulation realisations (shown in different colours). The dashed line indicates the baryon fraction where we performed the cut to remove disc fragments from the final sample.}
    \label{fig:baryonic_fraction}
\end{figure}

As shown in Figure~\ref{fig:baryonic_fraction}, the distribution of $f_b$ for all luminous subhaloes across the \textsc{hestia} runs reveals a distinct bimodality. The majority of subhaloes have $f_b \lesssim 0.2$, consistent with dark-matter-dominated satellite galaxies. A separate population is found at extremely high baryonic fractions ($f_b \gtrsim 0.9$). These objects lack a significant dark matter component and are located in close proximity to the host's central regions, so we consider them as fragments of the galactic disc rather than authentic dwarf galaxies. The vertical dashed line at $f_b = 0.5$ effectively separates these two populations across all \textsc{hestia} realisations. After applying this cleaning step, we get a final sample of $\sim 20-60$ satellites per host (see Table~\ref{tab:number_satellites}), which is broadly comparable to the number of confirmed satellites in the observational sample.

\begin{table}[ht]
\centering 
\caption{Number of satellite galaxies per host after sample selection.}
\label{tab:number_satellites}
\begin{tabular}{lcc}
\toprule
Dataset & MW & M31 \\ 
\midrule
Observations                                       & 51 & 39 \\ 
Observations ($M_\star > 10^4\, \mathrm{M_\odot}$) & 25 & 39 \\ 
\addlinespace[0.5em]
\textsc{hestia} 9\_18                              & 45 & 29 \\ 
\textsc{hestia} 17\_11                             & 26 & 58 \\ 
\textsc{hestia} 37\_11                             & 24 & 19 \\ 
\bottomrule
\end{tabular}
\end{table}


\section{Methods}\label{sec:methods}
Similarly to \cite{fattahi_galaxy_2013}, we start by finding which objects have nearest neighbours (NNs) closer than those expected from a random isotropic population of satellites around their host. To do this, we use Monte Carlo sampling to generate an isotropic population with the same number of satellites and the same radial distribution as the observed sample. 

The radial distribution is modelled directly from the data by constructing the cumulative distribution function (CDF) of the satellites' Galactocentric distances. For each realisation, radial distances are then drawn by sampling from this empirical CDF using either a spline interpolation or a kernel density estimate (KDE), ensuring that the synthetic populations follow the same radial profile as the observed system. Angular positions are assigned isotropically by drawing random directions uniformly over the sphere. For the MW satellites, we additionally impose an obscuration region of $15\degree$ around the Galactic plane to mimic observational incompleteness\footnote{Available at \href{https://github.com/marianajulio/will_it_pair}{https://github.com/marianajulio/will\_it\_pair}.}.

\begin{figure}[ht]
    \includegraphics[width=0.495\columnwidth]{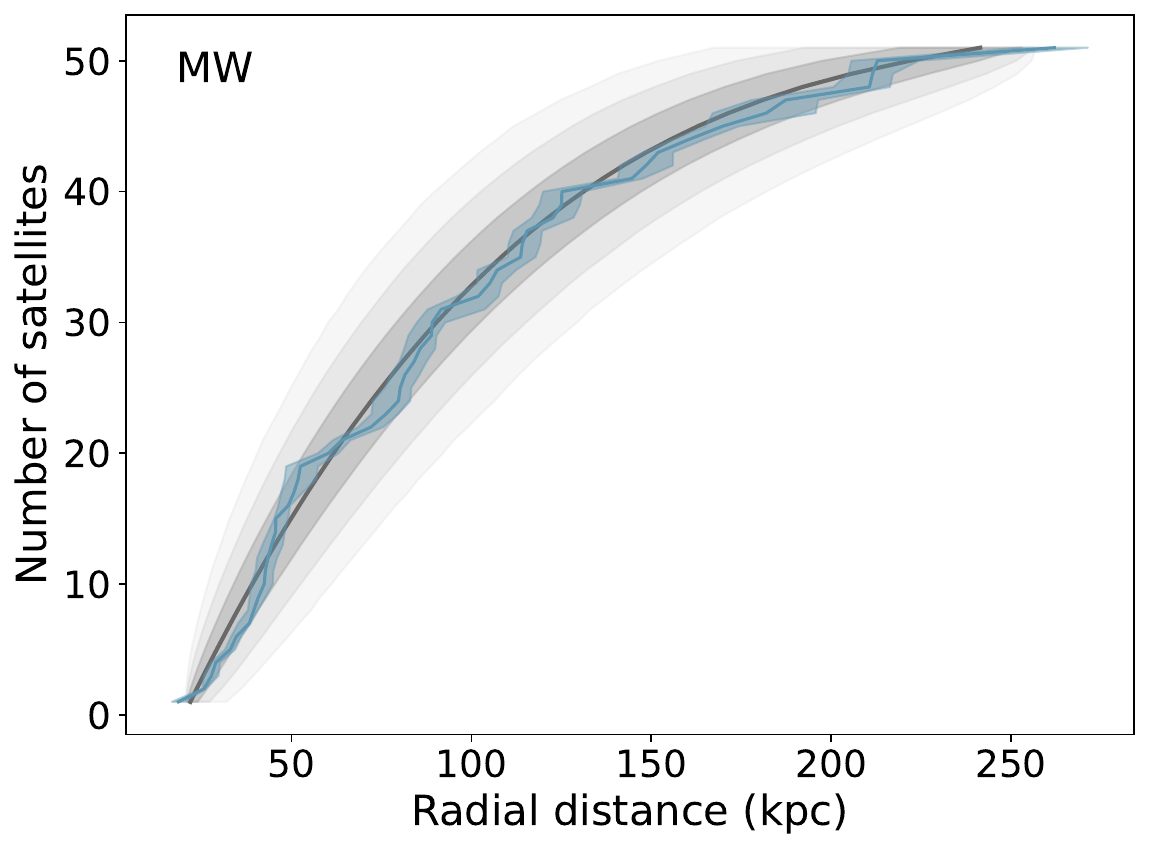}
    \includegraphics[width=0.495\columnwidth]{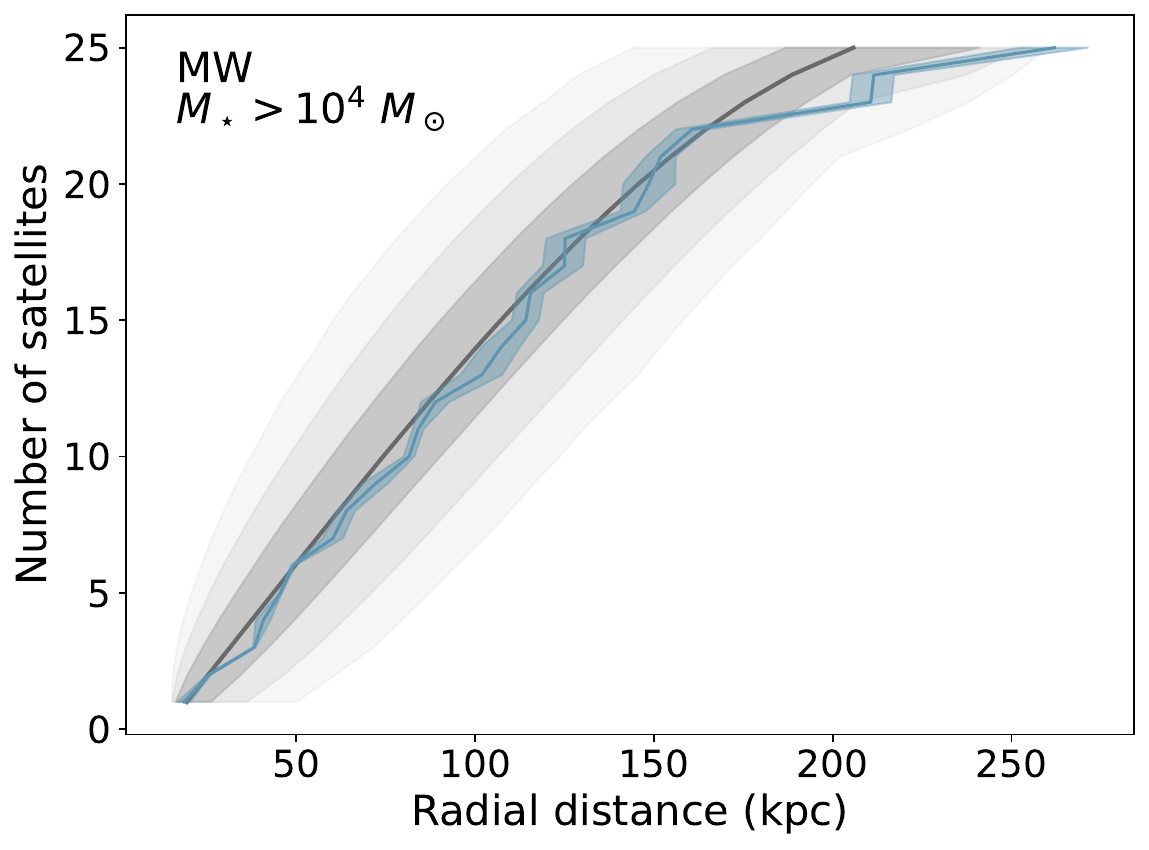}
    \includegraphics[width=0.5\columnwidth]{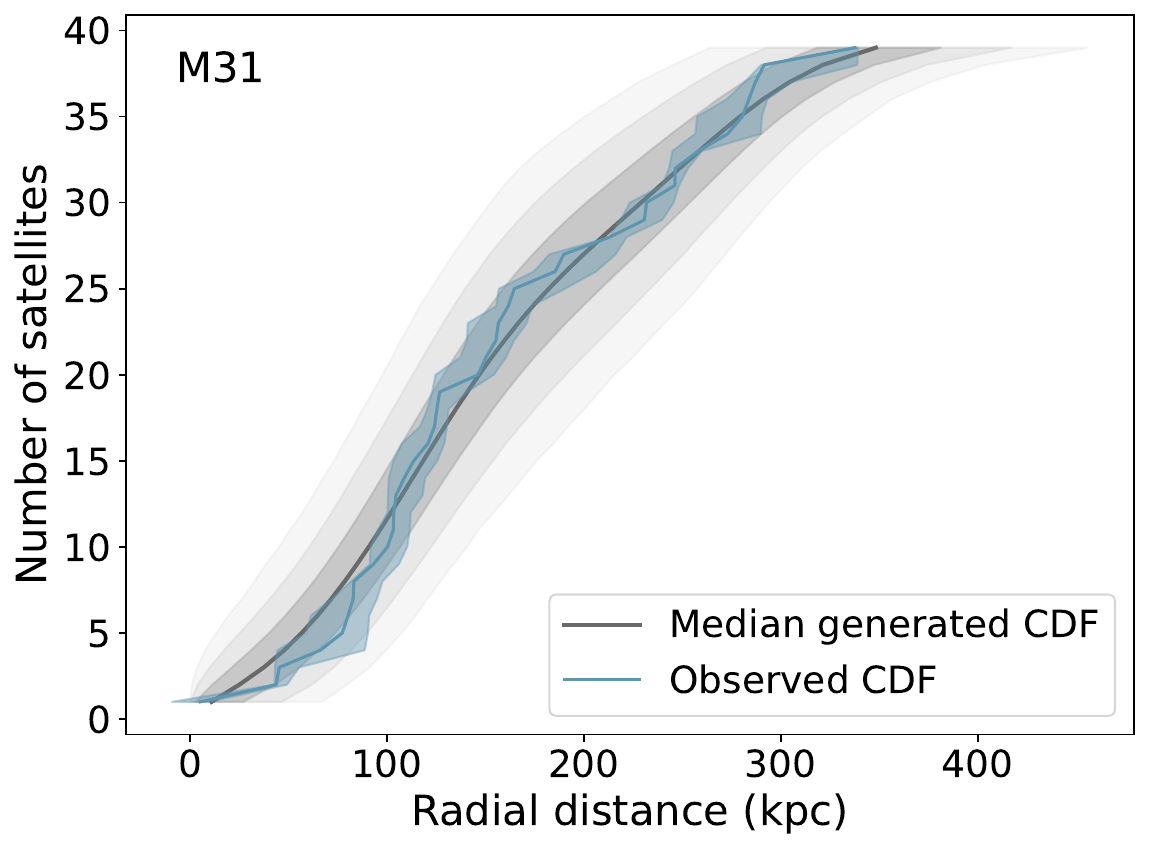}
    \caption{Cumulative radial distributions of satellites around the MW (top panels) and M31 (bottom panel). The solid blue line shows the cumulative number of satellites as a function of host-centric distance. The surrounding blue shaded region illustrates the variation in the cumulative profile obtained using the lower and upper bounds of the $1\sigma$ distance uncertainties. The solid dark-grey line shows the median cumulative distribution obtained from $10^5$ isotropic realisations. The grey shaded regions represent the $1\sigma$, $2\sigma$, and $3\sigma$ ranges of the generated isotropic populations.}
    \label{fig:cdf_obs}
\end{figure}

As illustrated in Figure~\ref{fig:cdf_obs}, the observed radial distribution of satellites is shown by the solid blue line, with the surrounding blue shaded region representing the variation in the cumulative profile when accounting for satellite distance uncertainties. We compare these observations against the $10^5$ isotropic realisations. The median of these realisations is represented by the solid dark grey line, while the shaded grey envelopes indicate the $1\sigma\,(68\%)$, $2\sigma\,(95\%)$, and $3\sigma\,(99.7\%)$ confidence intervals of the generated populations. This way, it is possible to see that by generating this amount of isotropic populations, we are able to reproduce the radial distribution of the observed ones, even when the distance uncertainties are taken into account (see below for details). The same analysis for the \textsc{hestia} simulated analogues can be found in Figure~\ref{fig:cdf_sims} of Appendix~\ref{app:cdfs_sims}.

We generate $10^5$ isotropic realisations for each dataset: the full MW sample, the stellar mass-limited MW sample, the M31 sample, and each of the individual \textsc{hestia} host analogues. In each iteration, we estimate, for every observed satellite, the distance to its NN within the synthetic isotropic population. The resulting distribution of NN distances provides the expectation for the separation of a satellite from its closest companion under isotropy. By comparing this distribution with the observed NN separation, we estimate the probability $P$ of that pair being that close by random chance.

To check the impact of distance uncertainties on the significance of the identified associations, we propagate the observational errors on the heliocentric distances of both satellites in each pair. For each Monte Carlo realisation, we assign a perturbed distance
\begin{equation}
d' = d + \mathcal{N}(0,\,\sigma_d),
\end{equation}
where $\mathcal{N}(0,\,\sigma_d)$ represents a normal distribution with mean zero and standard deviation $\sigma_d$, defined as the average of the upper and lower reported distance uncertainties. 

Using these perturbed distances while keeping the sky positions fixed, we recompute the three-dimensional separation between the satellite under study and its observed NN. Repeating this $10^5$ times produces a distribution of possible pair separations consistent with the distance uncertainties. The minimum and maximum separations obtained from this distribution are then compared to the isotropic NN-distance distribution to derive a corresponding range of probabilities, $P_{\mathrm{min}}-P_{\mathrm{max}}$.

For a fair comparison, we also assign mock distance uncertainties to the \textsc{hestia} satellites. We assign each simulated satellite a distance-dependent uncertainty $f(d)$ calibrated to match the errors of the observed sample. The lower and upper heliocentric distance bounds are then defined as
\begin{equation}
d_{\pm} = d \pm \sigma_d,
\end{equation}
where $d$ is the mock heliocentric distance and $\sigma_d = f(d)\, d$. These bounds are projected along the same line of sight and transformed into the corresponding host-centric distances.

\begin{figure}[ht]
    \centering
    \includegraphics[width=0.85\columnwidth]{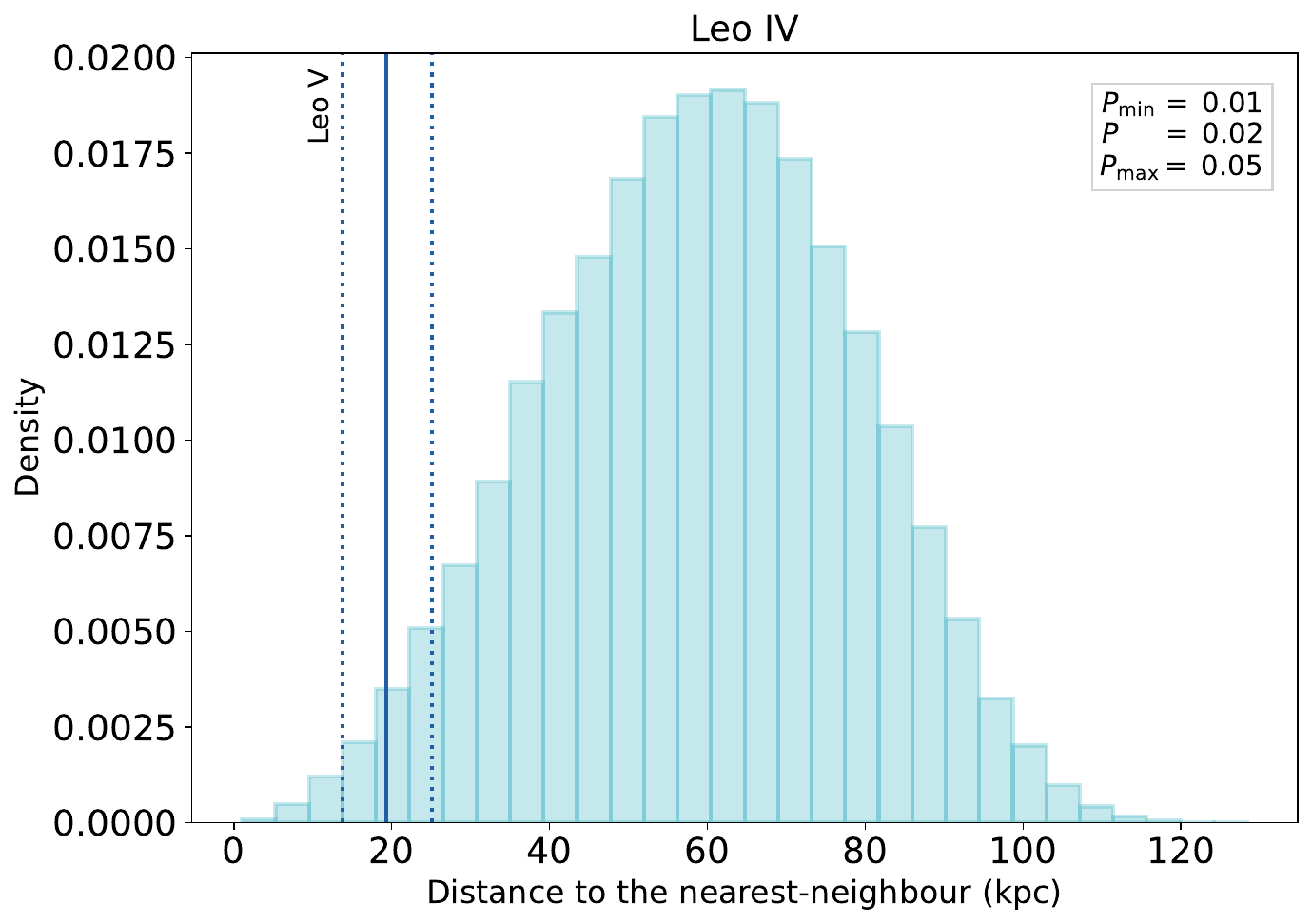}
    \includegraphics[width=0.85\columnwidth]{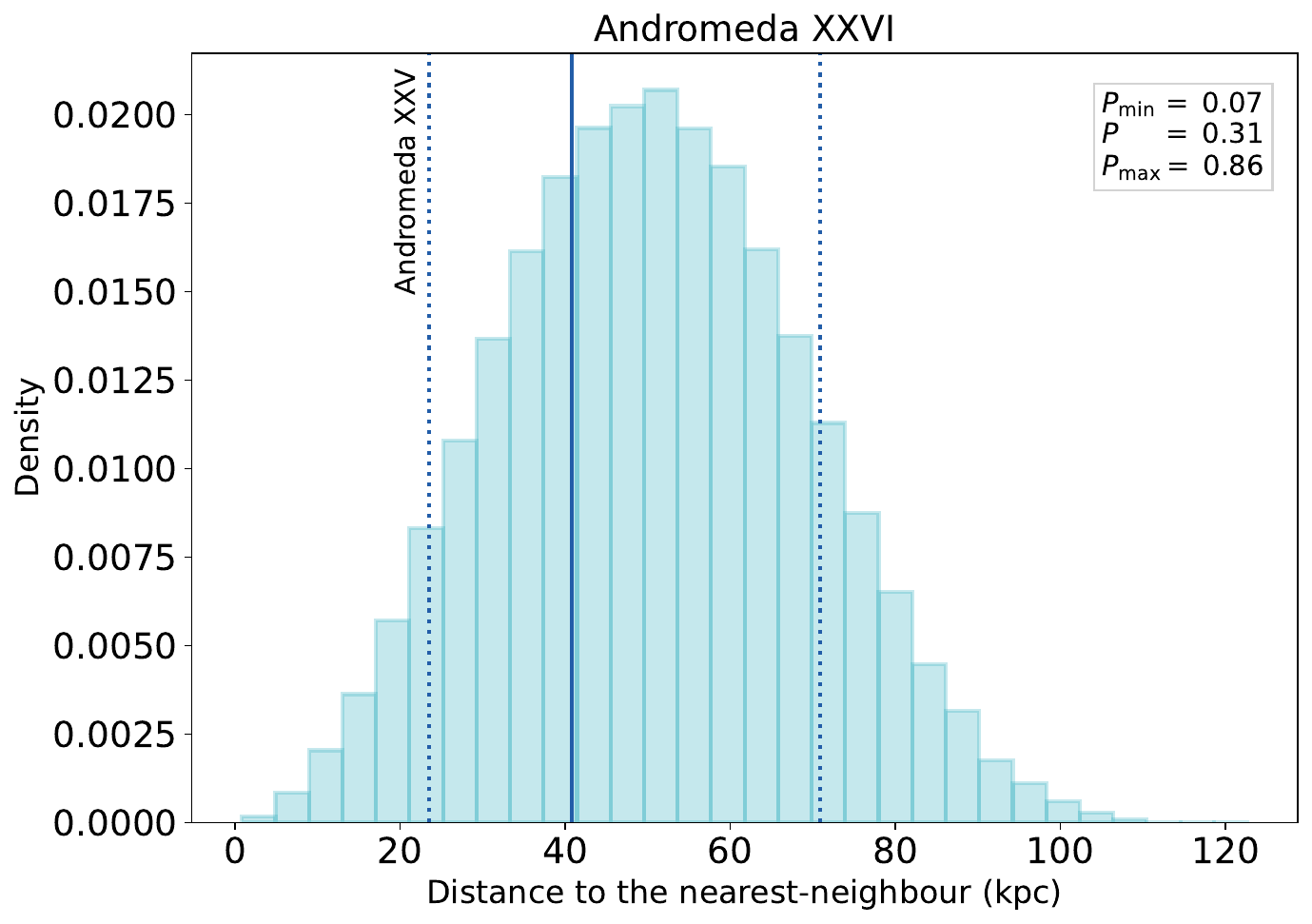}
    \caption{Examples of NN distance distributions obtained from $10^5$ isotropic realisations. The solid line marks the observed NN separation. The dashed lines show the minimum and maximum separations obtained by resampling the heliocentric distances within their uncertainties, corresponding to the range of allowed NN probabilities ($P_\mathrm{min}-P_\mathrm{max}$). Left: Leo IV identified as a highly significant pair candidate with its observed NN (Leo V), whose classification remains robust when distance uncertainties are considered. Right: Andromeda XXVI and its observed NN (Andromeda XXV), for which the inferred probability spans both sides of the $P=0.25$ threshold.}
    \label{fig:nn_distribution_example}
\end{figure}

The distinction between a significant pair and a random association, as well as the impact of including distance uncertainties, is illustrated in Figure \ref{fig:nn_distribution_example}. This figure shows two representative cases: one where the pair classification is robust to the inclusion of distance uncertainties, and another where the inferred significance changes once the uncertainties are taken into account.

We adopt $P < 0.25$ as a selection criterion to identify NN candidates for further analysis. This threshold provides a balance between selecting pairs with small 3D separations and retaining a sufficient number of candidates for subsequent analysis. A stricter threshold would result in too few candidates, potentially excluding real pairs, while a looser threshold would include increasingly distant pairs that are less likely to be dynamically associated. This statistical selection identifies spatially significant associations, but spatial proximity alone does not guarantee a physical association. For these candidates, we further investigate their physical properties for signs of common accretion. Theoretically, pairs that were accreted together should exhibit similar line-of-sight velocities and consistent directions of motion. While such kinematic data are often unavailable in observations, we can also look at their internal properties, such as the stellar mass.

To explore the dynamical state of the identified satellite associations, we use the gravitational binding criterion defined by \citet{geha_local_2010}. For a pair of satellites to be considered gravitationally bound, their mutual gravitational potential energy must exceed their relative kinetic energy \citep{davis_abell_1995}. This relationship is expressed through the dimensionless parameter $b$:
\begin{equation}
b \equiv \frac{2 G M_{\mathrm{pair}}}{\Delta r \Delta v^2},
\label{eq:b}
\end{equation}
where $\Delta r$ represents the total 3D physical separation between the two objects and $\Delta v$ is their relative velocity. In a perfectly bound system, $b > 1$. However, the application of this metric requires different approaches depending on whether it is applied to the simulated or observational datasets, primarily due to the availability of phase-space information and mass constraints.

For the \textsc{hestia} simulation suite, we have access to the full 6D phase-space information (3D positions and 3D velocities) and the total halo masses ($M_{\mathrm{vir}}$) for every subhalo. In this case, we calculate $b$ using the true 3D relative velocity, $\Delta v_{\mathrm{3D}} = \sqrt{\Delta v_x^2 + \Delta v_y^2 + \Delta v_z^2}$. With this, we are able to verify the binding state, allowing us to confirm whether the pairs identified by our $P$-value metric are indeed gravitationally trapped or are merely transient associations.

Conversely, for the majority of the observational sample, 3D velocity information is unavailable as PMs remain sparse or carry significant uncertainties. For these systems, we must rely on the line-of-sight velocity difference ($\Delta v_{\mathrm{los}}$). Because $\Delta v_{\mathrm{los}}$ captures only one component of the full velocity vector, it represents a lower limit of the true relative motion ($\Delta v_{\mathrm{los}} \leq \Delta v_{\mathrm{3D}}$). Consequently, the calculated $b$ factor for observations serves as an upper limit on the degree of binding. Furthermore, because the total dynamical masses of the observed satellites are not well-determined, we treat $M_{\mathrm{pair}}$ as an unknown. We invert the binding criterion to solve for the minimum binding mass ($M_{\mathrm{bound}}$) required to satisfy $b \geq 1$:
\begin{equation}
M_{\mathrm{bound}} = \frac{\Delta r \Delta v_{\mathrm{los}}^2}{2G}.
\label{eq:mbound}
\end{equation}
This $M_{\mathrm{bound}}$ value represents the lower limit of the total system mass necessary for the pair to constitute a bound system. 


\section{Results and discussion}\label{sec:results}

\subsection{Statistical analysis}\label{sec:statistical}
\begin{figure}[ht]
    \centering
    \includegraphics[width=0.9\columnwidth]{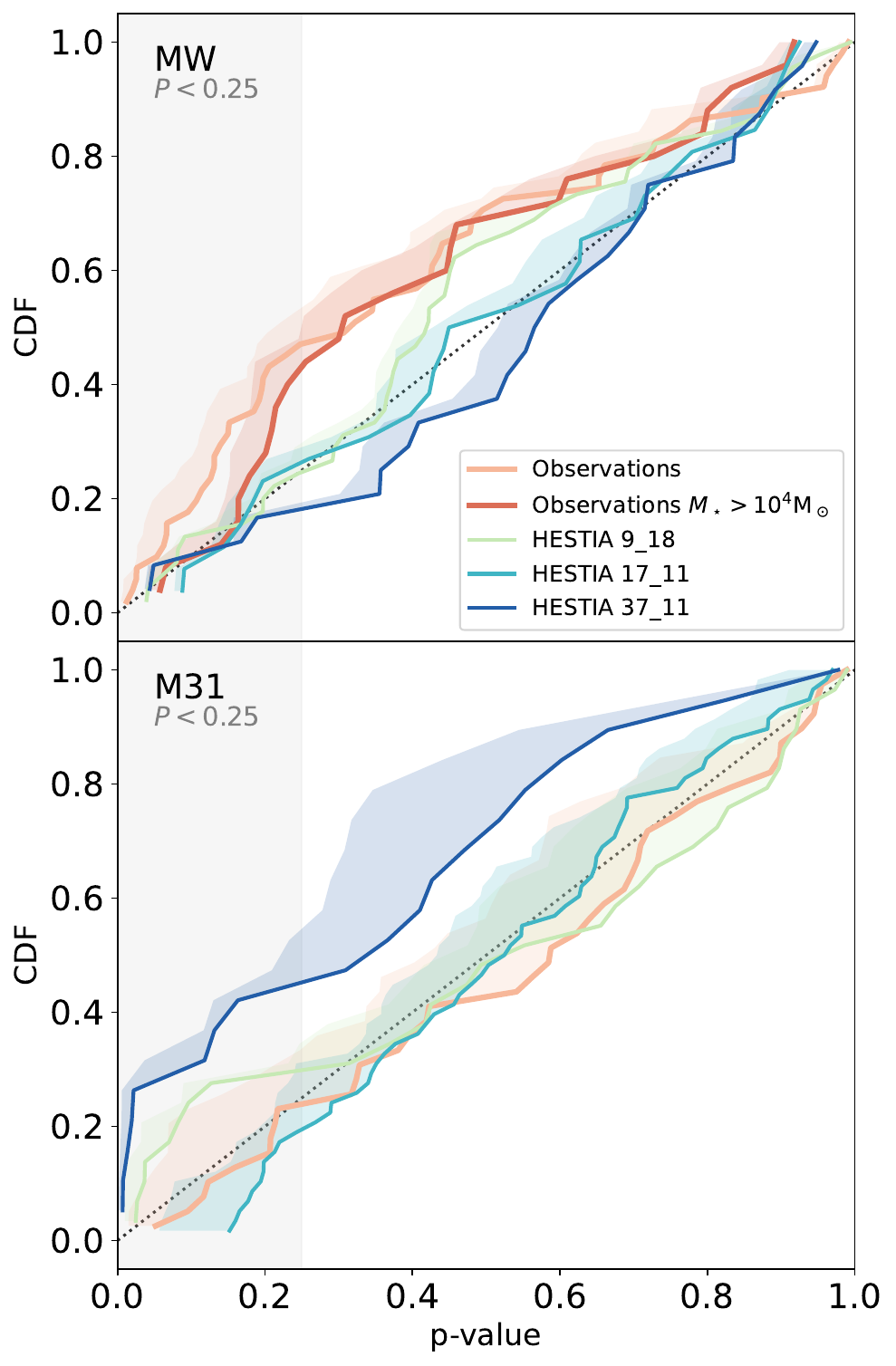}
    \caption{CDF of $p$-values for the NN distances in the MW (top) and M31 (bottom) satellite systems. The solid orange lines represent the full observational samples, while the red lines indicate the resolution-matched samples ($M_\star > 10^4\,\mathrm{M_\odot}$). Results from \textsc{hestia} simulated analogues are shown in shades of blue and green. Shaded regions indicate the most optimistic $P$-value distributions obtained by propagating distance uncertainties and adopting the minimum allowed separations, maximising the number of potential close associations. The black dotted 1:1 line represents the expectation for a perfectly isotropic distribution. The grey region marks the adopted $P < 0.25$ candidate selection threshold.}
    \label{fig:pvalues}
\end{figure}

We first evaluate the spatial clustering of the satellite populations by examining the distribution of $P$-values derived from our NN analysis. We assign a $P$-value to every individual satellite in the sample, which is defined as the probability of a satellite's NN being at the observed distance (or closer) if the satellite population was distributed isotropically. Figure~\ref{fig:pvalues} shows the CDF of these $P$-values for both the observational sample, with and without the stellar mass cut discussed above, and the \textsc{hestia} simulated suites. 

In a perfectly isotropic system, the $P$-value distribution is expected to be uniform, and therefore the CDF should follow the 1:1 diagonal line. Under this null hypothesis, 25\% of satellites are expected to have $P < 0.25$ by chance. An excess above the 1:1 line at low $P$ therefore indicates an excess of unusually close NNs relative to isotropic expectations.

To investigate the effect of distance uncertainties, we Monte Carlo sample the heliocentric distances of each satellite and its observed NN, keeping their sky positions fixed, and recompute their 3D separation. From the resulting distribution, we adopt the lower bound on the separation (corresponding to the smallest plausible distances within the uncertainties) to recompute the NN probability with respect to the isotropic distribution. This gives an optimistic estimate of the probability, $P_{\mathrm{min}}$, corresponding to the maximum number of close associations allowed by the data. These uncertainty ranges are shown as shaded bands around the CDFs in Figure~\ref{fig:pvalues}.

In the MW system (top panel), both the full observational sample and the resolution-matched sample ($M_\star > 10^4\,\mathrm{M_\odot}$) lie significantly above the 1:1 line. When we account for distance uncertainties, our results closely track the CDF that we get without uncertainties, suggesting that current observational errors are not large enough to artificially impact the clustering significance.
This indicates that the MW satellite system looks more clustered than expected from an isotropic distribution across all mass scales. 

Applying the stellar cut to match the simulation resolution limits shifts the observed distribution slightly towards higher $P$-values, reducing the fraction of highly significant close associations. This suggests that the ultra-faint dwarf (UFD) population increases the number of close associations. We note, however, that this signal may be partially driven by observational selection effects. Current surveys of UFDs are limited to specific sky footprints, resulting in uneven coverage. These spatial biases can artificially inflate the apparent angular clustering of faint satellites, even when they are widely separated along the line-of-sight \citep[e.g.][]{cameron_galaxy_2007, metz_discs_2009, pawlowski_alignment_2016, yan_kids-legacy_2025}. Consequently, the increased clustering seen in the full MW sample might partly reflect survey incompleteness rather than intrinsic physical associations.

Compared with the simulations, the observational MW curves lie well above all three \textsc{hestia} analogues. This implies that the real MW contains a higher incidence of close satellite associations than predicted by the simulations, a scenario also seen in \citet{fattahi_galaxy_2013}. The fact that the full sample is the most clustered suggests that the presence of UFDs creates a level of small-scale ``clumpiness'' that current simulations may not fully replicate due to resolution limits. Alternatively, it may point to differences in the accretion history or the survival of low-mass groups in the real MW halo, such as the LMC and its associated satellites \citep{patel_orbital_2020}, or the Crater-Leo group \citep{julio_satellite_2024}.

In contrast, the M31 system (bottom panel) remains broadly consistent with isotropy, showing no significant excess of close pairs. When distance uncertainties are propagated, the shaded band extends toward lower $P$-values, reflecting the larger distance errors of M31 satellites. In the most favourable case, the number of satellites satisfying $P < 0.25$ could increase from 9 to 13. Thus, while additional close associations cannot be excluded, the overall $P$-value distribution of M31 remains much closer to isotropic expectations than the MW. 

However, this behaviour is not reproduced by all the \textsc{hestia} M31 analogues. Two of the simulated M31 systems show substantially stronger clustering, with $P$-value distributions lying well above both the observational curve and the isotropic expectation. This suggests that, in these cases, the simulations overpredict the degree of small-scale ``clumpiness'' around M31.

Interestingly, the 37\_11 realisation, which contains the least massive hosts in our sample, shows the most extreme behaviour. In this case, the M31 analogue is less massive than the MW-like hosts in the other \textsc{hestia} runs, resulting in a smaller satellite population. Its $P$-value distribution instead closely resembles that of the observed MW system, even more so than the simulated MW analogues themselves. This behaviour is illustrated explicitly in Figure~\ref{fig:pvalues_v2} of Appendix~\ref{app:cdfs_opposite_hosts}, where we overplot the simulated CDFs against both observed hosts.

This suggests that the apparent tension in clustering significance may be partially driven by the total mass of the host halo and the resulting number of satellites ($N_{\mathrm{sats}}$), since systems with fewer satellites are statistically less likely to produce extremely small NN separations by chance. At the same time, stochastic variations in the accretion history, group infall, and the local environment of each simulated halo are also expected to contribute to the diversity in clustering signals. A larger suite of simulations spanning a broader range of host masses and assembly histories will therefore be required to robustly assess whether the observed MW--M31 contrast is statistically significant within $\Lambda$CDM.

We also note that the \textsc{hestia} satellite populations are generally more radially concentrated than the observed samples, with most simulated satellites lying within $d_{\mathrm{host}}\lesssim200\,\mathrm{kpc}$. In principle, a more compact satellite distribution would naturally increase the probability of finding small physical separations and could therefore bias NN statistics toward stronger clustering signals. However, this effect is largely accounted for in our methodology, since the isotropic realisations are constructed to reproduce the radial distribution of each individual host separately. The resulting $P$-value distributions therefore measure deviations from isotropy at fixed radial profile, rather than being directly driven by differences in the overall radial extent of the satellite populations.

Taken together, these results highlight a tension in opposite directions for the MW and M31. This divergence might result from differences in observational perspective and selection effects. While the MW satellite census is susceptible to the footprint limitations of deep surveys, the M31 satellite system is observed from the outside, resulting in a more uniform (though luminosity-limited) spatial distribution. In particular, the survey depth, distance completeness, and spatial coverage for the M31 satellite system have been rigorously tested and shown to be remarkably homogeneous across the footprint \citep[e.g.,][]{doliva-dolinsky_pandas_2022, savino_hubble_2022}, confirming that observed spatial features are not due to inhomogeneous search sensitivity. Interestingly, M31 is known to have its own form of spatial anisotropy, a ``lopsided'' distribution where the majority of satellites are located on the side of M31 facing the MW \citep{conn_three-dimensional_2013, savino_hubble_2022, kanehisa_andromedas_2025}. The fact that M31 appears consistent with isotropy in our NN analysis, despite this known lopsidedness, suggests that spatial asymmetries do not necessarily manifest as small-scale clustering or an excess of close pairs.


\subsection{Pair candidates}\label{sec:paircandidates}
While the CDF analysis may identify a statistical departure from isotropy in some cases, it cannot alone distinguish between physical associations and chance spatial proximity. To investigate the nature of these clustering signals, we now focus on the individual satellite associations that fall below our significance threshold of $P < 0.25$. 

Based on this threshold of $P < 0.25$, we identify a set of initial satellite pair candidates for both the observational and simulated samples. To avoid overcounting, we remove duplicate entries where both satellites in a pair were each other's NN, resulting in a census of unique candidate associations. However, it is important to note that these associations are not always two-sided pairs. It is common that several satellites share the same NN. For example, in the full observation MW sample, Tucana II, Tucana IV and the SMC are identified as being associated with Tucana V. This does not change the resulting number of pair candidates, which are summarised in Table~\ref{tab:number_pair_candidates}. It is also possible for an association to be asymmetrically significant: for example, Canes Venatici II is a significant neighbour to Canes Venatici I ($P < 0.25$), but Canes Venatici I is not a significant neighbour to Canes Venatici II ($P > 0.25$) due to differing local density expectations. In our census, we include any association where at least one member of the pair meets the significance threshold.

In the MW, we find 16 pair candidates in the full observational sample. However, this number drops to 7 when applying the $M_\star > 10^4\, \mathrm{M_\odot}$ resolution cut, suggesting that many of the MW pairs are composed of at least one UFD. The M31 sample remains unchanged at 6 candidates, since applying the stellar mass cut does not change the number of satellites associated with this host. This stability is a direct consequence of current observational limits: due to the distance to M31 ($\sim780\,\mathrm{kpc}$), its population of UFDs is undetectable, so the known satellite census of this galaxy is limited to the classical dwarf regime.

Within the \textsc{hestia} suites, the number of pair candidates varies across realisations, ranging from 3 to 7 for MW-like hosts and 5 to 8 for M31-like hosts.

To quantify the degree of phase-space clustering, we calculate the pairing efficiency as the fraction of all considered satellites ($N_{\mathrm{sats}}$) involved in at least one identified association ($N_{\mathrm{assoc}}$). The expected number of associations ($N_{\mathrm{exp,sat}}$) is defined as the number of satellites expected to identify a significant NN ($P < 0.25$) under the null hypothesis of isotropy. While a single physical association (e.g., a reciprocal pair) involves two satellites, $N_{\mathrm{exp,sat}}$ represents the frequency of significant directed associations. The comparison between $N_{\mathrm{assoc}}$ and $N_{\mathrm{exp,sat}}$ therefore tests whether the total number of satellites with unusually close NNs exceeds random expectations, regardless of the clustering configuration (reciprocal or otherwise).

\begin{table*}[ht]
\caption{Summary of satellite pair candidates and pairing efficiencies.}
\label{tab:number_pair_candidates}
\begin{center}
\begin{tabular}{lcccccc} 
\toprule
Dataset & $N_\mathrm{sats}$ & $N_\mathrm{pairs}$ & $N_{P<0.25}$ & $N_\mathrm{exp,sat}\pm \sigma_{N_\mathrm{exp}}$ & $N_\mathrm{assoc}$ & Eff.\% \\ 
\midrule
\textbf{MW} & & & & & & \\
Observations                                        & 51 & 16 & 24 & $12.8 \pm 3.6$ & 27 & 52.9\\ 
$M_\star > 10^4\, \mathrm{M_\odot}$                & 25 & 7  & 10 & $6.5 \pm 2.5$  & 13 & 52.0\\ 
\textsc{hestia} 9\_18                              & 45 & 7  & 11 & $11.3 \pm 3.4$ & 12 & 26.7\\ 
\textsc{hestia} 17\_11                             & 26 & 3  & 6  & $6.5 \pm 2.5$  & 6  & 23.1\\ 
\textsc{hestia} 37\_11                             & 24 & 3  & 6  & $6.0 \pm 2.4$  & 4  & 16.7\\ 
\addlinespace[0.5em] 

\textbf{M31} & & & & & & \\
Observations                                        & 39 & 6  & 9  & $9.8 \pm 3.1$  & 10 & 25.6\\ 
\textsc{hestia} 9\_18                              & 29 & 5  & 8  & $7.3 \pm 2.7$  & 9  & 31.0\\ 
\textsc{hestia} 17\_11                             & 58 & 8  & 14 & $14.5 \pm 3.8$ & 14 & 24.1\\ 
\textsc{hestia} 37\_11                             & 19 & 5  & 5  & $4.8 \pm 2.2$  & 8  & 42.1\\ 
\addlinespace[0.5em] 

\textbf{LG} & & & & & & \\
Observations                                        & 90 & 22 & 33 & $22.5 \pm 4.7$ & 37 & 41.1\\
$M_\star > 10^4\, \mathrm{M_\odot}$                & 64 & 13 & 19 & $16.0 \pm 4.0$ & 23 & 35.9\\
\textsc{hestia} 9\_18                              & 74 & 12 & 19 & $18.5 \pm 4.3$ & 21 & 28.4\\ 
\textsc{hestia} 17\_11                             & 84 & 11 & 20 & $21.0 \pm 4.6$ & 20 & 23.8\\ 
\textsc{hestia} 37\_11                             & 43 & 8  & 11 & $10.8 \pm 3.3$ & 12 & 27.9\\ 
\bottomrule
\end{tabular}
\end{center}

\footnotesize{\textbf{Columns from left to right:} 
(1) Dataset; 
(2) total number of satellites ($N_\mathrm{sats}$); 
(3) number of unique identified candidate associations after removing reciprocal duplicates ($N_\mathrm{pairs}$); 
(4) number of satellites satisfying the significance threshold $P < 0.25$;
(5) expected number of satellites with $P < 0.25$ under isotropy, with Poisson uncertainty ($N_\mathrm{exp,sat}$); 
(6) number of unique satellites involved in at least one association ($N_\mathrm{assoc}$); and 
(7) pairing efficiency ($N_{\text{assoc}} / N_{\text{sats}}$). 

We note that $N_{P < 0.25}$ and $N_\mathrm{exp,sat}$ refer to individual satellites satisfying the significance threshold, whereas $N_\mathrm{pairs}$ counts unique associations after removing reciprocal duplicates. These quantities are therefore not directly equivalent when several satellites share the same NN or when associations are not reciprocal.}
\end{table*}

As shown in Table~\ref{tab:number_pair_candidates}, and as already seen in Figure~\ref{fig:pvalues}, this metric reveals a significant difference between the observational MW and its simulated analogues. In the full MW observational sample, 27 out of 51 satellites are part of at least one association, corresponding to a pairing efficiency of $\sim 53\%$. Even when applying the stellar mass cut, the efficiency remains similarly high at $52\%$ (13 out of 25 satellites).

By contrast, the \textsc{hestia} MW-like hosts exhibit significantly lower efficiencies, ranging from $\sim 17\%$ to $\sim 27\%$. This level of clustering is consistent with expectations from a random isotropic distribution: given our threshold of $P < 0.25$, one would statistically expect approximately one-quarter of any satellites to be identified as ``associated'' by chance. The fact that the simulated hosts generally lie close to this $25\%$ expectation suggests they contain very few, if any, real satellite pairs. By contrast, the elevated pairing efficiency in the MW points to a real excess of close associations beyond random expectations. This indicates that the real MW is almost twice as ``clustered'' as its simulated counterparts, even when the absolute number of pairs ($N_\mathrm{pairs}$) is identical.

Once again, as previously noticed in Figure~\ref{fig:pvalues}, M31 appears more ``typical'' of the \textsc{hestia} environment. Its observational pairing efficiency of $25.6\%$ is broadly consistent with the simulated M31-like hosts (which range from $\sim 24\%$ to $\sim 42\%$). This suggests that the high clustering signal in the LG is primarily driven by the MW's satellite system. 

The number of unique pairs should be interpreted together with both $N_{P < 0.25}$ and $N_\mathrm{assoc}$, since $N_\mathrm{exp,sat}$ refers to individual satellites rather than pair counts. In the full MW sample, 24 satellites satisfy the $P < 0.25$ criterion, compared to $N_\mathrm{exp,sat}=12.8$ expected under isotropy, while 27 unique satellites participate in at least one identified association. This confirms that the MW contains an excess of satellites with unusually close NNs. In contrast, the M31 sample contains 9 satellites satisfying the $P < 0.25$ threshold, fully consistent with the isotropic expectation of $N_\mathrm{exp,sat}=9.8$ for a sample of 39 satellites. Although six unique M31 candidate pairs are identified, the overall incidence of significant NN associations is consistent with random expectations.

As discussed in the previous section, the comparatively large distance uncertainties of M31 satellites could have washed out some genuine close associations. Using the minimum separations permitted by the distance uncertainties, the number of M31 satellites satisfying $P < 0.25$ could increase from 9 to 13. This represents the most favourable case for maximising the clustering signal. Even under this optimistic scenario, the inferred clustering of the M31 satellite system remains substantially weaker than that of the MW and broadly consistent with expectations from an isotropic distribution.

When considering the LG as a whole, the observational pairing efficiency ($\sim 41\%$ for the full sample) consistently exceeds the values found across all \textsc{hestia} realisations ($\sim 24\%$ to $\sim 28\%$). This may suggest that the observed LG is currently likely dominated by recent, coherent infall of satellite groups.

These numbers represent the maximum possible number of associations based purely on spatial clustering. In the following sections, we move beyond simple spatial proximity to examine other indicators of physical association for each of these candidates. A true joint accretion event should leave traces in the satellites' broader phase-space distribution \citep[e.g.][]{lynden-bell_ghostly_1995}. \citet{fattahi_galaxy_2013} concluded that physically associated satellites should have small line-of-sight velocity differences and comparable luminosities. We follow the same logic, but use stellar mass rather than luminosity, since this quantity is available consistently for both the observations and the \textsc{hestia} simulations. 

Taking this into account, we evaluate these candidates based on: 
\begin{itemize}
    \item \textbf{Mass similarity:} associations of satellites typically exhibit comparable masses;
    \item \textbf{Kinematic coherence:} physical pairs are expected to share similar line-of-sight velocities;
    \item \textbf{Motion alignment:} when PMs are available, we examine the alignment of the velocity vectors;
    \item \textbf{Gravitational binding:} we calculate the minimum dynamical mass required for a pair to be bound and compare it to the expected total mass of the system.
\end{itemize}


\subsubsection{Mass similarity}\label{sec:pair_masses}

\begin{figure}[ht]
    \centering
    \includegraphics[width=\columnwidth]{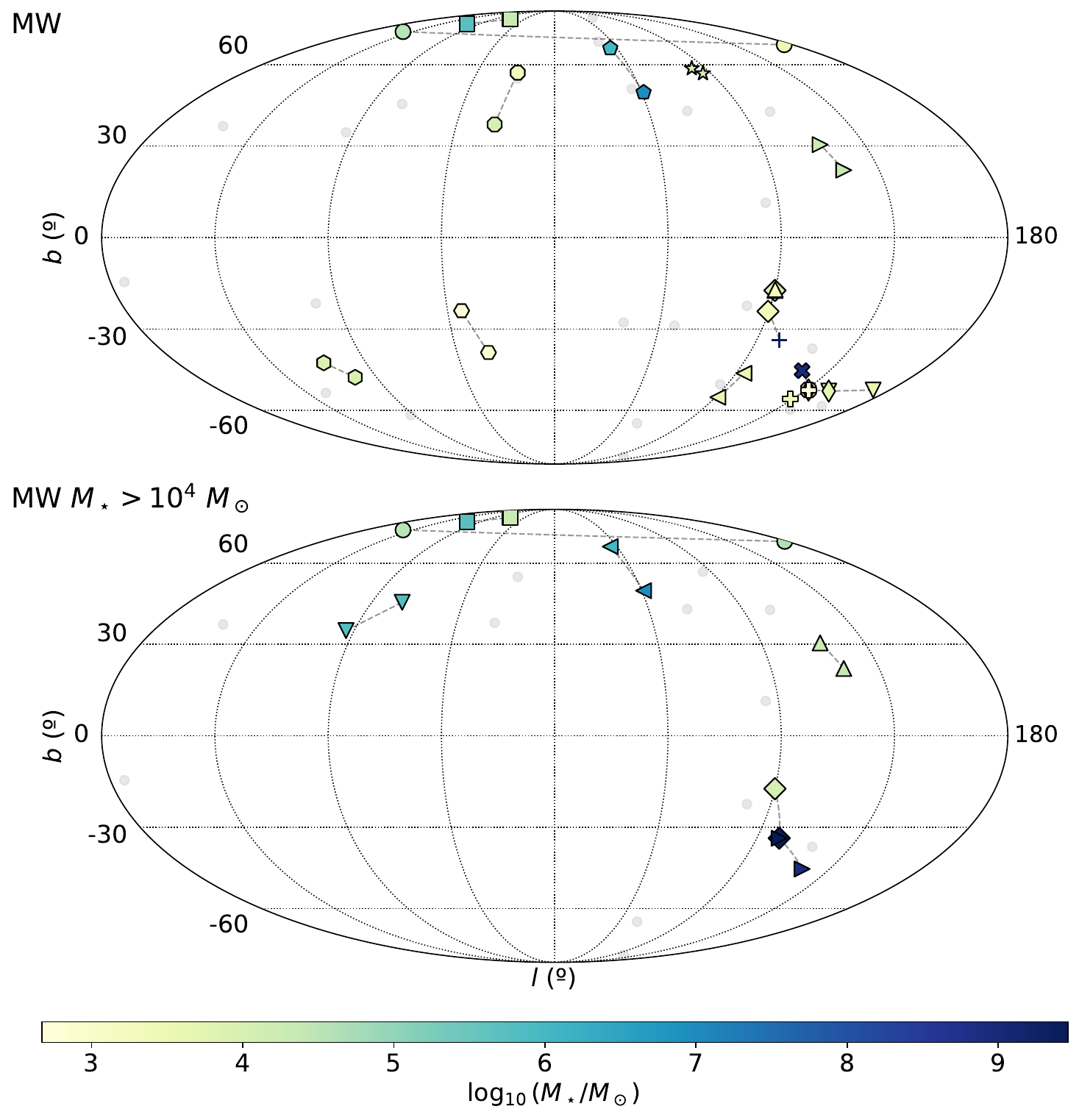}
    \includegraphics[width=\columnwidth]{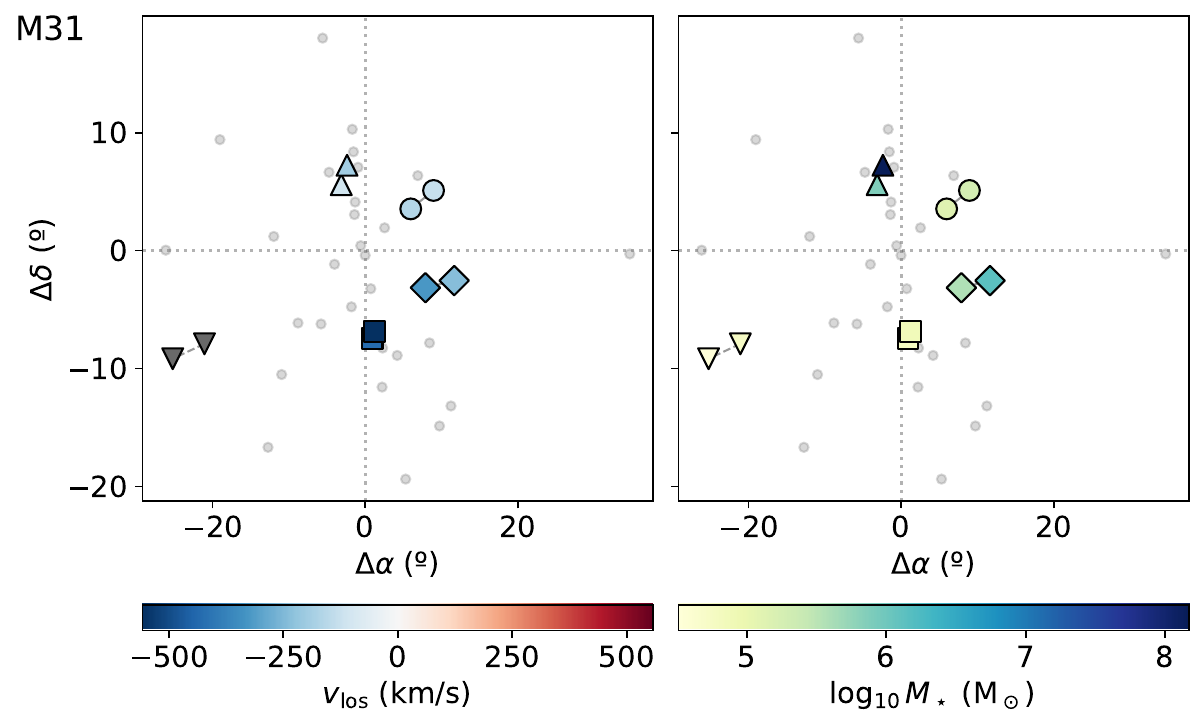}
    \caption{Comparative sky distribution and properties of satellite galaxy pairs in the MW and M31. The top panels show the distribution of MW satellites in Galactic coordinates ($l, b$) for the full sample (top) and the stellar-mass-cut sample ($M_\star > 10^4\,\mathrm{M}_\odot$, centre). The satellites in pairs are colour-coded by their stellar mass, $\log_{10}(M_\star/\mathrm{M}_\odot)$. The bottom panels display the M31 satellite distribution host-centric relative coordinates ($\Delta\alpha, \Delta\delta$). In the left panel, M31 pairs are colour-coded to indicate their relative line-of-sight velocity ($v_\mathrm{los}$), and in the right panel by their stellar mass. In all panels, satellites identified as pairs are connected by dashed lines and shared marker shapes, while the remaining satellites are represented as faint grey dots. Markers with a dark grey colour indicate satellites where the respective data is unavailable.}
    \label{fig:pairs_vlos_mass}
\end{figure}

To visualise the members of the identified satellite pairs and their corresponding physical properties, we show their distribution in Figure~\ref{fig:pairs_vlos_mass}. The MW samples are presented in Galactic coordinates ($l, b$), while the M31 sample is shown in host-centric relative coordinates ($\Delta\alpha, \Delta\delta$). Each connected symbol represents the two satellites belonging to a pair.

For the MW (top and centre panels), the pairs are broadly distributed across the sky, although a concentration of low-mass associations can be seen around the region of the Magellanic system. Although the full sample does not indicate that all identified pairs originate from a single accretion event, the spatial distribution suggests that a fraction of the excess clustering may be related to the recent infall of the LMC group. The role of these low-mass associations in shaping the pair population, and their connection to the Magellanic system, is discussed below.

We distinguish between ``major'' and ``hierarchical'' pairs based on their stellar mass ratios. We define ``major'' pairs as those consisting of satellites with comparable stellar masses, corresponding to $M_{\star,1}/M_{\star,2}<10$, while ``hierarchical'' pairs have a mass ratio $M_{\star,1}/M_{\star,2}>10$. This definition is adopted to distinguish pairs with broadly comparable masses from those with a substantially more massive primary and a lower-mass companion. In the full MW sample, approximately 69\% of our identified associations are major pairs, suggesting that our phase-space metric is highly effective at identifying satellites of comparable mass ($M_{\star,1}/M_{\star,2} \approx 1.0$ in Table~\ref{tab:properties_pair_candidates}). The remaining 31\% are hierarchical associations, which include extreme cases such as the LMC--Pictor~II and SMC--Tucana~V systems, where the mass ratio exceeds $10^6$. This is consistent with hierarchical structure formation: while some satellites seem to be accreted in pairs of similar mass, many seem to be part of larger groups dominated by a massive companion that has retained its own low-mass satellites \citep[see e.g.][]{donghia_small_2008}.

When the stellar mass cut is applied, the pair distribution changes, but the population remains dominated by mass symmetry. As indicated by the bolded entries in Table~\ref{tab:properties_pair_candidates}, only a subset of the original associations survive this threshold. These surviving systems are characterised by relatively low mass ratios (approximately 71\% of associations qualify as ``major'' pairs), such as the Draco--Ursa Minor and LMC--SMC systems. In contrast, the ``hierarchical'' pairs that are present in the full sample (typically consisting of an UFD associated with a more massive companion) are largely removed by this cut, as it excludes the faintest systems, which are responsible for these extreme mass ratios.

The fact that these hierarchical pairs disappear once the stellar mass cut is applied highlights that the clustering signal is strongly mass-dependent, and may significantly influence comparisons with both simulations due to resolution limits and M31, since its observations are restricted to relatively bright systems due to its distance. As a result, both cases are naturally biased towards detecting only higher-mass, ``major'' pairs, potentially underestimating the true abundance of hierarchical associations present in the underlying satellite population.

This interpretation is further supported by the spatial distribution of the identified pairs on the sky. In the full MW sample, a significant fraction of the low-mass associations are concentrated in the same region of the sky as the Magellanic Clouds (see Figure~\ref{fig:pairs_vlos_mass}), forming a visually apparent clustering of satellites around the LMC--SMC system. Many of these associations disappear once the stellar mass cut is applied, indicating that part of the excess clustering signal is driven by UFDs spatially associated with the Magellanic system. This behaviour is consistent with hierarchical accretion, in which a massive satellite such as the LMC brings its own population of satellites into the MW halo \citep[e.g.][]{donghia_small_2008, patel_orbital_2020, vasiliev_effect_2023}. However, the excess of close pairs remains significant even after applying the stellar mass cut, suggesting that the observed clustering cannot be explained solely by the faintest Magellanic companions. If the Magellanic system is indeed on a second pericentric passage, as suggested by \citet{vasiliev_dear_2024}, some of these companions may already have been tidally stripped from the LMC while still retaining correlated phase-space signatures. The observed excess of close associations in the MW may therefore partly reflect the accretion and subsequent disruption of the Magellanic group, as well as other infalling associations such as the Crater--Leo group \citep{julio_satellite_2024}.

For M31 (bottom panels), the number of identified pairs is smaller, and they appear more spatially concentrated due to the limited angular extent of the system on the sky. Despite this, the same qualitative trends are present. The velocity-coloured panel (bottom-left) suggests that some pairs have comparable line-of-sight velocities, although the small sample size limits any strong conclusions. The stellar mass distribution (bottom-right), similar to the MW, indicates a preference for ``major'' associations, with approximately 83\% (5 out of 6) of the pairs exhibiting mass ratios $M_{\star,1}/M_{\star,2} < 10$. However, as current observational data for M31 are largely restricted to relatively bright systems, the lack of hierarchical pairs in the M31 sample is expected.

The presence of hierarchical associations does not necessarily contradict the previous finding of \citet{fattahi_galaxy_2013} that satellite pairs tend to have similar luminosities. Their analysis was based on a brighter satellite population ($M_V \leq -8$) and a substantially smaller census of LG dwarfs, prior to the discovery of many UFDs. The inclusion of these recently discovered low-mass systems extends the identified pair population to larger stellar mass ratios, revealing associations between more massive dwarfs and their faint companions.

\subsubsection{Kinematic coherence}\label{sec:pair_kinematics}
\begin{figure*}[ht]
  \begin{minipage}[c]{0.62\textwidth}
    \includegraphics[width=\textwidth]{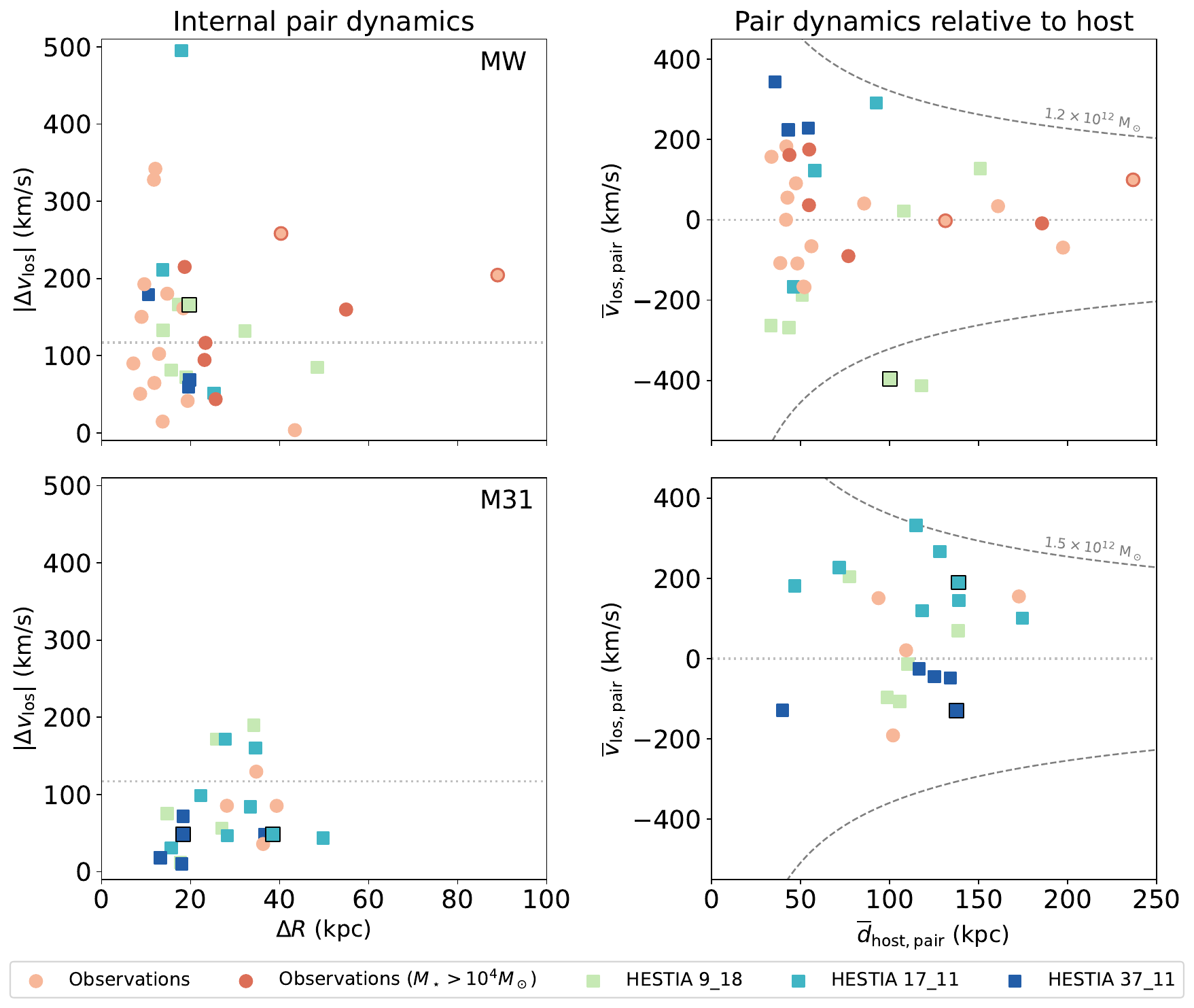}
  \end{minipage}\hfill
  \begin{minipage}[c]{0.37\textwidth}
        \caption{Phase-space diagnostics of satellite pair candidates in the MW (top) and M31 (bottom), compared to \textsc{hestia} analogues. Observed pairs are shown as circles and simulated pairs as squares; gravitationally bound systems are highlighted with black edges, while circles with red edges indicate observed pairs present in both the full and $M_\star > 10^4,\mathrm{M}\odot$ samples. Left: Internal pair dynamics, showing the 3D separation ($\Delta R$) versus line-of-sight velocity difference ($|\Delta v_{\mathrm{los}}|$). The horizontal dashed line at $117\,\mathrm{km\,s}^{-1}$ marks the present-day LMC--SMC velocity difference, adopted as the threshold for potential dynamical association. Right: Pair location and host-relative motion, showing the pair-averaged host distance ($\overline{d}_{\mathrm{host, pair}}$) versus pair-averaged line-of-sight velocity relative to the host ($\overline{v}_{\mathrm{los, pair}}$). Grey dashed curves indicate the escape velocity for the assumed host masses ($1.2 \times 10^{12}\,\mathrm{M_\odot}$ for the MW and $1.5 \times 10^{12}\,\mathrm{M_\odot}$ for M31) and are shown for reference only, as the plotted velocities are not full three-dimensional host-centric velocities.}
    \label{fig:kinematics_diagnostic}
  \end{minipage}
\end{figure*}

The phase-space distribution of the identified satellite pairs is shown in Figure~\ref{fig:kinematics_diagnostic}, where we compare the observational pair candidates of the MW (top panels) and M31 (bottom panels) against their respective analogues in the \textsc{hestia} simulations.

In the left-hand panels, which show the internal separation ($\Delta R$) versus the line-of-sight velocity difference ($|\Delta v_\mathrm{los}|$), both systems exhibit pairs that are predominantly clustered at small separations ($\Delta R < 40\,\mathrm{kpc}$). In the MW, the higher-mass subset (red points) spans a broader region of the phase-space than the low-mass population. In contrast, the low-mass satellites (represented by the orange points without coloured edges) are predominantly found at smaller internal separations ($< 50\,\mathrm{kpc}$). 

This concentration of low-mass satellites may again be influenced by observational selection effects. UFDs are increasingly difficult to detect at large Galactocentric distances, leading to a more radially concentrated observed sample. This higher spatial density in the inner halo increases the probability of finding satellites in close neighbours, which may artificially boost the clustering signal for the low-mass systems. Nevertheless, several observed pairs in both the MW and M31 occupy the region of small velocity differences. Following \citet{fattahi_galaxy_2013}, we use the line-of-sight velocity difference of the LMC--SMC pair as a reference scale when evaluating kinematic similarity. Using the systemic heliocentric velocities reported in the Local Volume Database \citep{pace_local_2025}, the present-day line-of-sight velocity difference between the LMC and SMC is $|\Delta v_\mathrm{los}| \sim 117\,\mathrm{km\,s}^{-1}$ (compared to $\sim 75\,\mathrm{km\,s}^{-1}$ used in \citealt{fattahi_galaxy_2013}). We note that this value is used as an upper limit for identifying satellite pairs with relatively small line-of-sight velocity differences, as the LMC and SMC are substantially more massive than the other satellite pairs.

Adopting this updated value as a less conservative threshold, we find that approximately 44\% (7 out of 16) of the full MW sample and 43\% (3 out of 7) of the resolution-matched sample exhibit velocity differences below this limit. In M31, 75\% (3 out of 4) of the pairs with available velocity measurements satisfy this criterion. Therefore, these systems represent the most promising candidates for true dynamically associated pairs, although full phase-space information is required to confirm whether they are truly bound.

A similar fraction of low $|\Delta v_\mathrm{los}|$ pairs is found in the \textsc{hestia} simulations (20 out of 31 across hosts, corresponding to approximately $65\%$), indicating that small line-of-sight velocity differences can arise naturally in cosmological satellite populations. This reinforces the need to combine kinematic information with spatial clustering when identifying genuinely associated systems.

The right-hand panels show the pairs' average distance from the host ($\overline{d}_\mathrm{host}$) against their average line-of-sight velocity relative to the systemic motion of the host ($\overline{v}_{\mathrm{los}}$). For the MW, the observational pairs are mostly concentrated at small host-centric distances ($\overline{d}_{\mathrm{host}} \lesssim 100\,\mathrm{kpc}$). In M31, the pairs are distributed further from the host on average, both in the observations and simulations, with most pair candidates around $100-150\,\mathrm{kpc}$. We also note that two M31 satellites (Pegasus V and Pegasus VII) lack measurements of $v_{\mathrm{los}}$, which prevents a complete characterisation of all pairs. As a result, the trends identified here likely represent a conservative view of the true level of kinematic coherence within the satellite pair population.

For the observations, the MW velocities are corrected to the Galactic standard-of-rest frame, while the M31 velocities are measured relative to the systemic velocity of M31. In the simulations, we place a mock observer at the Solar position and measure the line-of-sight velocities in an equivalent way, ensuring a consistent comparison between the observations and the mock-observed simulated systems. We emphasise that these quantities do not correspond to true three-dimensional host-centric radial velocities, since only the line-of-sight component of the motion is considered.

\subsubsection{Motion alignment}\label{sec:pair_pms}
While the line-of-sight velocities and 3D separations provide a statistical basis for identifying pairs, a more robust test of physical association lies in their three-dimensional velocity. However, proper motion measurements remain sparse in the LG and are often subject to high uncertainties. In the M31 system, PMs have been successfully derived for only a handful of the brightest satellites \citep[e.g.][]{sohn_hst_2020, casetti-dinescu_hst_2024, casetti-dinescu_hst_2025}. For the M31 pair candidates identified in this work, we currently lack PM data for both members of the pair. This prevents a direct comparison of their orbital paths, which would be required to confirm a common infall origin or a gravitational bond. Consequently, our kinematic assessment for the M31 system remains limited to the line-of-sight component.

\begin{figure*}[ht]
    \centering
    \includegraphics[width=\textwidth]{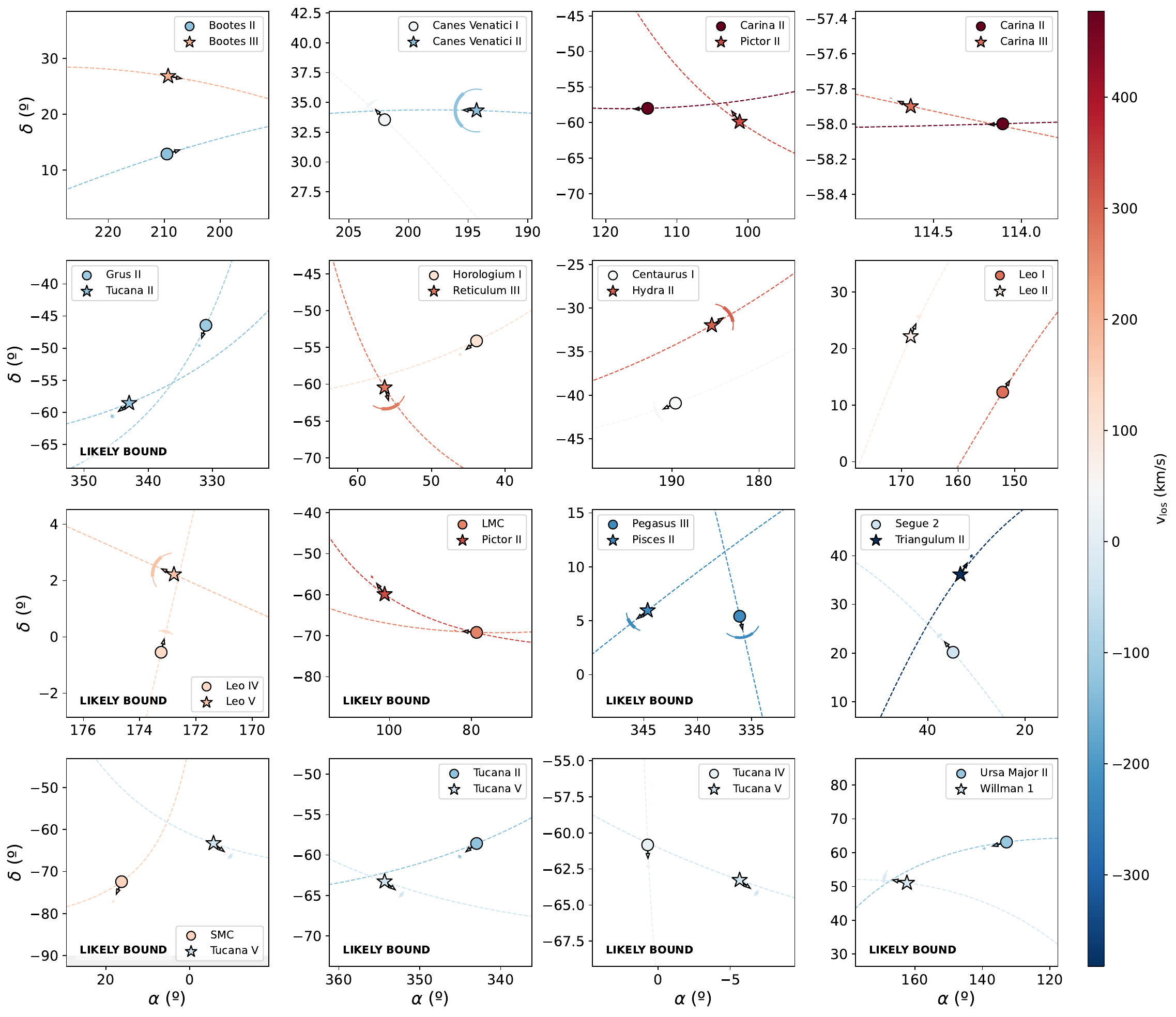}
    \caption{Spatial distribution and direction of motion of candidate satellite pairs with clustering significance $P < 0.25$ for the full observational sample of MW satellites. Each panel shows one pair, with markers coloured by $v_\mathrm{los}$. Circles denote the first satellite in the identified pair and stars its NN. Arrows indicate the inferred most likely PM direction after correcting for the Solar reflex motion, dashed curves show the corresponding great-circle trajectories, and the arcs indicate the $1\sigma$ and $2\sigma$ uncertainties. Grey regions mark $| \delta | > 90^\circ$ to denote regions beyond the physical limits of the equatorial coordinate system. Pairs labelled ``LIKELY BOUND'' have a minimum binding mass consistent with a plausible total dynamical mass for the system.}
    \label{fig:pairs_pm_mw}
\end{figure*}

\begin{figure*}[ht]
    \centering
    \includegraphics[width=\textwidth]{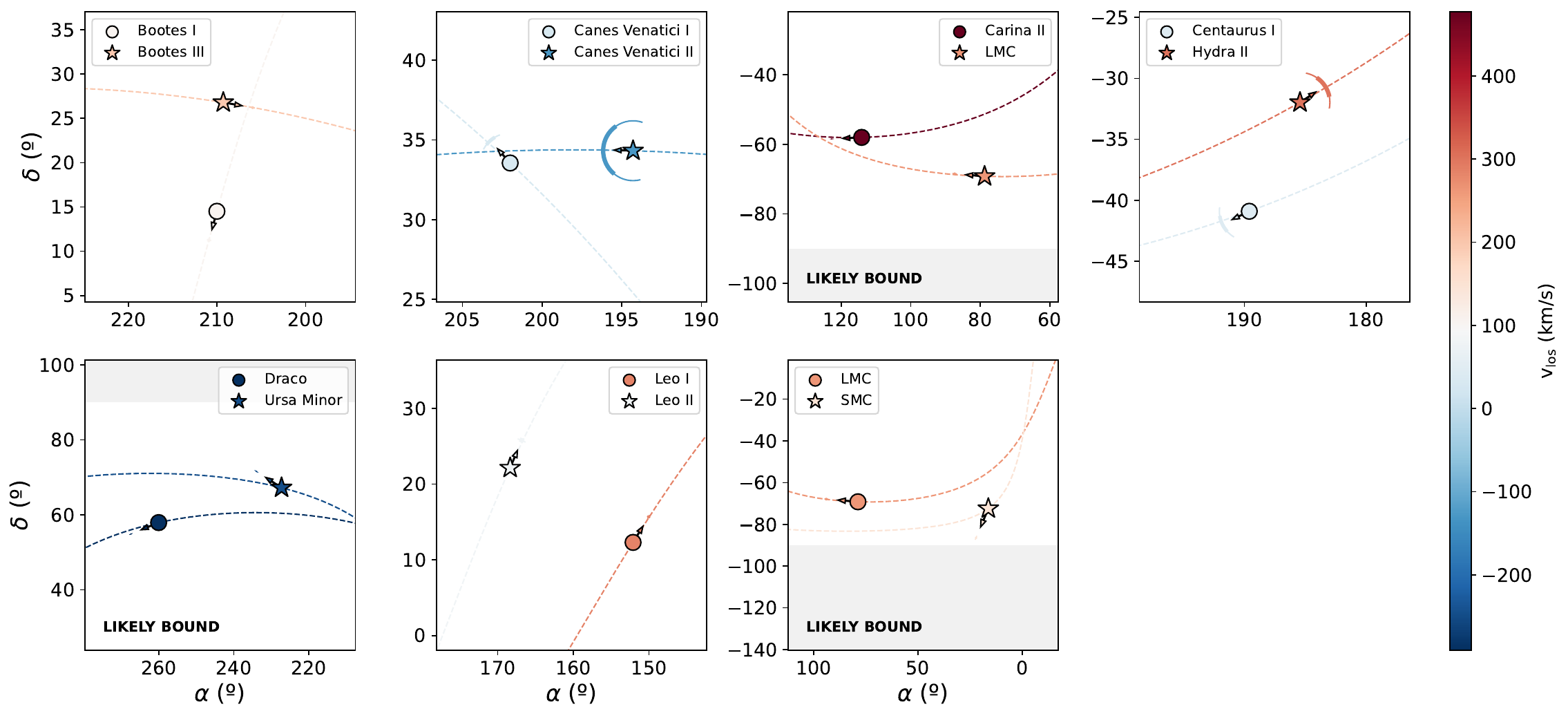}
    \caption{Same as Figure~\ref{fig:pairs_pm_mw}, but for the resolution-matched MW sample.}
    \label{fig:pairs_pm_mw_smcut}
\end{figure*}

\begin{figure*}[ht]
    \includegraphics[width=\textwidth]{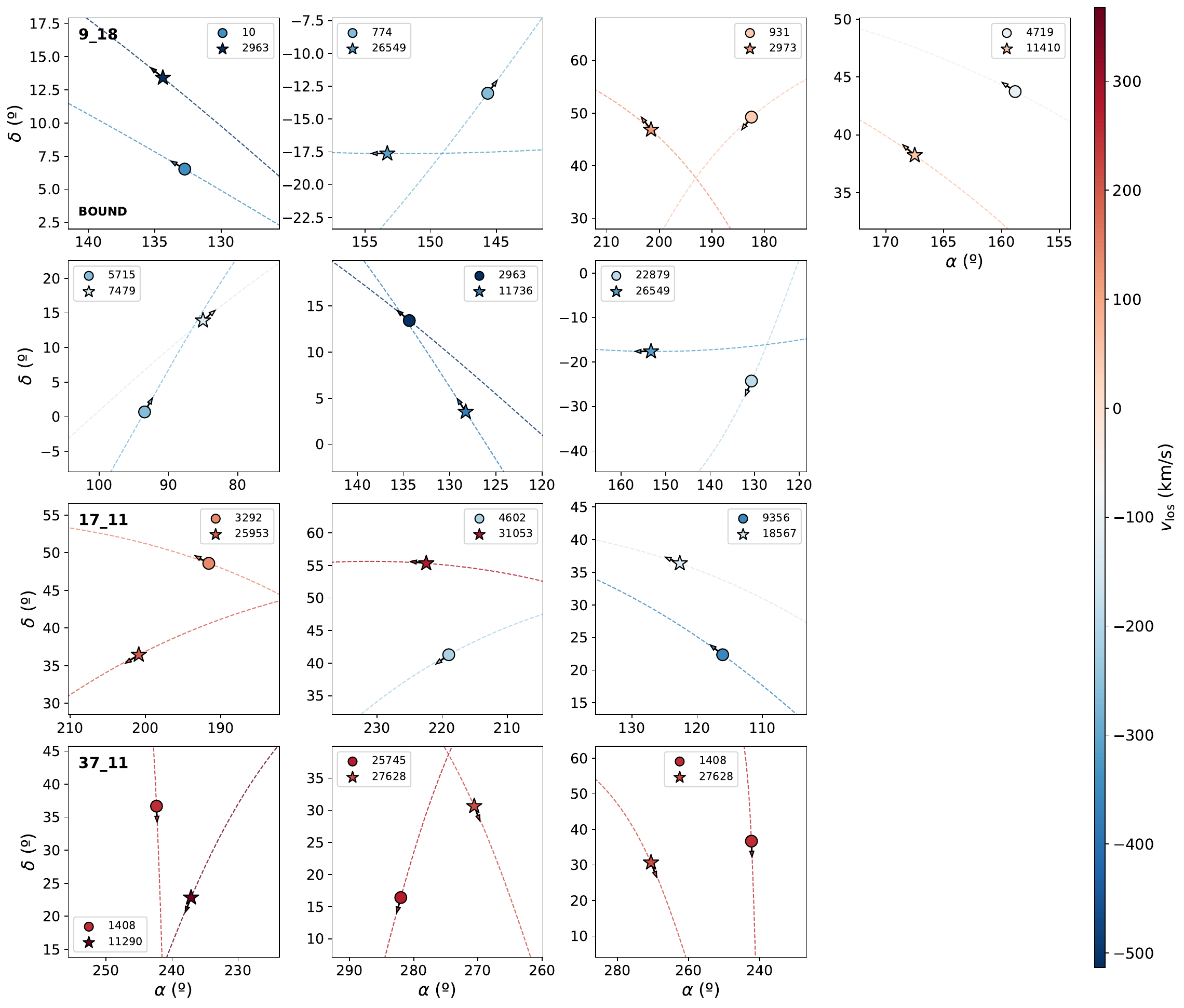}
    \caption{Same as Figure~\ref{fig:pairs_pm_mw}, but for the \textsc{hestia} MW analogues. Arrows indicate the PM direction as seen by a mock observer placed at the Solar position in the simulated MW analogue. Pairs labelled ``BOUND'' are gravitationally bound according to their full three-dimensional phase-space information.}
    \label{fig:pairs_pm_sims_mw}

\end{figure*}

\begin{figure*}[ht]
    \centering
    \includegraphics[width=\textwidth] {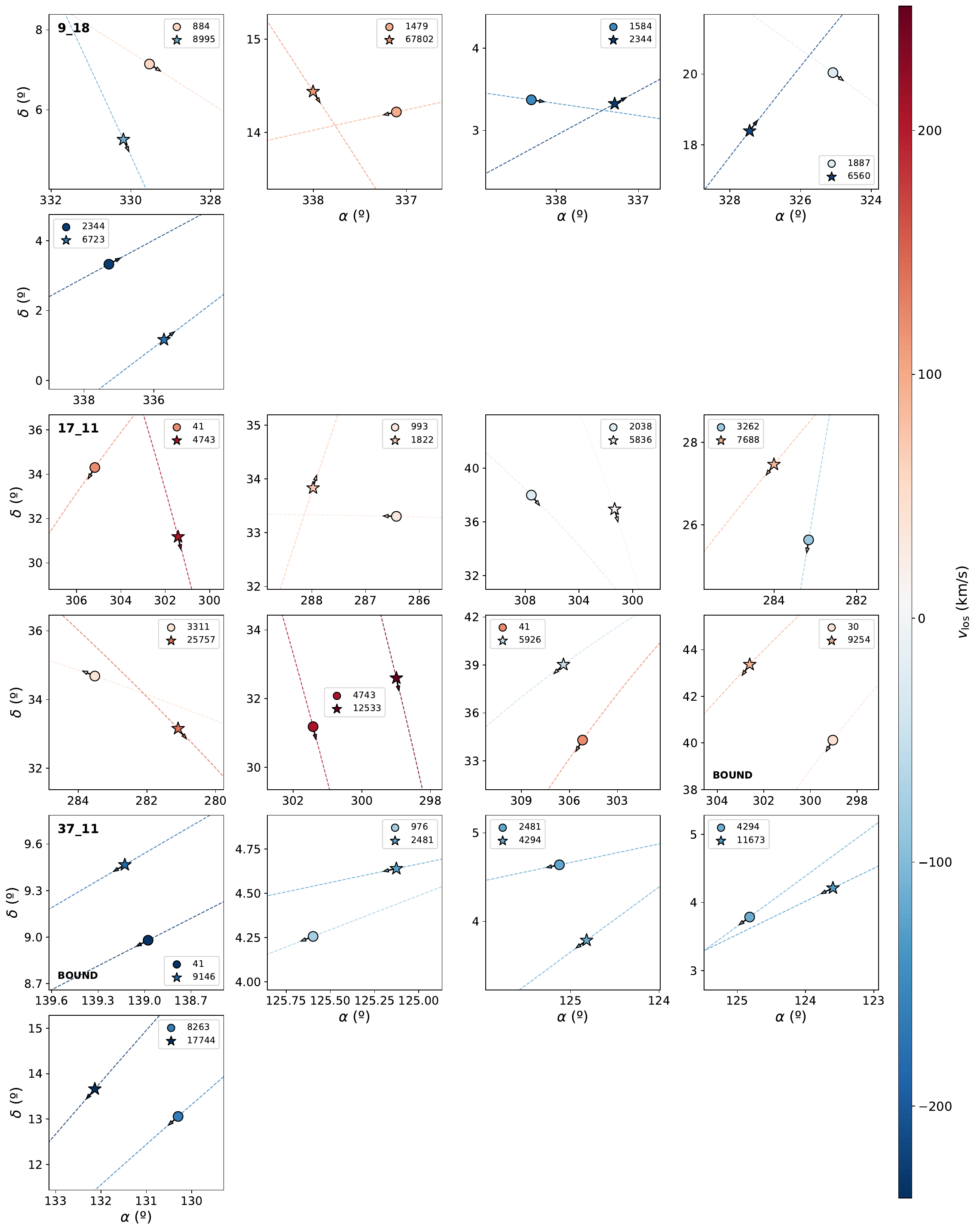} \\
    \caption{Same as Figure~\ref{fig:pairs_pm_sims_mw}, but for the \textsc{hestia} M31 analogues. Arrows indicate the PM direction as seen by a mock observer located at the position of the simulated MW analogue relative to M31.}
    \label{fig:pairs_pm_sims_m31}
\end{figure*}

In Figures~\ref{fig:pairs_pm_mw} and~\ref{fig:pairs_pm_mw_smcut}, we show the spatial distribution and direction of motion for pairs identified in both observational samples of the MW with high clustering significance ($P < 0.25$). Figures~\ref{fig:pairs_pm_sims_mw} and~\ref{fig:pairs_pm_sims_m31} show, respectively, the kinematics for the MW and M31 analogues in the \textsc{hestia} simulations. A qualitative classification of the motion alignment for each pair is provided in Appendix~\ref{app:kinematic_coherence}.

These figures illustrate the most likely direction of motion in the $(\alpha, \delta)$ plane. For the observational MW samples, arrows indicate the most likely proper motion direction inferred from the measured PMs after correcting for the Solar reflex motion. The associated dashed curves show the great-circle tracks on the sky defined by these velocity directions, corresponding to the orbital planes of the satellites. The arcs centred on each satellite illustrate the $1\sigma$ and $2\sigma$ uncertainties in the inferred velocity directions.

Several pairs show velocity vectors that become consistent once their uncertainties are taken into account. When we consider the full MW sample (Figure~\ref{fig:pairs_pm_mw}), we find that $\sim 44\%$ of the identified pair candidates exhibit velocity vectors that are statistically consistent within their $2\sigma$ uncertainties. For example, systems such as Canes Venatici I--Canes Venatici II, Leo IV--Leo V, and Pegasus III--Pisces II exhibit possible overlapping directions, indicating that their motions on the plane of the sky are compatible with a common trajectory within the current uncertainties. The association of Leo IV and Leo V is of particular interest, as their physical connection has been previously suggested based on their extreme spatial and kinematic proximity \citep{belokurov_cats_2007, jong_enigmatic_2010}, and was recently confirmed as a highly significant group in the dynamical analysis of \citet{julio_satellite_2024}. Similarly, the kinematic alignment of Pegasus III and Pisces II reinforces previous suggestions that these two UFDs form a physically bound pair accreted together \citep{kim_heros_2015, richstein_structural_2022}. In contrast, other pairs (e.g. Carina II--Pictor II) have really tight constraints on their PMs and show less overlap, suggesting that not all spatially close systems are kinematically coherent.

When the $M_\star > 10^4\, \mathrm{M_\odot}$ cut is applied (Figure~\ref{fig:pairs_pm_mw_smcut}), some of the most significant pairs based on 3D distance and $v_{\mathrm{los}}$ appear to show divergent directions of motion in the $(\alpha, \delta)$ plane. In particular, the LMC--SMC and Draco--Ursa Minor systems appear to be moving in different directions when considering only the projected proper motion vectors. However, this apparent misalignment arises from projecting curved trajectories on the celestial sphere onto the $(\alpha, \delta)$ plane. Proper motion vectors represent only the local tangent to the orbit, while the true motion of a satellite follows a great circle corresponding to its orbital plane. When these great-circle trajectories are considered, both the LMC--SMC and Draco--Ursa Minor pairs are found to be consistent with coherent orbital motion, despite their apparently divergent projected velocity vectors. For this subset sample of the MW, approximately $70\%$ of the pair candidates show velocity vectors that are statistically consistent within their $2\sigma$ uncertainties.

A case of high-confidence association is found in the Carina II--LMC pair, which shows near-perfect alignment in our analysis. This provides strong kinematic evidence that Carina II is a member of the Magellanic satellite system, a conclusion supported by several recent studies characterising LMC satellites brought in during the system's first infall \citep{fritz_gaia_2018, kallivayalil_missing_2018, patel_orbital_2020, battaglia_gaia_2022, pace_proper_2022}. We also identify Carina II--Carina III as pair candidates with significant kinematic overlap with the LMC system in our full sample, further illustrating the ``clumpliness'' of the Magellanic accretion event.

A similar behaviour is observed in the simulated host analogues (Figures~\ref{fig:pairs_pm_sims_mw} and \ref{fig:pairs_pm_sims_m31}). In these figures, arrows indicate the proper motion direction of the satellites as seen by the mock observer defined in the simulation, while the dashed curves again show the corresponding great-circle tracks. Unlike the observational case, these directions are derived directly from the simulated velocities and are therefore shown without uncertainties. A subset of pairs (14 of 31 candidates across all hosts, corresponding to approximately $45\%$) exhibit a clear alignment in their projected velocity vectors, indicating that coherent satellite motions arise naturally in cosmological simulations.

The presence of directional coherence in both the observations and simulations suggests that at least a fraction of the identified pairs are not chance projections or transient encounters, but instead trace satellites sharing a common orbital motion. When combined with the small line-of-sight velocity differences and spatial proximity identified in previous sections, this suggests that some of these systems are likely to represent genuine physical associations, rather than chance alignments or fly-by events.

\subsubsection{Gravitational binding}\label{sec:bound_pairs}

To move beyond the statistical likelihood of association, we now check the physical nature of the identified pairs using the dimensionless binding parameter $b$ (Equation~\ref{eq:b}) for the simulations where we have the 3D velocity vectors and the minimum binding mass, $M_\mathrm{bound}$ (Equation~\ref{eq:mbound}), for the observations. $M_{\mathrm{bound}}$ represents the lower limit of the total system mass required for the pair to be gravitationally bound at present time, given its current phase-space configuration. These results are presented in the last column of Table~\ref{tab:properties_pair_candidates} and~\ref{tab:properties_pair_candidates_simulations} for the observations and simulations, respectively.

When we apply the 3D binding criterion ($b > 1$) to the full \textsc{hestia} sample, we find that only a small fraction of the identified phase-space associations are currently gravitationally bound. Out of the numerous associations identified by our $P$-value metric (31 across hosts and runs), only three systems satisfy the requirement for gravitational binding: one in the MW 9\_18 analogue, and two in the M31 analogues (17\_11 and 37\_11). Our results indicate that approximately 10\% of candidate phase-space associations correspond to bound pairs at the present day. 

This result is consistent with the expectation that $\sim 25\%$ of phase-space associations arise from chance alignments rather than physical proximity. Furthermore, it is possible that some pairs were bound prior to infall and were disrupted by the host's tidal forces. Most ``true'' pairs likely enter the host as part of a bound group but are rapidly transformed into ``association'' objects that share a common orbital trajectory and phase-space signature but are no longer formally bound to one another.

Interestingly, all three bound systems identified in the \textsc{hestia} simulations also exhibit coherent orbital motions, suggesting that surviving gravitational pairs retain a clear kinematic signature of their common origin. However, the reverse is not true: several systems with aligned velocity vectors are not currently bound, indicating that some of these pairs likely trace common infall rather than surviving binaries.

For the observational sample, we evaluate the feasibility of binding by comparing $M_{\mathrm{bound}}$ to typical dwarf galaxy mass scales. In the full MW sample, we find that approximately 38\% (6 out of 16) of associations have $M_{\mathrm{bound}} \lesssim 10^{10}\, \mathrm{M_\odot}$, making them plausible bound systems within the expected mass range of dwarf galaxy haloes. Among these, Pegasus III--Pisces II ($M_{\mathrm{bound}} \approx 0.07 \times 10^9\, \mathrm{M_\odot}$) and Grus II--Tucana II ($\approx 0.34 \times 10^9\, \mathrm{M_\odot}$) stand out as the most likely candidates for physically bound systems. Conversely, pairs such as Leo I--Leo II require a total system mass exceeding $4 \times 10^{11}\, \mathrm{M_\odot}$ to be gravitationally bound. This extreme requirement reinforces the conclusion that their current proximity is a transient alignment within the Galactic halo rather than a physical association. This is further supported by recent dynamical studies, which suggest that Leo II is more likely associated with the Leo IV--Leo V pair \citep{julio_satellite_2024}, rather than Leo I. The $M_{\mathrm{bound}}$ we find for the Leo IV--Leo V pair ($\sim 4 \times 10^9\, \mathrm{M_\odot}$) is orders of magnitude lower than that of the Leo I--II alignment, falling well within the expected mass range of dwarf galaxy systems. The consistency of Leo II with the Leo IV--Leo V association in higher-dimensional phase-space further supports the interpretation that it is a member of this infalling group, rather than a companion of Leo I.

In the resolution-matched sample, approximately 43\% (3 out of 7) of the associations are physically plausible candidates. Specifically, Draco--Ursa Minor ($\sim 6 \times 10^9\, \mathrm{M_\odot}$) and the LMC--SMC ($\sim 37 \times 10^9\, \mathrm{M_\odot}$) require higher masses than the ultra-faint pairs, but these values remain well within (and in the case of the Magellanic Clouds, significantly below) the expected total masses for such systems. Given that the LMC's total mass is estimated to be $\sim 10^{11}\, \mathrm{M_\odot}$ \citep[e.g.][]{erkal_total_2019}, the required $M_{\mathrm{bound}}$ is easily satisfied, confirming their status as a bound pair. Similarly, the value for Draco--Ursa Minor is consistent with recent mass estimates for their respective dark matter haloes \citep[e.g.][]{read_dark_2019, julio_radial_2025}, supporting the idea that they were accreted together. 

The connection between gravitational binding and coherent motion observed in the \textsc{hestia} simulations is also reflected in the observational sample. The Leo IV--Leo V, Pegasus III--Pisces II, LMC--SMC, and Draco--Ursa Minor pairs all satisfy both the binding criterion and the motion-alignment analysis, strengthening their interpretation as true physical associations. Similarly, LMC-related associations such as LMC--Pictor II, Carina II--LMC, and LMC--SMC all exhibit plausible binding estimates and coherent velocity directions. While the identified LMC-related pairs change with the stellar mass cut, the persistence of coherent associations further supports a scenario in which the LMC was accreted together with a population of lower-mass satellites.

For the M31 system, 75\% (3 out of 4 with data) of the identified pairs are physically plausible based on their $M_{\mathrm{bound}}$ values. We find that Andromeda X--Andromeda XXIV ($\sim 5 \times 10^9\, \mathrm{M}_\odot$) stands out as a highly promising candidate. The low $M_{\mathrm{bound}}$ for this pair suggests that even with a moderate tangential velocity component, the system would remain gravitationally bound within the expected mass range of M31's dwarf population. Similarly, Andromeda XXV--NGC 147 ($\sim 24 \times 10^9\, \mathrm{M}_\odot$) exhibits a threshold easily satisfied by the presence of a massive dwarf like NGC 147.

\begin{sidewaystable*}[ht]
\caption{Properties of the pair candidates.}
\label{tab:properties_pair_candidates}
\begin{center}
\begin{tabular}{lcccccccccc}
\toprule
Pair & $\Delta R_\mathrm{3D}$ & $\Delta R_\mathrm{los}$ & $\Delta v_\mathrm{3D}$ & $\Delta v_\mathrm{los}$ & $\overline{d}_\mathrm{host}$ & $\overline{v}_\mathrm{host}$ & $\Delta\theta$ & $\frac{M_{\star,1}}{M_{\star,2}}$ & $M_\mathrm{bound}$ & $\frac{M_\mathrm{bound}}{L}$ \\
& (kpc) & (kpc) & (km\,s$^{-1}$) & (km\,s$^{-1}$) & (kpc) & (km\,s$^{-1}$) & (º) & & ($10^9$ M$_\odot$) & \\
\midrule
\textbf{MW} & & & & & & & & & & \\
Bootes II -- Bootes III & $11.8^{+0.4}_{-0.4}$ & $5.0^{+1.1}_{-1.3}$ & $432.7^{+11.3}_{-10.5}$ & $327.9^{+3.9}_{-3.8}$ & $42.6^{+0.6}_{-0.6}$ & $55.2^{+1.9}_{-2.0}$ & $13.9$ & $13.2$ & $147$ & $8.08\times10^6$ \\
Canes Ven. I -- Canes Ven. II* & $54.5^{+6.8}_{-6.9}$ & $50.5^{+7.2}_{-7.5}$ & $205.0^{+40.9}_{-25.8}$ & $159.9^{+1.2}_{-1.4}$ & $185.7^{+3.8}_{-3.5}$ & $-8.9^{+0.7}_{-0.7}$ & $6.4$ & $26.9$ & $162$ & $5.82\times10^5$ \\
Carina II -- Pictor II & $9.2^{+1.7}_{-1.8}$ & $7.7^{+1.9}_{-2.2}$ & $334.1^{+11.7}_{-11.1}$ & $150.4^{+1.5}_{-1.6}$ & $42.0^{+1.0}_{-1.0}$ & $182.8^{+0.8}_{-0.8}$ & $6.9$ & $5.9$ & $24$ & $3.59\times10^6$ \\
Carina II -- Carina III & $9.6^{+1.3}_{-1.3}$ & $9.6^{+1.3}_{-1.3}$ & $262.5^{+11.3}_{-11.1}$ & $192.5^{+3.3}_{-3.5}$ & $33.6^{+0.6}_{-0.7}$ & $157.1^{+1.7}_{-1.7}$ & $0.3$ & $7.4$ & $41$ & $6.34\times10^6$ \\
Grus II -- Tucana II & $14.3^{+1.8}_{-0.9}$ & $3.8^{+4.5}_{-2.7}$ & $144.2^{+12.3}_{-10.1}$ & $14.7^{+1.0}_{-1.1}$ & $51.4^{+3.0}_{-2.7}$ & $-166.3^{+0.6}_{-0.6}$ & $14.1$ & $1.3$ & $0.4$ & $7.43\times10^4$ \\
Horologium I -- Reticulum III & $18.3^{+11.2}_{-4.5}$ & $12.1^{+13.3}_{-7.9}$ & $278.3^{+62.9}_{-60.7}$ & $161.0^{+8.1}_{-7.4}$ & $85.9^{+6.6}_{-6.6}$ & $40.5^{+3.9}_{-3.9}$ & $9.2$ & $1.1$ & $55$ & $1.48\times10^7$ \\
Centaurus I -- Hydra II* & $40.5^{+7.7}_{-6.7}$ & $33.9^{+8.6}_{-8.1}$ & $366.6^{+65.3}_{-54.3}$ & $258.3^{+1.6}_{-1.5}$ & $131.7^{+4.1}_{-4.6}$ & $-2.3^{+0.7}_{-0.8}$ & $9.5$ & $1.3$ & $314$ & $1.49\times10^7$ \\
Leo I -- Leo II* & $88.9^{+7.4}_{-6.2}$ & $49.0^{+11.5}_{-11.2}$ & $168.4^{+5.5}_{-5.2}$ & $204.4^{+0.8}_{-0.8}$ & $237.2^{+5.1}_{-5.5}$ & $99.6^{+0.4}_{-0.4}$ & $18.4$ & $8.5$ & $432$ & $8.47\times10^4$ \\
Leo IV -- Leo V & $19.8^{+5.8}_{-5.9}$ & $18.2^{+6.2}_{-6.6}$ & $278.2^{+144.2}_{-127.3}$ & $41.6^{+1.4}_{-1.4}$ & $161.0^{+3.1}_{-3.3}$ & $34.0^{+0.7}_{-0.7}$ & $2.8$ & $1.7$ & $4$ & $3.06\times10^5$ \\
LMC -- Pictor II & $11.9^{+0.8}_{-0.5}$ & $4.9^{+2.0}_{-2.2}$ & $156.4^{+14.4}_{-14.1}$ & $64.7^{+3.3}_{-3.5}$ & $47.4^{+1.1}_{-1.1}$ & $90.9^{+1.8}_{-1.8}$ & $13.2$ & $1.48\times10^6$ & $6$ & $4$ \\
Pegasus III -- Pisces II & $43.4^{+14.7}_{-10.3}$ & $31.8^{+18.5}_{-17.2}$ & $662.2^{+311.2}_{-280.3}$ & $3.9^{+3.5}_{-2.9}$ & $197.1^{+9.4}_{-9.9}$ & $-69.2^{+1.8}_{-1.9}$ & $8.5$ & $1.1$ & $0.1$ & $9.21\times10^3$ \\
Segue II -- Triangulum II & $12.1^{+2.1}_{-1.9}$ & $8.1^{+2.7}_{-3.0}$ & $370.2^{+10.5}_{-9.6}$ & $342.1^{+1.2}_{-1.0}$ & $38.5^{+1.4}_{-1.4}$ & $-107.8^{+0.6}_{-0.5}$ & $16.1$ & $1.8$ & $164$ & $2.07\times10^8$ \\
SMC -- Tucana V & $14.8^{+3.5}_{-1.7}$ & $7.9^{+5.9}_{-5.3}$ & $223.6^{+25.2}_{-22.2}$ & $180.3^{+1.1}_{-1.1}$ & $56.0^{+2.9}_{-3.1}$ & $-65.9^{+0.5}_{-0.5}$ & $12.2$ & $1.91\times10^6$ & $56$ & $126$ \\
Tucana II -- Tucana V & $8.6^{+4.2}_{-1.4}$ & $5.0^{+5.6}_{-3.7}$ & $348.1^{+29.4}_{-28.4}$ & $90.0^{+1.3}_{-1.2}$ & $52.2^{+3.7}_{-3.5}$ & $-167.8^{+0.6}_{-0.6}$ & $7.2$ & $11.5$ & $8$ & $2.78\times10^6$ \\
Tucana IV -- Tucana V & $9.0^{+6.4}_{-4.5}$ & $8.3^{+6.7}_{-5.3}$ & $160.1^{+30.2}_{-26.9}$ & $50.6^{+1.9}_{-2.0}$ & $48.0^{+3.1}_{-3.3}$ & $-108.8^{+1.0}_{-0.9}$ & $3.9$ & $5.8$ & $3$ & $1.69\times10^6$ \\
Ursa Major II -- Willman I & $13.5^{+3.7}_{-1.5}$ & $5.6^{+5.5}_{-3.8}$ & $239.9^{+47.4}_{-44.4}$ & $102.4^{+2.1}_{-2.2}$ & $41.4^{+3.6}_{-3.6}$ & $0.1^{+1.2}_{-1.0}$ & $19.8$ & $5.6$ & $17$ & $2.80\times10^6$ \\
\addlinespace[0.5em]
\textbf{MW ($M_\star > 10^4\,\mathrm{M_\odot}$)} & & & & & & & & & & \\
Bootes I -- Bootes III & $23.1^{+2.2}_{-2.1}$ & $19.8^{+2.4}_{-2.4}$ & $267.9^{+8.3}_{-8.1}$ & $94.7^{+3.5}_{-3.8}$ & $54.8^{+1.4}_{-1.2}$ & $175.2^{+1.8}_{-1.9}$ & $12.3$ & $1.3$ & $24$ & $6.20\times10^5$ \\
Carina II -- LMC & $18.6^{+0.5}_{-0.4}$ & $12.2^{+0.7}_{-0.7}$ & $202.5^{+9.7}_{-9.4}$ & $214.8^{+3.7}_{-3.5}$ & $43.7^{+0.3}_{-0.3}$ & $161.6^{+1.6}_{-1.8}$ & $18.9$ & $2.51\times10^5$ & $100$ & $69$ \\
Draco -- Ursa Minor & $25.7^{+1.6}_{-1.1}$ & $11.5^{+3.9}_{-3.6}$ & $73.0^{+2.6}_{-2.7}$ & $43.8^{+0.8}_{-0.9}$ & $76.9^{+1.9}_{-1.9}$ & $-90.3^{+0.4}_{-0.4}$ & $17.4$ & $1.0$ & $6$ & $9.49\times10^3$ \\
LMC -- SMC & $23.3^{+2.1}_{-1.8}$ & $13.2^{+3.0}_{-2.8}$ & $96.1^{+18.5}_{-17.5}$ & $116.5^{+3.5}_{-3.4}$ & $54.7^{+1.4}_{-1.6}$ & $36.5^{+1.7}_{-1.8}$ & $19.8$ & $3.2$ & $37$ & $19$\\
\addlinespace[0.5em]
\textbf{M31} & & & & & & & & & & \\
And. X -- And. XXIV & $37.3^{+18.3}_{-7.9}$ & $24.2^{+23.8}_{-16.7}$ &  & $36.0^{+5.3}_{-5.6}$ & $172.7^{+11.5}_{-10.9}$ & $155.0^{+2.6}_{-2.9}$ & $2.6$ & $1.3$ & $6$ & $3.39\times10^4$ \\
And. XI -- And. XII & $36.7^{+34.8}_{-23.7}$ & $35.8^{+35.3}_{-25.6}$ &  & $129.6^{+4.2}_{-3.9}$ & $101.9^{+5.7}_{-5.3}$ & $-191.3^{+2.0}_{-2.0}$ & $0.6$ & $1.2$ & $72$ & $1.06\times10^6$ \\
And. XV -- And. XXIII & $44.6^{+14.1}_{-4.8}$ & $21.4^{+22.4}_{-15.1}$ &  & $85.3^{+2.0}_{-2.0}$ & $109.3^{+2.0}_{-2.2}$ & $20.7^{+0.9}_{-1.0}$ & $3.0$ & $3.6$ & $38$ & $4.29\times10^4$ \\
And. XXV -- NGC 147 & $31.5^{+21.0}_{-8.0}$ & $21.9^{+25.3}_{-15.1}$ &  & $85.4^{+1.3}_{-1.2}$ & $93.9^{+3.3}_{-3.6}$ & $150.6^{+0.6}_{-0.6}$ & $1.7$ & $199.5$ & $27$ & $358$ \\
Pegasus V -- Pegasus VII & $81.2^{+57.1}_{-31.2}$ & $66.7^{+63.0}_{-46.9}$ &  &  & $268.5^{+4.7}_{-4.2}$ &  & $3.7$ & $1.7$ &  &  \\
\bottomrule
\end{tabular}
\end{center}
\small{\textbf{Columns from left to right:} (1) Names of the satellites part of a pair candidate; (2) 3D physical separation; (3) line-of-sight separation; (4) 3D velocity difference; (5) line-of-sight velocity difference; (6) mean distance from the host; (7) mean pair velocity relative to the host; (8) angular separation; (9) stellar mass ratio defined as $\max(M_{\star,1}, M_{\star,2}) / \min(M_{\star,1}, M_{\star,2})$; (10) the total mass that the system would have to have in order to be bound; and (11) the dynamic $M/L$ ratios that the system would have to have in order to be bound. The pairs marked with an asterisk are identified in both the full observational sample and the stellar mass limited sample.}
\end{sidewaystable*}

\begin{table*}[ht]
\caption{Properties of the pair candidates found in the simulations.}
\label{tab:properties_pair_candidates_simulations}
\begin{center}
\begin{tabular}{lcccccccccc} 
\toprule
Pair & Sim. & $\Delta R_\mathrm{3D}$ & $\Delta R_\mathrm{los}$ &  $\Delta v_\mathrm{3D}$ & $\Delta v_\mathrm{los}$ & ${\overline{d}}_\mathrm{host}$ & ${\overline{v}}_\mathrm{host}$ & $\Delta\theta$ & $M_{\star,1}/M_{\star,2}$ & $b$\\ 
& number & (kpc) & (kpc) &  (km\,s$^{-1}$) & (km\,s$^{-1}$) & (kpc) & (km\,s$^{-1}$) & (º) & & \\ 
\midrule
\textbf{MW analogues} & & & & & & & & & & \\
10--2963 & 9\_18 & 19.7 & 14.8 & 192.2 & 165.6 & 100.0 & -396.3 & 6.8 & 853.7 & 2.1822 \\
774--26549 & 9\_18 & 19.0 & 17.8 & 315.9 & 72.3 & 43.5 & -268.3 & 7.0 & 214.3 & 0.0052 \\
931--2973 & 9\_18 & 48.5 & 33.4 & 347.1 & 85.1 & 150.9 & 127.3 & 12.6 & 3.0 & 0.0023 \\
4719--11410 & 9\_18 & 17.2 & 2.1 & 196.1 & 166.1 & 108.0 & 21.7 & 8.6 & 1.30 & 0.0033 \\
5715--7479 & 9\_18 & 13.8 & 3.7 & 111.6 & 133.1 & 50.8 & -187.0 & 14.5 & 9.6 & 0.0110 \\
2963--11736 & 9\_18 & 32.2 & 21.1 & 265.1 & 131.9 & 118.0 & -413.2 & 11.6 & 26.3 & 0.0008 \\
22879--26549 & 9\_18 & 15.7 & 2.6 & 350.2 & 81.5 & 33.3 & -263.7 & 22.1 & 3.5 & 0.0002 \\
\addlinespace[0.5em]
3292--25953 & 17\_11 & 25.3 & 7.8 & 137.8 & 51.3 & 92.5 & 290.8 & 14.3 & 18.5 & 0.0042 \\
4602--31053 & 17\_11 & 18.0 & 8.1 & 518.2 & 495.2 & 57.9 & 122.6 & 15.2 & 8.9 & 0.0003 \\
9356--18567 & 17\_11 & 13.7 & 5.0 & 185.0 & 211.2 & 46.0 & -167.2 & 13.8 & 8.1 & 0.0017 \\
\addlinespace[0.5em]
1408--11290 & 37\_11 & 10.5 & 5.1 & 142.4 & 179.2 & 35.6 & 343.1 & 18.9 & 167.7 & 0.0151 \\
25745--27628 & 37\_11 & 19.7 & 12.8 & 233.7 & 68.7 & 54.4 & 228.5 & 15.2 & 1.1 & 0.0003 \\
1408--27628 & 37\_11 & 19.5 & 9.9 & 111.0 & 39.3 & 59.4 & 223.8 & 24.8 & 1377.4 & 0.0122 \\
\textbf{M31 analogues} & & & & & & & & & & \\
884-8995 & 9\_18 & 25.9 & 9.04 & 257.4 & 171.5 & 138.5 & 69.7 & 2.0 & 343.3 & 0.0041 \\
1479-67802 & 9\_18 & 17.7 & 15.8 & 278.0 & 11.9 & 77.3 & 204.1 & 0.9 & 1.5 & 0.0045 \\  
1584-2344 & 9\_18 & 14.7 & 10.9 & 254.2 & 75.3 & 98.7 & -96.9 & 1.0 & 83.0 & 0.0078 \\
1887-6560 & 9\_18 & 34.2 & 16.6 & 283.7 & 189.8 & 110.2 & -13.8 & 2.8 & 397.8 & 0.0015 \\
2344-6723 & 9\_18 & 27.0 & 8.3 & 84.4 & 56.2 & 105.6 & -106.5 & 2.7 & 8.8 & 0.0240 \\
\addlinespace[0.5em]
41--4743 & 17\_11 & 33.4 & 8.5 & 206.9 & 84.1 & 128.2 & 266.6 & 4.5 & 26191.9 & 0.1160 \\
993--1822 & 17\_11 & 15.7 & 11.8 & 408.0 & 31.0 & 46.8 & 180.7 & 1.4 & 19.2 & 0.0038 \\
2038--5836 & 17\_11 & 49.8 & 9.1 & 187.3 & 43.5 & 174.7 & 100.6 & 5.0 & 25.6 & 0.0022 \\
3262--7688 & 17\_11 & 27.7 & 20.6 & 242.0 & 171.6 & 118.3 & 119.1 & 2.0 & 7.0 & 0.0016 \\
3311--25757 & 17\_11 & 22.3 & 14.1 & 225.4 & 98.5 & 71.6 & 226.4 & 2.6 & 23.9 & 0.0018 \\
4743--12533 & 17\_11 & 28.2 & 22.2 & 171.2 & 46.5 & 114.9 & 331.9 & 2.5 & 1.8 & 0.0020 \\
41--5926 & 17\_11 & 34.6 & 2.2 & 188.5 & 160.7 & 138.8 & 144.5 & 4.8 & 151349.7 & 0.1344 \\
30--9254 & 17\_11 & 38.5 & 0.7 & 66.7 & 48.4 & 138.6 & 189.6 & 4.2 & 5675.6 & 1.6527 \\
\addlinespace[0.5em]
41--9146 & 37\_11 & 18.3 & 17.3 & 53.6 & 48.2 & 137.5 & -130. & 0.5 & 1065.3 & 4.0451 \\
976--2481 & 37\_11 & 36.7 & 36.2 & 90.0 & 48.2 & 116.5 & -25.5 & 0.6 & 2.3 & 0.0315 \\
2481--4294 & 37\_11 & 18.0 & 15.9 & 108.5 & 10.0 & 125.1 & -44.6 & 0.9 & 1.1 & 0.0188 \\
4294--11673 & 37\_11 & 13.2 & 6.1 & 57.4 & 18.1 & 134.1 & -48.6 & 1.3 & 34.4 & 0.0427 \\
8263--17744 & 37\_11 & 18.3 & 3.9 & 87.8 & 71.6 & 39.8 & -129.1 & 1.9 & 43.4 & 0.0059 \\
\bottomrule
\end{tabular}
\end{center}
\small{\textbf{Columns from left to right:} (1) Names of the satellites part of a pair candidate; (2) Number of the simulation; (3) 3D physical separation; (4) line-of-sight separation; (5) 3D velocity difference; (6) line-of-sight velocity difference; (7) mean distance from the host; (8) mean pair velocity relative to the host; (9) angular separation; (10) stellar mass ratio defined as $\max(M_{\star,1}, M_{\star,2}) / \min(M_{\star,1}, M_{\star,2})$; and (11) the binding criteria $b$.}
\end{table*}

\section{Conclusion}\label{sec:conclusion}

In this work, we have revisited the incidence and nature of satellite galaxy pairs in the LG by combining updated observational data with \textsc{hestia} cosmological simulations within a consistent statistical framework.

We first identify candidate associations using NN statistics and quantify their significance through $P$-values relative to isotropic expectations. The MW satellite system exhibits a clear excess of close associations compared to both isotropic realisations and $\Lambda$CDM simulations. This is reflected in its high pairing efficiency ($\sim 53\%$), which remains high even when applying the stellar mass cut motivated by the simulation's resolution ($\sim 52\%$). In contrast, the M31 satellite system shows no comparable excess and remains broadly consistent with an isotropic spatial distribution.

The properties of the identified pairs reveal a dependence on stellar mass, although the population is consistently dominated by ``major'' pairs. In the full MW sample, $\sim 69\%$ of the associations correspond to systems with comparable stellar masses ($M_{\star,1}/M_{\star,2} < 10$), while the remaining $\sim 31\%$ are ``hierarchical'' pairs involving an ultra-faint companion. After applying the stellar mass cut, the overall number of pairs decreases, but the relative contribution of major pairs remains similar ($\sim 71\%$), as most hierarchical associations are removed.

This indicates that, while the clustering signal includes a contribution from low-mass companions, it is not dominated by them. However, both simulations and M31 observations (limited by resolution and distance, respectively) primarily probe the major-pair regime since they lack low-mass satellites. As a result, comparisons across these samples may underestimate the true abundance of hierarchical associations in the underlying satellite population.

While spatial proximity provides an effective method to search for pair candidates, additional phase-space information is required to understand their physical nature. Using a revised reference scale based on the present-day LMC--SMC line-of-sight velocity difference ($|\Delta v_{\mathrm{los}}| \sim 117\,\mathrm{km\,s^{-1}}$), we find that $\sim 44\%$ (7 out of 16) of MW pairs in the full sample, and $\sim 43\%$ (3 out of 7) in the resolution-matched sample, exhibit small velocity differences consistent with potential dynamical association. For M31, $\sim 75\%$ (3 out of 4) of the pairs with available velocity measurements satisfy this criterion, although this estimate is based on a small sample. Because the LMC--SMC system occupies a substantially different mass regime from the dwarf-galaxy pairs considered here, a velocity difference below this value does not by itself imply dynamical association. We therefore use it only as an upper limit for pair candidates, combining it with full phase-space metrics and proper-motion information wherever available.

Proper motion information filters this picture further. Among the MW pairs with available measurements, we find that $\sim 44\%$ (7 out of 16) exhibit velocity vectors that are consistent within their uncertainties, indicating that a non-negligible subset of pairs is compatible with coherent orbital motion. Several of these systems also satisfy the line-of-sight velocity criterion, demonstrating that the two diagnostics identify overlapping subsets of candidate associations. However, the overlap is not complete, highlighting that line-of-sight velocities and projected proper motion alignments probe complementary aspects of phase-space coherence. For the more massive satellites ($M_\star > 10^4\,\mathrm{M_\odot}$), this fraction increases to $\sim 70\%$ (5 out of 7). The lack of proper motion measurements for M31 satellites prevents a comparable analysis for that system.

In the simulations, where full phase-space information is available, a comparable analysis shows that only a subset of spatially selected pairs exhibit coherent kinematics. We find that $\sim 65\%$ of simulated pairs satisfy the line-of-sight velocity criterion. A visual inspection of the proper motion directions indicates that a significant fraction of these systems ($\sim 45\%$) show aligned motions in projection. Together, these results confirm that many close NN associations do not correspond to long-lived bound binaries, but instead arise from transient configurations or remnants of disrupted groups.

Our dynamical analysis further shows that only a fraction of pairs are consistent with being currently gravitationally bound. In the observations, we find that $\sim 38\%$ of the MW pair candidates satisfy this criterion, increasing to $\sim 43\%$ when applying the stellar mass cut. For M31, this fraction rises to $\sim 75\%$, although this likely reflects a selection bias towards more massive and tightly bound systems. In the simulations, however, only $\sim 10\%$ of pairs are identified as bound, indicating that most pairs are not long-lived bound systems but rather transient configurations or remnants of disrupted groups. This is consistent with expectations from hierarchical structure formation, where satellites are accreted in groups but rapidly lose their mutual binding within the host halo.

Although NN statistics provide an efficient way to identify candidate satellite associations, combining all available diagnostics significantly reduces the number of systems consistent with genuine physical pairs. Requiring simultaneous spatial clustering, kinematic coherence, aligned motions, and compatibility with gravitational binding leaves only a small fraction of the original candidates. In the full MW observational sample, only $\sim 19\%$ (3 out of 16) of the spatially selected candidates satisfy all three criteria simultaneously, increasing to $\sim 29\%$ (2 out of 7) in the resolution-matched sample. In the \textsc{hestia} simulations, the fraction is even smaller, with only $\sim 8\%$ (1 out of 13) of the MW analogue candidates and $\sim 11\%$ (2 out of 18) of the M31 analogue candidates satisfying all criteria simultaneously. This demonstrates that most close NN associations do not correspond to long-lived physical binaries, but instead trace transient configurations, tidal remnants, or satellites that were accreted together but are no longer gravitationally bound today.

Taken together, these results suggest that the excess of satellite pairs in the MW is not primarily due to an abundance of bound binaries, but instead reflects the presence of correlated phase-space structures associated with recent or ongoing group infall. The Magellanic system provides a clear example of this process.

Finally, the contrasting behaviour of the MW and M31 highlights the importance of individual accretion histories, as well as observational and numerical limitations, in shaping the inferred clustering properties of satellite systems. While the MW exhibits an excess of close associations, M31 remains consistent with isotropic expectations even when accounting for distance uncertainties. At the same time, some simulated M31 analogues display stronger clustering than observed, suggesting that environmental factors and stochastic assembly histories can produce a wide range of outcomes.

Future progress will require improved completeness for faint satellites, particularly around M31, together with more accurate distance measurements and precise proper motion to better constrain the full 6D phase-space information of candidate associations. On the theoretical side, a larger suite of high-resolution cosmological simulations spanning a broader range of host masses and assembly histories will be needed to robustly determine the statistical significance of the observed MW--M31 differences within the $\Lambda$CDM framework.

\begin{acknowledgements}
MPJ and MSP acknowledge funding via a Leibniz-Junior Research Group (project number J94/2020). This research has made use of Astropy \citep{astropy_collaboration_astropy_2018}, galpy \citep{bovy_galpy_2015}, matplotlib \citep{hunter_matplotlib_2007}, NASA's Astrophysics Data System Bibliographic Services, NumPy \citep{harris_array_2020} and SciPy \citep{virtanen_scipy_2020}
\end{acknowledgements}


\bibliography{pairs}{}

@article{kanehisa_andromedas_2025,
	title = {Andromeda's asymmetric satellite system as a challenge to cold dark matter cosmology},
	volume = {9},
	issn = {2397-3366},
	url = {https://ui.adsabs.harvard.edu/abs/2025NatAs...9..692K},
	doi = {10.1038/s41550-025-02480-3},
	urldate = {2026-02-10},
	journal = {Nature Astronomy},
	author = {Kanehisa, Kosuke Jamie and Pawlowski, Marcel S. and Libeskind, Noam},
	month = may,
	year = {2025},
	pages = {692--701},
}

@article{savino_hubble_2022,
	title = {The {Hubble} {Space} {Telescope} {Survey} of {M31} {Satellite} {Galaxies}. {I}. {RR} {Lyrae}-based {Distances} and {Refined} {3D} {Geometric} {Structure}},
	volume = {938},
	issn = {0004-637X},
	url = {https://ui.adsabs.harvard.edu/abs/2022ApJ...938..101S},
	doi = {10.3847/1538-4357/ac91cb},
	urldate = {2026-02-10},
	journal = {ApJ},
	publisher = {IOP},
	author = {Savino, Alessandro and Weisz, Daniel R. and Skillman, Evan D. and Dolphin, Andrew and Kallivayalil, Nitya and Wetzel, Andrew and Anderson, Jay and Besla, Gurtina and Boylan-Kolchin, Michael and Bullock, James S. and Cole, Andrew A. and Collins, Michelle L. M. and Cooper, M. C. and Deason, Alis J. and Dotter, Aaron L. and Fardal, Mark and Ferguson, Annette M. N. and Fritz, Tobias K. and Geha, Marla C. and Gilbert, Karoline M. and Guhathakurta, Puragra and Ibata, Rodrigo and Irwin, Michael J. and Jeon, Myoungwon and Kirby, Evan and Lewis, Geraint F. and Mackey, Dougal and Majewski, Steven R. and Martin, Nicolas and McConnachie, Alan and Patel, Ekta and Rich, R. Michael and Simon, Joshua D. and Sohn, Sangmo Tony and Tollerud, Erik J. and van der Marel, Roeland P.},
	month = oct,
	year = {2022},
	pages = {101},
}

@article{astropy_collaboration_astropy_2018,
	title = {The {Astropy} {Project}: {Building} an {Open}-science {Project} and {Status} of the v2.0 {Core} {Package}},
	volume = {156},
	doi = {10.3847/1538-3881/aabc4f},
	number = {3},
	journal = {AJ},
	author = {{Astropy Collaboration} and Price-Whelan, A. M. and Sipocz, B. M. and Günther, H. M. and Lim, P. L. and Crawford, S. M. and Conseil, S. and Shupe, D. L. and Craig, M. W. and Dencheva, N. and Ginsburg, A. and VanderPlas, J. T. and Bradley, L. D. and Pérez-Suárez, D. and de Val-Borro, M. and Aldcroft, T. L. and Cruz, K. L. and Robitaille, T. P. and Tollerud, E. J. and Ardelean, C. and Babej, T. and Bach, Y. P. and Bachetti, M. and Bakanov, A. V. and Bamford, S. P. and Barentsen, G. and Barmby, P. and Baumbach, A. and Berry, K. L. and Biscani, F. and Boquien, M. and Bostroem, K. A. and Bouma, L. G. and Brammer, G. B. and Bray, E. M. and Breytenbach, H. and Buddelmeijer, H. and Burke, D. J. and Calderone, G. and Cano Rodríguez, J. L. and Cara, M. and Cardoso, J. V. M. and Cheedella, S. and Copin, Y. and Corrales, L. and Crichton, D. and D'Avella, D. and Deil, C. and Depagne, É. and Dietrich, J. P. and Donath, A. and Droettboom, M. and Earl, N. and Erben, T. and Fabbro, S. and Ferreira, L. A. and Finethy, T. and Fox, R. T. and Garrison, L. H. and Gibbons, S. L. J. and Goldstein, D. A. and Gommers, R. and Greco, J. P. and Greenfield, P. and Groener, A. M. and Grollier, F. and Hagen, A. and Hirst, P. and Homeier, D. and Horton, A. J. and Hosseinzadeh, G. and Hu, L. and Hunkeler, J. S. and Ivezić, Ž. and Jain, A. and Jenness, T. and Kanarek, G. and Kendrew, S. and Kern, N. S. and Kerzendorf, W. E. and Khvalko, A. and King, J. and Kirkby, D. and Kulkarni, A. M. and Kumar, A. and Lee, A. and Lenz, D. and Littlefair, S. P. and Ma, Z. and Macleod, D. M. and Mastropietro, M. and McCully, C. and Montagnac, S. and Morris, B. M. and Mueller, M. and Mumford, S. J. and Muna, D. and Murphy, N. A. and Nelson, S. and Nguyen, G. H. and Ninan, J. P. and Nöthe, M. and Ogaz, S. and Oh, S. and Parejko, J. K. and Parley, N. and Pascual, S. and Patil, R. and Patil, A. A. and Plunkett, A. L. and Prochaska, J. X. and Rastogi, T. and Reddy Janga, V. and Sabater, J. and Sakurikar, P. and Seifert, M. and Sherbert, L. E. and Sherwood-Taylor, H. and Shih, A. Y. and Sick, J. and Silbiger, M. T. and Singanamalla, S. and Singer, L. P. and Sladen, P. H. and Sooley, K. A. and Sornarajah, S. and Streicher, O. and Teuben, P. and Thomas, S. W. and Tremblay, G. R. and Turner, J. E. H. and Terrón, V. and van Kerkwijk, M. H. and de la Vega, A. and Watkins, L. L. and Weaver, B. A. and Whitmore, J. B. and Woillez, J. and Zabalza, V. and {Astropy Contributors}},
	month = sep,
	year = {2018},
	pages = {123},
}

@article{pace_local_2025,
	title = {The {Local} {Volume} {Database}: a library of the observed properties of nearby dwarf galaxies and star clusters},
	volume = {8},
	shorttitle = {The {Local} {Volume} {Database}},
	url = {https://astro.theoj.org/article/144859},
	doi = {10.33232/001c.144859},
	language = {en},
	urldate = {2025-10-06},
	journal = {OJAp},
	author = {Pace, Andrew B.},
	month = sep,
	year = {2025},
}

@article{fritz_gaia_2018,
	title = {Gaia {DR2} proper motions of dwarf galaxies within 420 kpc. {Orbits}, {Milky} {Way} mass, tidal influences, planar alignments, and group infall},
	volume = {619},
	doi = {10.1051/0004-6361/201833343},
	journal = {A\&A},
	author = {Fritz, T. K. and Battaglia, G. and Pawlowski, M. S. and Kallivayalil, N. and van der Marel, R. and Sohn, S. T. and Brook, C. and Besla, G.},
	month = nov,
	year = {2018},
	pages = {A103},
}

@article{patel_orbital_2020,
	title = {The {Orbital} {Histories} of {Magellanic} {Satellites} {Using} {Gaia} {DR2} {Proper} {Motions}},
	volume = {893},
	doi = {10.3847/1538-4357/ab7b75},
	number = {2},
	journal = {ApJ},
	author = {Patel, Ekta and Kallivayalil, Nitya and Garavito-Camargo, Nicolas and Besla, Gurtina and Weisz, Daniel R. and van der Marel, Roeland P. and Boylan-Kolchin, Michael and Pawlowski, Marcel S. and Gómez, Facundo A.},
	month = apr,
	year = {2020},
	pages = {121},
}

@article{torrealba_feeble_2016,
	title = {The feeble giant. {Discovery} of a large and diffuse {Milky} {Way} dwarf galaxy in the constellation of {Crater}},
	volume = {459},
	doi = {10.1093/mnras/stw733},
	number = {3},
	journal = {MNRAS},
	author = {Torrealba, G. and Koposov, S. E. and Belokurov, V. and Irwin, M.},
	month = jul,
	year = {2016},
	pages = {2370--2378},
}

@article{wetzel_satellite_2015,
	title = {Satellite {Dwarf} {Galaxies} in a {Hierarchical} {Universe}: {Infall} {Histories}, {Group} {Preprocessing}, and {Reionization}},
	volume = {807},
	doi = {10.1088/0004-637X/807/1/49},
	number = {1},
	journal = {ApJ},
	author = {Wetzel, Andrew R. and Deason, Alis J. and Garrison-Kimmel, Shea},
	month = jul,
	year = {2015},
	pages = {49},
}

@article{jong_enigmatic_2010,
	title = {{THE} {ENIGMATIC} {PAIR} {OF} {DWARF} {GALAXIES} {LEO} {IV} {AND} {LEO} {V}: {COINCIDENCE} {OR} {COMMON} {ORIGIN}?},
	volume = {710},
	issn = {0004-637X},
	shorttitle = {{THE} {ENIGMATIC} {PAIR} {OF} {DWARF} {GALAXIES} {LEO} {IV} {AND} {LEO} {V}},
	url = {https://dx.doi.org/10.1088/0004-637X/710/2/1664},
	doi = {10.1088/0004-637X/710/2/1664},
	number = {2},
	urldate = {2023-05-17},
	journal = {ApJ},
	author = {Jong, Jelte T. A. de and Martin, Nicolas F. and Rix, Hans-Walter and Smith, Kester W. and Jin, Shoko and Macciò, Andrea V.},
	month = feb,
	year = {2010},
	pages = {1664},
}

@article{belokurov_cats_2007,
	title = {Cats and {Dogs}, {Hair} and {A} {Hero}: {A} {Quintet} of {New} {Milky} {Way} {Companions}},
	volume = {654},
	issn = {0004-637X, 1538-4357},
	url = {http://arxiv.org/abs/astro-ph/0608448},
	doi = {10.1086/509718},
	number = {2},
	urldate = {2023-05-17},
	journal = {ApJ},
	author = {Belokurov, V. and Zucker, D. B. and Evans, N. W. and Kleyna, J. T. and Koposov, S. and Hodgkin, S. T. and Irwin, M. J. and Gilmore, G. and Wilkinson, M. I. and Fellhauer, M. and Bramich, D. M. and Hewett, P. C. and Vidrih, S. and De Jong, J. T. A. and Smith, J. A. and Rix, H.-W. and Bell, E. F. and Wyse, R. F. G. and Newberg, H. J. and Mayeur, P. A. and Yanny, B. and Rockosi, C. M. and Gnedin, O. Y. and Schneider, D. P. and Beers, T. C. and Barentine, J. C. and Brewington, H. and Brinkmann, J. and Harvanek, M. and Kleinman, S. J. and Krzesinski, J. and Long, D. and Nitta, A. and Snedden, S. A.},
	month = jan,
	year = {2007},
	pages = {897--906},
}

@article{pawlowski_phase-space_2021,
	title = {Phase-{Space} {Correlations} among {Systems} of {Satellite} {Galaxies}},
	volume = {9},
	issn = {2075-4434},
	url = {https://www.mdpi.com/2075-4434/9/3/66},
	doi = {10.3390/galaxies9030066},
	language = {en},
	number = {3},
	urldate = {2022-07-12},
	journal = {Galaxies},
	author = {Pawlowski, Marcel S.},
	month = sep,
	year = {2021},
	pages = {66},
}

@article{pawlowski_planes_2018,
	title = {The planes of satellite galaxies problem, suggested solutions, and open questions},
	volume = {33},
	issn = {0217-7323, 1793-6632},
	url = {https://www.worldscientific.com/doi/abs/10.1142/S0217732318300045},
	doi = {10.1142/S0217732318300045},
	language = {en},
	number = {06},
	urldate = {2022-07-12},
	journal = {Mod. Phys. Lett. A.},
	author = {Pawlowski, Marcel S.},
	month = feb,
	year = {2018},
	pages = {1830004},
}

@article{lynden-bell_ghostly_1995,
	title = {Ghostly streams from the formation of the {Galaxy}'s halo},
	volume = {275},
	issn = {0035-8711},
	url = {https://ui.adsabs.harvard.edu/abs/1995MNRAS.275..429L},
	doi = {10.1093/mnras/275.2.429},
	urldate = {2022-07-12},
	journal = {MNRAS},
	author = {Lynden-Bell, D. and Lynden-Bell, R. M.},
	month = jul,
	year = {1995},
	pages = {429--442},
}

@article{pawlowski_vast_2014,
	title = {The {Vast} {Polar} {Structure} of the {Milky} {Way} {Attains} {New} {Members}},
	volume = {790},
	issn = {0004-637X},
	url = {https://ui.adsabs.harvard.edu/abs/2014ApJ...790...74P},
	doi = {10.1088/0004-637X/790/1/74},
	urldate = {2023-05-17},
	journal = {ApJ},
	author = {Pawlowski, Marcel S. and Kroupa, Pavel},
	month = jul,
	year = {2014},
	pages = {74},
}

@article{pawlowski_vpos_2012,
	title = {The {VPOS}: a vast polar structure of satellite galaxies, globular clusters and streams around the {Milky} {Way}},
	volume = {423},
	issn = {0035-8711},
	shorttitle = {The {VPOS}},
	url = {https://doi.org/10.1111/j.1365-2966.2012.20937.x},
	doi = {10.1111/j.1365-2966.2012.20937.x},
	number = {2},
	urldate = {2023-05-17},
	journal = {MNRAS},
	author = {Pawlowski, M. S. and Pflamm-Altenburg, J. and Kroupa, P.},
	month = jun,
	year = {2012},
	pages = {1109--1126},
}

@article{pawlowski_co-orbiting_2014,
	title = {{CO}-{ORBITING} {PLANES} {OF} {SUB}-{HALOS} {ARE} {SIMILARLY} {UNLIKELY} {AROUND} {PAIRED} {AND} {ISOLATED} {HOSTS}},
	volume = {789},
	issn = {2041-8205},
	url = {https://dx.doi.org/10.1088/2041-8205/789/1/L24},
	doi = {10.1088/2041-8205/789/1/L24},
	language = {en},
	number = {1},
	urldate = {2023-05-17},
	journal = {ApJ Letters},
	author = {Pawlowski, Marcel S. and McGaugh, Stacy S.},
	month = jun,
	year = {2014},
	pages = {L24},
}

@article{fattahi_galaxy_2013,
	title = {Galaxy pairs in the local group.},
	volume = {431},
	issn = {0035-8711},
	url = {https://ui.adsabs.harvard.edu/abs/2013MNRAS.431L..73F},
	doi = {10.1093/mnrasl/slt011},
	urldate = {2023-05-22},
	journal = {MNRAS},
	author = {Fattahi, A. and Navarro, J. F. and Starkenburg, E. and Barber, C. R. and McConnachie, A. W.},
	month = apr,
	year = {2013},
	pages = {L73--L77},
}

@article{white_core_1978,
	title = {Core condensation in heavy halos: a two-stage theory for galaxy formation and clustering.},
	volume = {183},
	doi = {10.1093/mnras/183.3.341},
	journal = {MNRAS},
	author = {White, S. D. M. and Rees, M. J.},
	month = may,
	year = {1978},
	pages = {341--358},
}

@article{springel_aquarius_2008,
	title = {The {Aquarius} {Project}: the subhaloes of galactic haloes},
	volume = {391},
	issn = {0035-8711},
	shorttitle = {The {Aquarius} {Project}},
	url = {https://doi.org/10.1111/j.1365-2966.2008.14066.x},
	doi = {10.1111/j.1365-2966.2008.14066.x},
	number = {4},
	urldate = {2023-05-23},
	journal = {MNRAS},
	author = {Springel, V. and Wang, J. and Vogelsberger, M. and Ludlow, A. and Jenkins, A. and Helmi, A. and Navarro, J. F. and Frenk, C. S. and White, S. D. M.},
	month = dec,
	year = {2008},
	pages = {1685--1711},
}

@article{lynden-bell_fornax-leo-sculptor_1982,
	title = {The {Fornax}-{Leo}-{Sculptor} system},
	volume = {102},
	journal = {The Observatory},
	author = {Lynden-Bell, D.},
	month = oct,
	year = {1982},
	pages = {202--208},
}

@article{majewski_fornax-leo-sculptor_1994,
	title = {The {Fornax}-{Leo}-{Sculptor} {Stream} {Revisited}},
	volume = {431},
	issn = {0004-637X},
	url = {https://ui.adsabs.harvard.edu/abs/1994ApJ...431L..17M},
	doi = {10.1086/187462},
	urldate = {2023-08-09},
	journal = {ApJ},
	author = {Majewski, Steven R.},
	month = aug,
	year = {1994},
	pages = {L17},
}

@article{kim_heros_2015,
	title = {A {HERO}’{S} {DARK} {HORSE}: {DISCOVERY} {OF} {AN} {ULTRA}-{FAINT} {MILKY} {WAY} {Satellite} {IN} {PEGASUS}},
	volume = {804},
	issn = {2041-8205},
	shorttitle = {A {HERO}’{S} {DARK} {HORSE}},
	url = {https://dx.doi.org/10.1088/2041-8205/804/2/L44},
	doi = {10.1088/2041-8205/804/2/L44},
	language = {en},
	number = {2},
	urldate = {2023-08-09},
	journal = {ApJ Letters},
	author = {Kim, Dongwon and Jerjen, Helmut and Mackey, Dougal and Costa, Gary S. Da and Milone, Antonino P.},
	month = may,
	year = {2015},
	pages = {L44},
}

@article{garofalo_born_2021,
	title = {Born in a {Pair} (?): {Pisces} {II} and {Pegasus} {III}},
	volume = {916},
	issn = {0004-637X},
	shorttitle = {Born in a {Pair} (?},
	url = {https://ui.adsabs.harvard.edu/abs/2021ApJ...916...10G},
	doi = {10.3847/1538-4357/ac0253},
	urldate = {2023-08-09},
	journal = {ApJ},
	author = {Garofalo, Alessia and Tantalo, Maria and Cusano, Felice and Clementini, Gisella and Calura, Francesco and Muraveva, Tatiana and Paris, Diego and Speziali, Roberto},
	month = jul,
	year = {2021},
	pages = {10},
}

@article{arias_ngc_2016,
	title = {{NGC} 147, {NGC} 185 and {CassII}: a genetic approach to orbital properties, star formation and tidal debris},
	volume = {456},
	issn = {0035-8711},
	shorttitle = {{NGC} 147, {NGC} 185 and {CassII}},
	url = {https://doi.org/10.1093/mnras/stv2781},
	doi = {10.1093/mnras/stv2781},
	number = {2},
	urldate = {2023-08-09},
	journal = {MNRAS},
	author = {Arias, Veronica and Guglielmo, Magda and Fernando, Nuwanthika and Lewis, Geraint F. and Bland-Hawthorn, Joss and Bate, Nicholas F. and Conn, Anthony and Irwin, Mike J. and Ferguson, Annette M. N. and Ibata, Rodrigo A. and McConnachie, Alan W. and Martin, Nicolas},
	month = feb,
	year = {2016},
	pages = {1654--1665},
}

@article{tully_associations_2006,
	title = {Associations of {Dwarf} {Galaxies}},
	volume = {132},
	issn = {1538-3881},
	url = {https://iopscience.iop.org/article/10.1086/505466/meta},
	doi = {10.1086/505466},
	language = {en},
	number = {2},
	urldate = {2023-08-09},
	journal = {AJ},
	author = {Tully, R. Brent and Rizzi, L. and Dolphin, A. E. and Karachentsev, I. D. and Karachentseva, V. E. and Makarov, D. I. and Makarova, L. and Sakai, S. and Shaya, E. J.},
	month = jun,
	year = {2006},
	pages = {729},
}

@article{tully_squelched_2002,
	title = {Squelched {Galaxies} and {Dark} {Halos}},
	volume = {569},
	issn = {0004-637X},
	url = {https://iopscience.iop.org/article/10.1086/339425/meta},
	doi = {10.1086/339425},
	language = {en},
	number = {2},
	urldate = {2023-08-09},
	journal = {ApJ},
	author = {Tully, R. Brent and Somerville, Rachel S. and Trentham, Neil and Verheijen, Marc A. W.},
	month = apr,
	year = {2002},
	pages = {573},
}

@article{richstein_structural_2022,
	title = {Structural {Parameters} and {Possible} {Association} of the {Ultra}-faint {Dwarfs} {Pegasus} {III} and {Pisces} {II} from {Deep} {Hubble} {Space} {Telescope} {Photometry}},
	volume = {933},
	issn = {0004-637X},
	url = {https://ui.adsabs.harvard.edu/abs/2022ApJ...933..217R},
	doi = {10.3847/1538-4357/ac7226},
	urldate = {2023-08-09},
	journal = {ApJ},
	author = {Richstein, Hannah and Patel, Ekta and Kallivayalil, Nitya and Simon, Joshua D. and Zivick, Paul and Tollerud, Erik and Fritz, Tobias and Warfield, Jack T. and Besla, Gurtina and van der Marel, Roeland P. and Wetzel, Andrew and Choi, Yumi and Deason, Alis and Geha, Marla and Guhathakurta, Puragra and Jeon, Myoungwon and Kirby, Evan N. and Libralato, Mattia and Sacchi, Elena and Sohn, Sangmo Tony},
	month = jul,
	year = {2022},
	pages = {217},
}

@article{van_den_bergh_binary_1998,
	title = {The {Binary} {Galaxies} {NGC} 147 and {NGC} 185},
	volume = {116},
	issn = {0004-6256},
	url = {https://ui.adsabs.harvard.edu/abs/1998AJ....116.1688V},
	doi = {10.1086/300576},
	urldate = {2023-08-10},
	journal = {AJ},
	author = {van den Bergh, Sidney},
	month = oct,
	year = {1998},
	pages = {1688--1689},
}

@article{bell_ultrafaint_2022,
	title = {Ultrafaint {Dwarf} {Galaxy} {Candidates} in the {M81} {Group}: {Signatures} of {Group} {Accretion}},
	volume = {937},
	issn = {0004-637X},
	shorttitle = {Ultrafaint {Dwarf} {Galaxy} {Candidates} in the {M81} {Group}},
	url = {https://ui.adsabs.harvard.edu/abs/2022ApJ...937L...3B},
	doi = {10.3847/2041-8213/ac8e5e},
	urldate = {2023-08-09},
	journal = {ApJ},
	author = {Bell, Eric F. and Smercina, Adam and Price, Paul A. and D'Souza, Richard and Bailin, Jeremy and de Jong, Roelof S. and Gozman, Katya and Jang, In Sung and Monachesi, Antonela and Gnedin, Oleg Y. and Slater, Colin T.},
	month = sep,
	year = {2022},
	pages = {L3},
}

@article{geha_local_2010,
	title = {Local {Group} {Dwarf} {Elliptical} {Galaxies}. {II}. {Stellar} {Kinematics} to {Large} {Radii} in {NGC} 147 and {NGC} 185},
	volume = {711},
	issn = {0004-637X},
	url = {https://ui.adsabs.harvard.edu/abs/2010ApJ...711..361G},
	doi = {10.1088/0004-637X/711/1/361},
	urldate = {2023-08-29},
	journal = {ApJ},
	author = {Geha, M. and van der Marel, R. P. and Guhathakurta, P. and Gilbert, K. M. and Kalirai, J. and Kirby, E. N.},
	month = mar,
	year = {2010},
	pages = {361--373},
}

@article{sohn_hst_2020,
	title = {{HST} {Proper} {Motions} of {NGC} 147 and {NGC} 185: {Orbital} {Histories} and {Tests} of a {Dynamically} {Coherent} {Andromeda} {Satellite} {Plane}},
	volume = {901},
	issn = {0004-637X},
	shorttitle = {{HST} {Proper} {Motions} of {NGC} 147 and {NGC} 185},
	url = {https://dx.doi.org/10.3847/1538-4357/abaf49},
	doi = {10.3847/1538-4357/abaf49},
	language = {en},
	number = {1},
	urldate = {2023-08-29},
	journal = {ApJ},
	author = {Sohn, Sangmo Tony and Patel, Ekta and Fardal, Mark A. and Besla, Gurtina and Marel, Roeland P. van der and Geha, Marla and Guhathakurta, Puragra},
	month = sep,
	year = {2020},
	pages = {43},
}

@article{davis_abell_1995,
	title = {Abell 548: an {X}-{Ray} and {Optical} {Analysis} of {Substructure}},
	volume = {440},
	issn = {0004-637X},
	shorttitle = {Abell 548},
	url = {https://ui.adsabs.harvard.edu/abs/1995ApJ...440...48D},
	doi = {10.1086/175246},
	urldate = {2023-08-29},
	journal = {ApJ},
	author = {Davis, David S. and Bird, Christina M. and Mushotzky, Richard F. and Odewahn, S. C.},
	month = feb,
	year = {1995},
	pages = {48},
}

@article{kallivayalil_missing_2018,
	title = {The {Missing} {Satellites} of the {Magellanic} {Clouds}? {Gaia} {Proper} {Motions} of the {Recently} {Discovered} {Ultra}-faint {Galaxies}},
	volume = {867},
	issn = {0004-637X},
	shorttitle = {The {Missing} {Satellites} of the {Magellanic} {Clouds}?},
	url = {https://ui.adsabs.harvard.edu/abs/2018ApJ...867...19K},
	doi = {10.3847/1538-4357/aadfee},
	urldate = {2024-03-11},
	journal = {ApJ},
	author = {Kallivayalil, Nitya and Sales, Laura V. and Zivick, Paul and Fritz, Tobias K. and Del Pino, Andrés and Sohn, Sangmo Tony and Besla, Gurtina and van der Marel, Roeland P. and Navarro, Julio F. and Sacchi, Elena},
	month = nov,
	year = {2018},
	pages = {19},
}

@article{stierwalt_direct_2017,
	title = {Direct evidence of hierarchical assembly at low masses from isolated dwarf galaxy groups},
	volume = {1},
	issn = {2397-3366},
	url = {https://ui.adsabs.harvard.edu/abs/2017NatAs...1E..25S},
	doi = {10.1038/s41550-016-0025},
	urldate = {2024-03-13},
	journal = {Nature Astronomy},
	author = {Stierwalt, S. and Liss, S. E. and Johnson, K. E. and Patton, D. R. and Privon, G. C. and Besla, G. and Kallivayalil, N. and Putman, M.},
	month = jan,
	year = {2017},
	pages = {0025},
}

@article{julio_radial_2025,
	title = {The radial acceleration relation at the {EDGE} of galaxy formation: {Testing} its universality in low-mass dwarf galaxies},
	volume = {704},
	issn = {0004-6361},
	shorttitle = {The radial acceleration relation at the {EDGE} of galaxy formation},
	url = {https://ui.adsabs.harvard.edu/abs/2025A&A...704A.330J},
	doi = {10.1051/0004-6361/202557106},
	urldate = {2026-02-03},
	journal = {A\&A},
	publisher = {EDP},
	author = {Júlio, Mariana P. and Read, Justin I. and Pawlowski, Marcel S. and Li, Pengfei and Vaz, Daniel and Brinchmann, Jarle and Rey, Martin P. and Agertz, Oscar and Holmes, Tom},
	month = dec,
	year = {2025},
	pages = {A330},
}

@article{mcconnachie_observed_2012,
	title = {The {Observed} {Properties} of {Dwarf} {Galaxies} in and around the {Local} {Group}},
	volume = {144},
	doi = {10.1088/0004-6256/144/1/4},
	number = {1},
	journal = {\aj},
	author = {McConnachie, Alan W.},
	month = jul,
	year = {2012},
	pages = {4},
}

@article{blumenthal_formation_1984,
	title = {Formation of galaxies and large-scale structure with cold dark matter.},
	volume = {311},
	doi = {10.1038/311517a0},
	journal = {Nature},
	author = {Blumenthal, G. R. and Faber, S. M. and Primack, J. R. and Rees, M. J.},
	month = oct,
	year = {1984},
	pages = {517--525},
}

@article{read_dark_2019,
	title = {Dark matter heats up in dwarf galaxies},
	volume = {484},
	doi = {10.1093/mnras/sty3404},
	number = {1},
	journal = {MNRAS},
	author = {Read, J. I. and Walker, M. G. and Steger, P.},
	month = mar,
	year = {2019},
	pages = {1401--1420},
}

@article{hunter_matplotlib_2007,
	title = {Matplotlib: {A} {2D} {Graphics} {Environment}},
	volume = {9},
	doi = {10.1109/MCSE.2007.55},
	number = {3},
	journal = {CSE},
	author = {Hunter, John D.},
	month = may,
	year = {2007},
	pages = {90--95},
}

@article{collins_observational_2022,
	title = {Observational constraints on stellar feedback in dwarf galaxies},
	volume = {6},
	doi = {10.1038/s41550-022-01657-4},
	number = {6},
	journal = {Nature Astronomy},
	author = {Collins, Michelle L. M. and Read, Justin I.},
	month = may,
	year = {2022},
	note = {Publisher: Springer Science and Business Media LLC},
	pages = {647--658},
}

@article{battaglia_gaia_2022,
	title = {Gaia early {DR3} systemic motions of {Local} {Group} dwarf galaxies and orbital properties with a massive {Large} {Magellanic} {Cloud}},
	volume = {657},
	doi = {10.1051/0004-6361/202141528},
	journal = {A\&A},
	author = {Battaglia, G. and Taibi, S. and Thomas, G. F. and Fritz, T. K.},
	month = jan,
	year = {2022},
	pages = {A54},
}

@article{casetti-dinescu_hst_2024,
	title = {{HST} {Proper} {Motion} of {Andromeda} {III}. {Another} {Satellite} {Coorbiting} the {M31} {Satellite} {Plane}},
	volume = {975},
	issn = {0004-637X},
	url = {https://ui.adsabs.harvard.edu/abs/2024ApJ...975..138C},
	doi = {10.3847/1538-4357/ad7b10},
	urldate = {2026-02-05},
	journal = {ApJ},
	publisher = {IOP},
	author = {Casetti-Dinescu, Dana I. and Pawlowski, Marcel S. and Girard, Terrence M. and Kanehisa, Kosuke Jamie and Petroski, Alexander and Martone, Max and Kozhurina-Platais, Vera and Platais, Imants},
	month = nov,
	year = {2024},
	pages = {138},
}

@article{casetti-dinescu_hst_2025,
	title = {{HST} {Proper} {Motions} of {Andromeda} {V} and {Andromeda} {VI}},
	volume = {992},
	issn = {0004-637X},
	url = {https://ui.adsabs.harvard.edu/abs/2025ApJ...992...54C},
	doi = {10.3847/1538-4357/ae0479},
	urldate = {2026-01-22},
	journal = {ApJ},
	author = {Casetti-Dinescu, Dana I. and Pawlowski, Marcel S. and Girard, Terrence M. and Kanehisa, Kosuke Jamie and Martone, Max and Petroski, Alexander},
	month = oct,
	year = {2025},
	pages = {54},
}

@article{pawlowski_satellite_2024,
	title = {The satellite galaxy plane of {NGC} 4490 in light of {\textbackslash}{ensuremathŁambdaCDM}. {Sparsity} of similarly extreme analogs and a possible role of satellite pairs},
	volume = {688},
	doi = {10.1051/0004-6361/202449954},
	journal = {A\&A},
	author = {Pawlowski, Marcel S. and Müller, Oliver and Taibi, Salvatore and Júlio, Mariana P. and Kanehisa, Kosuke Jamie and Heesters, Nick},
	month = aug,
	year = {2024},
	pages = {A153},
}

@article{taibi_portrait_2024,
	title = {A portrait of the vast polar structure as a young phenomenon: {Hints} from its member satellites},
	volume = {681},
	doi = {10.1051/0004-6361/202347473},
	journal = {A\&A},
	author = {Taibi, S. and Pawlowski, M. S. and Khoperskov, S. and Steinmetz, M. and Libeskind, N. I.},
	month = jan,
	year = {2024},
	pages = {A73},
}

@article{ibata_vast_2013,
	title = {A vast, thin plane of corotating dwarf galaxies orbiting the {Andromeda} galaxy},
	volume = {493},
	number = {7430},
	journal = {Nature},
	author = {Ibata, Rodrigo A and Lewis, Geraint F and Conn, Anthony R and Irwin, Michael J and McConnachie, Alan W and Chapman, Scott C and Collins, Michelle L and Fardal, Mark and Ferguson, Annette MN and Ibata, Neil G and {others}},
	year = {2013},
	pages = {62--65},
    doi = {10.1038/nature11717}
}

@article{conn_three-dimensional_2013,
	title = {The three-dimensional structure of the {M31} satellite system; strong evidence for an inhomogeneous distribution of satellites},
	volume = {766},
	number = {2},
	journal = {ApJ},
	author = {Conn, Anthony R and Lewis, Geraint F and Ibata, Rodrigo A and Parker, Quentin A and Zucker, Daniel B and McConnachie, Alan W and Martin, Nicolas F and Valls-Gabaud, David and Tanvir, Nial and Irwin, Michael J and {others}},
	year = {2013},
	pages = {120},
    doi = {10.1088/0004-637X/766/2/120}
}

@article{bovy_galpy_2015,
	title = {galpy: {A} python {Library} for {Galactic} {Dynamics}},
	volume = {216},
	issn = {0067-0049},
	shorttitle = {galpy},
	doi = {10.1088/0067-0049/216/2/29},
	journal = {The Astrophysical Journal Supplement Series},
	publisher = {IOP},
	author = {Bovy, Jo},
	month = feb,
	year = {2015},
}

@article{vasiliev_effect_2023,
	title = {The effect of the {LMC} on the {Milky} {Way} system},
	volume = {11},
	number = {2},
	journal = {Galaxies},
	author = {Vasiliev, Eugene},
	year = {2023},
	pages = {59},
    doi = {10.3390/galaxies11020059},
}

@article{li_infall_2008,
	title = {Infall of substructures on to a {Milky} {Way}-like dark halo},
	volume = {385},
	number = {3},
	journal = {MNRAS},
	author = {Li, Yang-Shyang and Helmi, Amina},
	year = {2008},
	pages = {1365--1373},
}

@article{donghia_small_2008,
	title = {Small dwarf galaxies within larger dwarfs: why some are luminous while most go dark},
	volume = {686},
	number = {2},
	journal = {ApJ},
	author = {D’Onghia, Elena and Lake, George},
	year = {2008},
	pages = {L61},
    doi = {10.1086/592995}
}

@article{wang_spatial_2013,
	title = {The spatial distribution of galactic satellites in the $\Lambda$ cold dark matter cosmology},
	volume = {429},
	number = {2},
	journal = {MNRAS},
	author = {Wang, Jie and Frenk, Carlos S and Cooper, Andrew P},
	year = {2013},
	pages = {1502--1513},
    doi = {10.1093/mnras/sts442}
}

@article{harris_array_2020,
	title = {Array programming with {NumPy}},
	volume = {585},
	url = {https://doi.org/10.1038/s41586-020-2649-2},
	doi = {10.1038/s41586-020-2649-2},
	number = {7825},
	journal = {Nature},
	author = {Harris, Charles R. and Millman, K. Jarrod and Walt, Stéfan J. van der and Gommers, Ralf and Virtanen, Pauli and Cournapeau, David and Wieser, Eric and Taylor, Julian and Berg, Sebastian and Smith, Nathaniel J. and Kern, Robert and Picus, Matti and Hoyer, Stephan and Kerkwijk, Marten H. van and Brett, Matthew and Haldane, Allan and Río, Jaime Fernández del and Wiebe, Mark and Peterson, Pearu and Gérard-Marchant, Pierre and Sheppard, Kevin and Reddy, Tyler and Weckesser, Warren and Abbasi, Hameer and Gohlke, Christoph and Oliphant, Travis E.},
	month = sep,
	year = {2020},
	pages = {357--362},
}

@article{virtanen_scipy_2020,
	title = {{SciPy} 1.0: {Fundamental} {Algorithms} for {Scientific} {Computing} in {Python}},
	volume = {17},
	doi = {10.1038/s41592-019-0686-2},
	journal = {Nature Methods},
	author = {Virtanen, Pauli and Gommers, Ralf and Oliphant, Travis E. and Haberland, Matt and Reddy, Tyler and Cournapeau, David and Burovski, Evgeni and Peterson, Pearu and Weckesser, Warren and Bright, Jonathan and van der Walt, Stéfan J. and Brett, Matthew and Wilson, Joshua and Millman, K. Jarrod and Mayorov, Nikolay and Nelson, Andrew R. J. and Jones, Eric and Kern, Robert and Larson, Eric and Carey, C J and Polat, İlhan and Feng, Yu and Moore, Eric W. and VanderPlas, Jake and Laxalde, Denis and Perktold, Josef and Cimrman, Robert and Henriksen, Ian and Quintero, E. A. and Harris, Charles R. and Archibald, Anne M. and Ribeiro, Antônio H. and Pedregosa, Fabian and van Mulbregt, Paul and {SciPy 1.0 Contributors}},
	year = {2020},
	pages = {261--272},
}

@article{pace_proper_2022,
	title = {Proper motions, orbits, and tidal influences of milky way dwarf spheroidal galaxies},
	volume = {940},
	number = {2},
	journal = {ApJ},
	author = {Pace, Andrew B and Erkal, Denis and Li, Ting S},
	year = {2022},
	pages = {136},
    doi = {10.3847/1538-4357/ac997b}
}

@article{julio_satellite_2024,
	title = {Satellite group infall into the {Milky} {Way}: {Exploring} the {Crater}-{Leo} case with new {HST} proper motions},
	volume = {687},
	issn = {0004-6361},
	shorttitle = {Satellite group infall into the {Milky} {Way}},
	url = {https://ui.adsabs.harvard.edu/abs/2024A&A...687A.212J},
	doi = {10.1051/0004-6361/202449985},
	urldate = {2025-03-20},
	journal = {A\&A},
	publisher = {EDP},
	author = {Júlio, Mariana P. and Pawlowski, Marcel S. and Tony Sohn, Sangmo and Taibi, Salvatore and van der Marel, Roeland P. and McGaugh, Stacy S.},
	month = jul,
	year = {2024},
	pages = {A212},
}

@article{gu_younger_2024,
	title = {A younger {Universe} implied by satellite pair correlations from {SDSS} observations of massive galaxy groups},
	doi = {10.1038/s41550-023-02192-6},
	journal = {Nature Astronomy},
	author = {Gu, Qing and Guo, Qi and Cautun, Marius and Shao, Shi and Pei, Wenxiang and Wang, Wenting and Gao, Liang and Wang, Jie},
	month = jan,
	year = {2024},
	note = {\_eprint: 2401.10342},
}

@article{ibata_velocity_2014,
	title = {Velocity anti-correlation of diametrically opposed galaxy satellites in the low-redshift {Universe}},
	volume = {511},
	doi = {10.1038/nature13481},
	number = {7511},
	journal = {\nat},
	author = {Ibata, Neil G. and Ibata, Rodrigo A. and Famaey, Benoit and Lewis, Geraint F.},
	month = jul,
	year = {2014},
	note = {\_eprint: 1407.8178},
	pages = {563--566},
}

@article{pawlowski_alignment_2016,
	title = {The alignment of {SDSS} satellites with the {VPOS}: effects of the survey footprint shape},
	volume = {456},
	doi = {10.1093/mnras/stv2673},
	number = {1},
	journal = {MNRAS},
	author = {Pawlowski, Marcel S.},
	month = feb,
	year = {2016},
	pages = {448--458},
}

@article{vasiliev_dear_2024,
	title = {Dear {Magellanic} {Clouds}, welcome back!},
	volume = {527},
	doi = {10.1093/mnras/stad2612},
	number = {1},
	journal = {MNRAS},
	author = {Vasiliev, Eugene},
	month = jan,
	year = {2024},
	note = {\_eprint: 2306.04837},
	pages = {437--456},
}

@article{watkins_masses_2010,
	title = {The masses of the {Milky} {Way} and {Andromeda} galaxies},
	volume = {406},
	issn = {0035-8711},
	url = {https://ui.adsabs.harvard.edu/abs/2010MNRAS.406..264W},
	doi = {10.1111/j.1365-2966.2010.16708.x},
	urldate = {2026-01-30},
	journal = {MNRAS},
	publisher = {OUP},
	author = {Watkins, Laura L. and Evans, N. Wyn and An, Jin H.},
	month = jul,
	year = {2010},
	note = {ADS Bibcode: 2010MNRAS.406..264W},
	pages = {264--278},
}

@article{deason_satellite_2014,
	title = {Satellite {Dwarf} {Galaxies} in a {Hierarchical} {Universe}: {The} {Prevalence} of {Dwarf}-{Dwarf} {Major} {Mergers}},
	volume = {794},
	issn = {0004-637X},
	shorttitle = {Satellite {Dwarf} {Galaxies} in a {Hierarchical} {Universe}},
	url = {https://ui.adsabs.harvard.edu/abs/2014ApJ...794..115D},
	doi = {10.1088/0004-637X/794/2/115},
	urldate = {2026-02-23},
	journal = {ApJ},
	publisher = {IOP},
	author = {Deason, Alis and Wetzel, Andrew and Garrison-Kimmel, Shea},
	month = oct,
	year = {2014},
	pages = {115},
}

@article{sales_satellites_2013,
	title = {Satellites and haloes of dwarf galaxies},
	volume = {428},
	issn = {0035-8711},
	url = {https://doi.org/10.1093/mnras/sts054},
	doi = {10.1093/mnras/sts054},
	number = {1},
	urldate = {2026-02-23},
	journal = {MNRAS},
	author = {Sales, Laura V. and Wang, Wenting and White, Simon D. M. and Navarro, Julio F.},
	month = jan,
	year = {2013},
	pages = {573--578},
}

@article{patton_dynamically_2002,
	title = {Dynamically {Close} {Galaxy} {Pairs} and {Merger} {Rate} {Evolution} in the {CNOC2} {Redshift} {Survey}},
	volume = {565},
	issn = {0004-637X},
	url = {https://ui.adsabs.harvard.edu/abs/2002ApJ...565..208P},
	doi = {10.1086/324543},
	urldate = {2026-02-18},
	journal = {ApJ},
	publisher = {IOP},
	author = {Patton, D. R. and Pritchet, C. J. and Carlberg, R. G. and Marzke, R. O. and Yee, H. K. C. and Hall, P. B. and Lin, H. and Morris, S. L. and Sawicki, M. and Shepherd, C. W. and Wirth, G. D.},
	month = jan,
	year = {2002},
	pages = {208--222},
}

@article{duncan_observational_2019,
	title = {Observational {Constraints} on the {Merger} {History} of {Galaxies} since z ≈ 6: {Probabilistic} {Galaxy} {Pair} {Counts} in the {CANDELS} {Fields}},
	volume = {876},
	issn = {0004-637X},
	shorttitle = {Observational {Constraints} on the {Merger} {History} of {Galaxies} since z ≈ 6},
	url = {https://ui.adsabs.harvard.edu/abs/2019ApJ...876..110D},
	doi = {10.3847/1538-4357/ab148a},
	urldate = {2026-02-18},
	journal = {ApJ},
	publisher = {IOP},
	author = {Duncan, Kenneth and Conselice, Christopher J. and Mundy, Carl and Bell, Eric and Donley, Jennifer and Galametz, Audrey and Guo, Yicheng and Grogin, Norman A. and Hathi, Nimish and Kartaltepe, Jeyhan and Kocevski, Dale and Koekemoer, Anton M. and Pérez-González, Pablo G. and Mantha, Kameswara B. and Snyder, Gregory F. and Stefanon, Mauro},
	month = may,
	year = {2019},
	pages = {110},
}

@article{callingham_group_2025,
	title = {Group accretion in {Milky} {Way}-like stellar haloes},
	volume = {700},
	issn = {0004-6361},
	url = {https://ui.adsabs.harvard.edu/abs/2025A&A...700A.175C},
	doi = {10.1051/0004-6361/202554966},
	urldate = {2026-02-23},
	journal = {A\&A},
	publisher = {EDP},
	author = {Callingham, Thomas M. and Helmi, Amina},
	month = aug,
	year = {2025},
	pages = {A175},
}

@article{stierwalt_tiny_2015,
	title = {{TiNy} {Titans}: {The} {Role} of {Dwarf}-{Dwarf} {Interactions} in {Low}-mass {Galaxy} {Evolution}},
	volume = {805},
	issn = {0004-637X},
	shorttitle = {{TiNy} {Titans}},
	url = {https://ui.adsabs.harvard.edu/abs/2015ApJ...805....2S},
	doi = {10.1088/0004-637X/805/1/2},
	urldate = {2026-02-23},
	journal = {ApJ},
	publisher = {IOP},
	author = {Stierwalt, S. and Besla, G. and Patton, D. and Johnson, K. and Kallivayalil, N. and Putman, M. and Privon, G. and Ross, G.},
	month = may,
	year = {2015},
	note = {ADS Bibcode: 2015ApJ...805....2S},
	pages = {2},
}

@article{besla_are_2007,
	title = {Are the {Magellanic} {Clouds} on {Their} {First} {Passage} about the {Milky} {Way}?},
	volume = {668},
	issn = {0004-637X},
	url = {https://ui.adsabs.harvard.edu/abs/2007ApJ...668..949B},
	doi = {10.1086/521385},
	urldate = {2026-03-12},
	journal = {ApJ},
	publisher = {IOP},
	author = {Besla, Gurtina and Kallivayalil, Nitya and Hernquist, Lars and Robertson, Brant and Cox, T. J. and van der Marel, Roeland P. and Alcock, Charles},
	month = oct,
	year = {2007},
	pages = {949--967},
}

@article{libeskind_hestia_2020,
	title = {The {Hestia} project: simulations of the {Local} {Group}},
	volume = {498},
	issn = {0035-8711, 1365-2966},
	shorttitle = {The {Hestia} project},
	url = {http://arxiv.org/abs/2008.04926},
	doi = {10.1093/mnras/staa2541},
	number = {2},
	urldate = {2025-01-09},
	journal = {MNRAS},
	author = {Libeskind, Noam I. and Carlesi, Edoardo and Grand, Rob J. J. and Khalatyan, Arman and Knebe, Alexander and Pakmor, Ruediger and Pilipenko, Sergey and Pawlowski, Marcel S. and Sparre, Martin and Tempel, Elmo and Wang, Peng and Courtois, Helene M. and Gottloeber, Stefan and Hoffman, Yehuda and Minchev, Ivan and Pfrommer, Christoph and Sorce, Jenny G. and Springel, Volker and Steinmetz, Matthias and Tully, R. Brent and Vogelsberger, Mark and Yepes, Gustavo},
	month = sep,
	year = {2020},
	pages = {2968--2983},
}

@article{besla_frequency_2018,
	title = {The frequency of dwarf galaxy multiples at low redshift in {SDSS} versus cosmological expectations},
	volume = {480},
	issn = {0035-8711},
	url = {https://ui.adsabs.harvard.edu/abs/2018MNRAS.480.3376B},
	doi = {10.1093/mnras/sty2041},
	urldate = {2024-08-19},
	journal = {MNRAS},
	publisher = {OUP},
	author = {Besla, Gurtina and Patton, David R. and Stierwalt, Sabrina and Rodriguez-Gomez, Vicente and Patel, Ekta and Kallivayalil, Nitya J. and Johnson, Kelsey E. and Pearson, Sarah and Privon, George C. and Putman, Mary E.},
	month = nov,
	year = {2018},
	pages = {3376--3396},
}

@article{planck_collaboration_planck_2014,
	title = {Planck 2013 results. {XVI}. {Cosmological} parameters},
	volume = {571},
	issn = {0004-6361},
	url = {https://ui.adsabs.harvard.edu/abs/2014A&A...571A..16P},
	doi = {10.1051/0004-6361/201321591},
	urldate = {2025-02-19},
	journal = {A\&A},
	author = {{Planck Collaboration} and Ade, P. A. R. and Aghanim, N. and Armitage-Caplan, C. and Arnaud, M. and Ashdown, M. and Atrio-Barandela, F. and Aumont, J. and Baccigalupi, C. and Banday, A. J. and Barreiro, R. B. and Bartlett, J. G. and Battaner, E. and Benabed, K. and Benoît, A. and Benoit-Lévy, A. and Bernard, J. -P. and Bersanelli, M. and Bielewicz, P. and Bobin, J. and Bock, J. J. and Bonaldi, A. and Bond, J. R. and Borrill, J. and Bouchet, F. R. and Bridges, M. and Bucher, M. and Burigana, C. and Butler, R. C. and Calabrese, E. and Cappellini, B. and Cardoso, J. -F. and Catalano, A. and Challinor, A. and Chamballu, A. and Chary, R. -R. and Chen, X. and Chiang, H. C. and Chiang, L. -Y. and Christensen, P. R. and Church, S. and Clements, D. L. and Colombi, S. and Colombo, L. P. L. and Couchot, F. and Coulais, A. and Crill, B. P. and Curto, A. and Cuttaia, F. and Danese, L. and Davies, R. D. and Davis, R. J. and de Bernardis, P. and de Rosa, A. and de Zotti, G. and Delabrouille, J. and Delouis, J. -M. and Désert, F. -X. and Dickinson, C. and Diego, J. M. and Dolag, K. and Dole, H. and Donzelli, S. and Doré, O. and Douspis, M. and Dunkley, J. and Dupac, X. and Efstathiou, G. and Elsner, F. and Enßlin, T. A. and Eriksen, H. K. and Finelli, F. and Forni, O. and Frailis, M. and Fraisse, A. A. and Franceschi, E. and Gaier, T. C. and Galeotta, S. and Galli, S. and Ganga, K. and Giard, M. and Giardino, G. and Giraud-Héraud, Y. and Gjerløw, E. and González-Nuevo, J. and Górski, K. M. and Gratton, S. and Gregorio, A. and Gruppuso, A. and Gudmundsson, J. E. and Haissinski, J. and Hamann, J. and Hansen, F. K. and Hanson, D. and Harrison, D. and Henrot-Versillé, S. and Hernández-Monteagudo, C. and Herranz, D. and Hildebrandt, S. R. and Hivon, E. and Hobson, M. and Holmes, W. A. and Hornstrup, A. and Hou, Z. and Hovest, W. and Huffenberger, K. M. and Jaffe, A. H. and Jaffe, T. R. and Jewell, J. and Jones, W. C. and Juvela, M. and Keihänen, E. and Keskitalo, R. and Kisner, T. S. and Kneissl, R. and Knoche, J. and Knox, L. and Kunz, M. and Kurki-Suonio, H. and Lagache, G. and Lähteenmäki, A. and Lamarre, J. -M. and Lasenby, A. and Lattanzi, M. and Laureijs, R. J. and Lawrence, C. R. and Leach, S. and Leahy, J. P. and Leonardi, R. and León-Tavares, J. and Lesgourgues, J. and Lewis, A. and Liguori, M. and Lilje, P. B. and Linden-Vørnle, M. and López-Caniego, M. and Lubin, P. M. and Macías-Pérez, J. F. and Maffei, B. and Maino, D. and Mandolesi, N. and Maris, M. and Marshall, D. J. and Martin, P. G. and Martínez-González, E. and Masi, S. and Massardi, M. and Matarrese, S. and Matthai, F. and Mazzotta, P. and Meinhold, P. R. and Melchiorri, A. and Melin, J. -B. and Mendes, L. and Menegoni, E. and Mennella, A. and Migliaccio, M. and Millea, M. and Mitra, S. and Miville-Deschênes, M. -A. and Moneti, A. and Montier, L. and Morgante, G. and Mortlock, D. and Moss, A. and Munshi, D. and Murphy, J. A. and Naselsky, P. and Nati, F. and Natoli, P. and Netterfield, C. B. and Nørgaard-Nielsen, H. U. and Noviello, F. and Novikov, D. and Novikov, I. and O'Dwyer, I. J. and Osborne, S. and Oxborrow, C. A. and Paci, F. and Pagano, L. and Pajot, F. and Paladini, R. and Paoletti, D. and Partridge, B. and Pasian, F. and Patanchon, G. and Pearson, D. and Pearson, T. J. and Peiris, H. V. and Perdereau, O. and Perotto, L. and Perrotta, F. and Pettorino, V. and Piacentini, F. and Piat, M. and Pierpaoli, E. and Pietrobon, D. and Plaszczynski, S. and Platania, P. and Pointecouteau, E.},
	month = nov,
	year = {2014},
	pages = {A16},
}

@article{grand_auriga_2017,
	title = {The {Auriga} {Project}: the properties and formation mechanisms of disc galaxies across cosmic time},
	volume = {467},
	issn = {0035-8711},
	shorttitle = {The {Auriga} {Project}},
	url = {https://doi.org/10.1093/mnras/stx071},
	doi = {10.1093/mnras/stx071},
	number = {1},
	urldate = {2025-02-19},
	journal = {MNRAS},
	author = {Grand, Robert J. J. and Gómez, Facundo A. and Marinacci, Federico and Pakmor, Rüdiger and Springel, Volker and Campbell, David J. R. and Frenk, Carlos S. and Jenkins, Adrian and White, Simon D. M.},
	month = may,
	year = {2017},
	pages = {179--207},
}

@article{springel_e_2010,
	title = {E pur si muove: {Galilean}-invariant cosmological hydrodynamical simulations on a moving mesh},
	volume = {401},
	issn = {0035-8711},
	shorttitle = {E pur si muove},
	url = {https://doi.org/10.1111/j.1365-2966.2009.15715.x},
	doi = {10.1111/j.1365-2966.2009.15715.x},
	number = {2},
	urldate = {2025-02-19},
	journal = {MNRAS},
	author = {Springel, Volker},
	month = jan,
	year = {2010},
	pages = {791--851},
}

@article{metz_discs_2009,
	title = {Discs of {Satellites}: the new dwarf spheroidals},
	volume = {394},
	issn = {00358711, 13652966},
	shorttitle = {Discs of {Satellites}},
	url = {http://arxiv.org/abs/0901.1658},
	doi = {10.1111/j.1365-2966.2009.14489.x},
	number = {4},
	urldate = {2026-03-30},
	journal = {MNRAS},
	author = {Metz, Manuel and Kroupa, Pavel and Jerjen, Helmut},
	month = apr,
	year = {2009},
	pages = {2223--2228},
}

@article{yan_kids-legacy_2025,
	title = {{KiDS}-{Legacy}: {Angular} galaxy clustering from deep surveys with complex selection effects},
	volume = {694},
	copyright = {© The Authors 2025},
	issn = {0004-6361, 1432-0746},
	shorttitle = {{KiDS}-{Legacy}},
	url = {https://www.aanda.org/articles/aa/abs/2025/02/aa52808-24/aa52808-24.html},
	doi = {10.1051/0004-6361/202452808},
	language = {en},
	urldate = {2026-03-30},
	journal = {A\&A},
	publisher = {EDP Sciences},
	author = {Yan, Ziang and Wright, Angus H. and Chisari, Nora Elisa and Georgiou, Christos and Joudaki, Shahab and Loureiro, Arthur and Reischke, Robert and Asgari, Marika and Bilicki, Maciej and Dvornik, Andrej and Heymans, Catherine and Hildebrandt, Hendrik and Jalan, Priyanka and Joachimi, Benjamin and Lesci, Giorgio Francesco and Li, Shun-Sheng and Linke, Laila and Mahony, Constance and Moscardini, Lauro and Napolitano, Nicola R. and Stölzner, Benjamin and Wietersheim-Kramsta, Maximilian Von and Yoon, Mijin},
	month = feb,
	year = {2025},
	pages = {A259},
}

@article{cameron_galaxy_2007,
	title = {The galaxy luminosity–size relation and selection biases in the {Hubble} {Ultra} {Deep} {Field}},
	volume = {377},
	issn = {0035-8711},
	url = {https://doi.org/10.1111/j.1365-2966.2007.11507.x},
	doi = {10.1111/j.1365-2966.2007.11507.x},
	number = {2},
	urldate = {2026-03-30},
	journal = {MNRAS},
	author = {Cameron, E. and Driver, S. P.},
	month = may,
	year = {2007},
	pages = {523--534},
}

@article{erkal_total_2019,
	title = {The total mass of the {Large} {Magellanic} {Cloud} from its perturbation on the {Orphan} stream},
	volume = {487},
	issn = {0035-8711},
	url = {https://ui.adsabs.harvard.edu/abs/2019MNRAS.487.2685E},
	doi = {10.1093/mnras/stz1371},
	urldate = {2026-04-07},
	journal = {MNRAS},
	publisher = {OUP},
	author = {Erkal, D. and Belokurov, V. and Laporte, C. F. P. and Koposov, S. E. and Li, T. S. and Grillmair, C. J. and Kallivayalil, N. and Price-Whelan, A. M. and Evans, N. W. and Hawkins, K. and Hendel, D. and Mateu, C. and Navarro, J. F. and del Pino, A. and Slater, C. T. and Sohn, S. T. and {Orphan Aspen Treasury Collaboration}},
	month = aug,
	year = {2019},
	pages = {2685--2700},
}

@article{doliva-dolinsky_pandas_2022,
	title = {The {PAndAS} {View} of the {Andromeda} {Satellite} {System}. {III}. {Dwarf} {Galaxy} {Detection} {Limits}},
	volume = {933},
	issn = {0004-637X},
	url = {https://ui.adsabs.harvard.edu/abs/2022ApJ...933..135D},
	doi = {10.3847/1538-4357/ac6fd5},
	urldate = {2026-08-24},
	journal = {The Astrophysical Journal},
	publisher = {IOP},
	author = {Doliva-Dolinsky, Amandine and Martin, Nicolas F. and Thomas, Guillaume F. and Ferguson, Annette M. N. and Ibata, Rodrigo A. and Lewis, Geraint F. and Mackey, Dougal and McConnachie, Alan W. and Yuan, Zhen},
	month = jul,
	year = {2022},
	note = {ADS Bibcode: 2022ApJ...933..135D},
	pages = {135}
}

@article{pearson_local_2016,
	title = {Local {Volume} {TiNy} {Titans}: gaseous dwarf-dwarf interactions in the {Local} {Universe}},
	volume = {459},
	issn = {0035-8711},
	shorttitle = {Local {Volume} {TiNy} {Titans}},
	url = {https://ui.adsabs.harvard.edu/abs/2016MNRAS.459.1827P},
	doi = {10.1093/mnras/stw757},
	urldate = {2024-08-22},
	journal = {Monthly Notices of the Royal Astronomical Society},
	publisher = {OUP},
	author = {Pearson, Sarah and Besla, Gurtina and Putman, Mary E. and Lutz, Katharina A. and Fernández, Ximena and Stierwalt, Sabrina and Patton, David R. and Kim, Jinhyub and Kallivayalil, Nitya and Johnson, Kelsey and Sung, Eon-Chang},
	month = jun,
	year = {2016},
	note = {ADS Bibcode: 2016MNRAS.459.1827P},
	pages = {1827--1846},
}

@article{paudel_discovery_2024,
	title = {Discovery of a {Rare} {Group} of {Dwarf} {Galaxies} in the {Local} {Universe}},
	volume = {976},
	issn = {2041-8205, 2041-8213},
	url = {http://arxiv.org/abs/2411.10045},
	doi = {10.3847/2041-8213/ad8f3c},
	number = {1},
	urldate = {2025-05-23},
	journal = {The Astrophysical Journal Letters},
	author = {Paudel, Sanjaya and Sabiu, Cristiano G. and Yoon, Suk-Jin and Duc, Pierre-Alain and Yoo, Jaewon and Müller, Oliver},
	month = nov,
	year = {2024},
	note = {arXiv:2411.10045 [astro-ph]},
	pages = {L18},
}

@misc{paudel_groups_2026,
	title = {Groups of {Dwarf} {Galaxies} in the {Local} {Universe}},
	url = {https://ui.adsabs.harvard.edu/abs/2026arXiv260616299P},
	doi = {10.48550/arXiv.2606.16299},
	urldate = {2026-08-26},
	publisher = {arXiv},
	author = {Paudel, Sanjaya and Sabiu, Cristiano G. and Yoon, Suk-Jin and Nidhi Chhatkuli, Daya and Yoo, Jaewon and Pokhrel, Nau Raj},
	month = jun,
	year = {2026},
	note = {ADS Bibcode: 2026arXiv260616299P},
}

@ARTICLE{Jenkins2021ApJ...920...92J,
       author = {{Jenkins}, Sydney A. and {Li}, Ting S. and {Pace}, Andrew B. and {Ji}, Alexander P. and {Koposov}, Sergey E. and {Mutlu-Pakdil}, Bur{\c{c}}in},
        title = "{Very Large Telescope Spectroscopy of Ultra-faint Dwarf Galaxies. I. Bo{\"o}tes I, Leo IV, and Leo V}",
      journal = {\apj},
         year = 2021,
        month = oct,
       volume = {920},
       number = {2},
          eid = {92},
        pages = {92},
          doi = {10.3847/1538-4357/ac1353},
archivePrefix = {arXiv},
       eprint = {2101.00013},
 primaryClass = {astro-ph.GA},
       adsurl = {https://ui.adsabs.harvard.edu/abs/2021ApJ...920...92J}
}

@ARTICLE{Koposov2015ApJ...805..130K,
   author = {{Koposov}, S.~E. and {Belokurov}, V. and {Torrealba}, G. and 
	{Evans}, N.~W.},
    title = "{Beasts of the Southern Wild : Discovery of nine Ultra Faint satellites in the vicinity of the Magellanic Clouds.}",
  journal = {\apj},
archivePrefix = "arXiv",
   eprint = {1503.02079},
     year = 2015,
    month = jun,
   volume = 805,
      eid = {130},
    pages = {130},
      doi = {10.1088/0004-637X/805/2/130},
   adsurl = {http://adsabs.harvard.edu/abs/2015ApJ...805..130K}
}

@ARTICLE{Cerny2021ApJ...920L..44C,
       author = {{Cerny}, W. and {Pace}, A.~B. and {Drlica-Wagner}, A. and {Koposov}, S.~E. and {Vivas}, A.~K. and {Mau}, S. and {Riley}, A.~H. and {Bom}, C.~R. and {Carlin}, J.~L. and {Choi}, Y. and {Erkal}, D. and {Ferguson}, P.~S. and {James}, D.~J. and {Li}, T.~S. and {Mart{\'\i}nez-Delgado}, D. and {Mart{\'\i}nez-V{\'a}zquez}, C.~E. and {Munoz}, R.~R. and {Mutlu-Pakdil}, B. and {Olsen}, K.~A.~G. and {Pieres}, A. and {Sakowska}, J.~D. and {Sand}, D.~J. and {Simon}, J.~D. and {Smercina}, A. and {Stringfellow}, G.~S. and {Tollerud}, E.~J. and {Adam{\'o}w}, M. and {Hernandez-Lang}, D. and {Kuropatkin}, N. and {Santana-Silva}, L. and {Tucker}, D.~L. and {Zenteno}, A. and {Delve Collaboration}},
        title = "{Eridanus IV: an Ultra-faint Dwarf Galaxy Candidate Discovered in the DECam Local Volume Exploration Survey}",
      journal = {\apjl},
         year = 2021,
        month = oct,
       volume = {920},
       number = {2},
          eid = {L44},
        pages = {L44},
          doi = {10.3847/2041-8213/ac2d9a},
archivePrefix = {arXiv},
       eprint = {2107.09080},
 primaryClass = {astro-ph.GA},
       adsurl = {https://ui.adsabs.harvard.edu/abs/2021ApJ...920L..44C}
}

@ARTICLE{Koposov2018MNRAS.479.5343K,
   author = {{Koposov}, S.~E. and {Walker}, M.~G. and {Belokurov}, V. and 
	{Casey}, A.~R. and {Geringer-Sameth}, A. and {Mackey}, D. and 
	{Da Costa}, G. and {Erkal}, D. and {Jethwa}, P. and {Mateo}, M. and 
	{Olszewski}, E.~W. and {Bailey}, J.~I.},
    title = "{Snake in the Clouds: a new nearby dwarf galaxy in the Magellanic bridge*}",
  journal = {\mnras},
archivePrefix = "arXiv",
   eprint = {1804.06430},
     year = 2018,
    month = oct,
   volume = 479,
    pages = {5343-5361},
      doi = {10.1093/mnras/sty1772},
   adsurl = {http://adsabs.harvard.edu/abs/2018MNRAS.479.5343K}
}

@ARTICLE{Cerny2023ApJ...942..111C,
       author = {{Cerny}, W. and {Simon}, J.~D. and {Li}, T.~S. and {Drlica-Wagner}, A. and {Pace}, A.~B. and {Mart{\'\i}nez-V{\'a}zquez}, C.~E. and {Riley}, A.~H. and {Mutlu-Pakdil}, B. and {Mau}, S. and {Ferguson}, P.~S. and {Erkal}, D. and {Munoz}, R.~R. and {Bom}, C.~R. and {Carlin}, J.~L. and {Carollo}, D. and {Choi}, Y. and {Ji}, A.~P. and {Manwadkar}, V. and {Mart{\'\i}nez-Delgado}, D. and {Miller}, A.~E. and {No{\"e}l}, N.~E.~D. and {Sakowska}, J.~D. and {Sand}, D.~J. and {Stringfellow}, G.~S. and {Tollerud}, E.~J. and {Vivas}, A.~K. and {Carballo-Bello}, J.~A. and {Hernandez-Lang}, D. and {James}, D.~J. and {Nidever}, D.~L. and {Castellon}, J.~L. Nilo and {Olsen}, K.~A.~G. and {Zenteno}, A. and {Delve Collaboration}},
        title = "{Pegasus IV: Discovery and Spectroscopic Confirmation of an Ultra-faint Dwarf Galaxy in the Constellation Pegasus}",
      journal = {\apj},
         year = 2023,
        month = jan,
       volume = {942},
       number = {2},
          eid = {111},
        pages = {111},
          doi = {10.3847/1538-4357/aca1c3},
archivePrefix = {arXiv},
       eprint = {2203.11788},
 primaryClass = {astro-ph.GA},
       adsurl = {https://ui.adsabs.harvard.edu/abs/2023ApJ...942..111C}
}

@ARTICLE{Richardson2011ApJ...732...76R,
       author = {{Richardson}, Jenny C. and {Irwin}, Mike J. and {McConnachie}, Alan W. and {Martin}, Nicolas F. and {Dotter}, Aaron L. and {Ferguson}, Annette M.~N. and {Ibata}, Rodrigo A. and {Chapman}, Scott C. and {Lewis}, Geraint F. and {Tanvir}, Nial R. and {Rich}, R. Michael},
        title = "{PAndAS' Progeny: Extending the M31 Dwarf Galaxy Cabal}",
      journal = {\apj},
         year = 2011,
        month = may,
       volume = {732},
       number = {2},
          eid = {76},
        pages = {76},
          doi = {10.1088/0004-637X/732/2/76},
archivePrefix = {arXiv},
       eprint = {1102.2902},
 primaryClass = {astro-ph.CO},
       adsurl = {https://ui.adsabs.harvard.edu/abs/2011ApJ...732...76R}
}

@ARTICLE{Lee2009ApJ...703..692L,
       author = {{Lee}, Myung Gyoon and {Yuk}, In-Soo and {Park}, Hong Soo and {Harris}, Jason and {Zaritsky}, Dennis},
        title = "{Star Formation History and Chemical Evolution of the Sextans Dwarf Spheroidal Galaxy}",
      journal = {\apj},
         year = 2009,
        month = sep,
       volume = {703},
       number = {1},
        pages = {692-701},
          doi = {10.1088/0004-637X/703/1/692},
archivePrefix = {arXiv},
       eprint = {0907.5102},
 primaryClass = {astro-ph.CO},
       adsurl = {https://ui.adsabs.harvard.edu/abs/2009ApJ...703..692L}
}

@ARTICLE{Oakes2022ApJ...929..116O,
       author = {{Oakes}, Elias K. and {Hoyt}, Taylor J. and {Freedman}, Wendy L. and {Madore}, Barry F. and {Tran}, Quang H. and {Cerny}, William and {Beaton}, Rachael L. and {Seibert}, Mark},
        title = "{Distances to Local Group Galaxies via Population II, Stellar Distance Indicators. II. The Fornax Dwarf Spheroidal}",
      journal = {\apj},
         year = 2022,
        month = apr,
       volume = {929},
       number = {2},
          eid = {116},
        pages = {116},
          doi = {10.3847/1538-4357/ac5b07},
archivePrefix = {arXiv},
       eprint = {2204.09699},
 primaryClass = {astro-ph.GA},
       adsurl = {https://ui.adsabs.harvard.edu/abs/2022ApJ...929..116O}
}

@ARTICLE{Ji2021ApJ...921...32J,
       author = {{Ji}, Alexander P. and {Koposov}, Sergey E. and {Li}, Ting S. and {Erkal}, Denis and {Pace}, Andrew B. and {Simon}, Joshua D. and {Belokurov}, Vasily and {Cullinane}, Lara R. and {Da Costa}, Gary S. and {Kuehn}, Kyler and {Lewis}, Geraint F. and {Mackey}, Dougal and {Shipp}, Nora and {Simpson}, Jeffrey D. and {Zucker}, Daniel B. and {Hansen}, Terese T. and {Bland-Hawthorn}, Joss and {S5 Collaboration}},
        title = "{Kinematics of Antlia 2 and Crater 2 from the Southern Stellar Stream Spectroscopic Survey (S$^{5}$)}",
      journal = {\apj},
         year = 2021,
        month = nov,
       volume = {921},
       number = {1},
          eid = {32},
        pages = {32},
          doi = {10.3847/1538-4357/ac1869},
archivePrefix = {arXiv},
       eprint = {2106.12656},
 primaryClass = {astro-ph.GA},
       adsurl = {https://ui.adsabs.harvard.edu/abs/2021ApJ...921...32J}
}

@ARTICLE{Munoz2018ApJ...860...66M,
   author = {{Mu{\~n}oz}, R.~R. and {C{\^o}t{\'e}}, P. and {Santana}, F.~A. and 
	{Geha}, M. and {Simon}, J.~D. and {Oyarz{\'u}n}, G.~A. and {Stetson}, P.~B. and 
	{Djorgovski}, S.~G.},
    title = "{A MegaCam Survey of Outer Halo Satellites. III. Photometric and Structural Parameters}",
  journal = {\apj},
     year = 2018,
    month = jun,
   volume = 860,
      eid = {66},
    pages = {66},
      doi = {10.3847/1538-4357/aac16b},
   adsurl = {http://adsabs.harvard.edu/abs/2018ApJ...860...66M}
}

@ARTICLE{Torrealba2018MNRAS.475.5085T,
   author = {{Torrealba}, G. and {Belokurov}, V. and {Koposov}, S.~E. and 
	{Bechtol}, K. and {Drlica-Wagner}, A. and {Olsen}, K.~A.~G. and 
	{Vivas}, A.~K. and {Yanny}, B. and {Jethwa}, P. and {Walker}, A.~R. and 
	{Li}, T.~S. and {Allam}, S. and {Conn}, B.~C. and {Gallart}, C. and 
	{Gruendl}, R.~A. and {James}, D.~J. and {Johnson}, M.~D. and 
	{Kuehn}, K. and {Kuropatkin}, N. and {Martin}, N.~F. and {Martinez-Delgado}, D. and 
	{Nidever}, D.~L. and {No{\"e}l}, N.~E.~D. and {Simon}, J.~D. and 
	{Stringfellow}, G.~S. and {Tucker}, D.~L.},
    title = "{Discovery of two neighbouring satellites in the Carina constellation with MagLiteS}",
  journal = {\mnras},
archivePrefix = "arXiv",
   eprint = {1801.07279},
     year = 2018,
    month = apr,
   volume = 475,
    pages = {5085-5097},
      doi = {10.1093/mnras/sty170},
   adsurl = {http://adsabs.harvard.edu/abs/2018MNRAS.475.5085T}
}

@ARTICLE{Spencer2018AJ....156..257S,
       author = {{Spencer}, Meghin E. and {Mateo}, Mario and {Olszewski}, Edward W. and
         {Walker}, Matthew G. and {McConnachie}, Alan W. and {Kirby}, Evan N.},
        title = "{The Binary Fraction of Stars in Dwarf Galaxies: The Cases of Draco and Ursa Minor}",
      journal = {\aj},
         year = "2018",
        month = "Dec",
       volume = {156},
          eid = {257},
        pages = {257},
          doi = {10.3847/1538-3881/aae3e4},
archivePrefix = {arXiv},
       eprint = {1811.06597},
 primaryClass = {astro-ph.GA},
       adsurl = {https://ui.adsabs.harvard.edu/\#abs/2018AJ....156..257S}
}

@ARTICLE{Chiti2023AJ....165...55C,
       author = {{Chiti}, Anirudh and {Frebel}, Anna and {Ji}, Alexander P. and {Mardini}, Mohammad K. and {Ou}, Xiaowei and {Simon}, Joshua D. and {Jerjen}, Helmut and {Kim}, Dongwon and {Norris}, John E.},
        title = "{Detailed Chemical Abundances of Stars in the Outskirts of the Tucana II Ultrafaint Dwarf Galaxy}",
      journal = {\aj},
         year = 2023,
        month = feb,
       volume = {165},
       number = {2},
          eid = {55},
        pages = {55},
          doi = {10.3847/1538-3881/aca416},
archivePrefix = {arXiv},
       eprint = {2205.01740},
 primaryClass = {astro-ph.GA},
       adsurl = {https://ui.adsabs.harvard.edu/abs/2023AJ....165...55C}
}

@ARTICLE{Walker2015ApJ...808..108W,
   author = {{Walker}, M.~G. and {Mateo}, M. and {Olszewski}, E.~W. and {Bailey}, III, J.~I. and 
	{Koposov}, S.~E. and {Belokurov}, V. and {Evans}, N.~W.},
    title = "{Magellan/M2FS Spectroscopy of the Reticulum 2 Dwarf Spheroidal Galaxy}",
  journal = {\apj},
archivePrefix = "arXiv",
   eprint = {1504.03060},
     year = 2015,
    month = aug,
   volume = 808,
      eid = {108},
    pages = {108},
      doi = {10.1088/0004-637X/808/2/108},
   adsurl = {http://adsabs.harvard.edu/abs/2015ApJ...808..108W}
}

@ARTICLE{Spencer2017ApJ...836..202S,
   author = {{Spencer}, M.~E. and {Mateo}, M. and {Walker}, M.~G. and {Olszewski}, E.~W.
	},
    title = "{A Multi-epoch Kinematic Study of the Remote Dwarf Spheroidal Galaxy Leo II}",
  journal = {\apj},
archivePrefix = "arXiv",
   eprint = {1702.08836},
     year = 2017,
    month = feb,
   volume = 836,
      eid = {202},
    pages = {202},
      doi = {10.3847/1538-4357/836/2/202},
   adsurl = {http://adsabs.harvard.edu/abs/2017ApJ...836..202S}
}

@ARTICLE{Vivas2022ApJ...926...78V,
       author = {{Vivas}, A. Katherina and {Mart{\'\i}nez-V{\'a}zquez}, Clara E. and {Walker}, Alistair R. and {Belokurov}, Vasily and {Li}, Ting S. and {Erkal}, Denis},
        title = "{Variable Stars in the Giant Satellite Galaxy Antlia 2}",
      journal = {\apj},
         year = 2022,
        month = feb,
       volume = {926},
       number = {1},
          eid = {78},
        pages = {78},
          doi = {10.3847/1538-4357/ac43bd},
       adsurl = {https://ui.adsabs.harvard.edu/abs/2022ApJ...926...78V}
}

@ARTICLE{DallOra2006ApJ...653L.109D,
   author = {{Dall'Ora}, M. and {Clementini}, G. and {Kinemuchi}, K. and 
	{Ripepi}, V. and {Marconi}, M. and {Di Fabrizio}, L. and {Greco}, C. and 
	{Rodgers}, C.~T. and {Kuehn}, C. and {Smith}, H.~A.},
    title = "{Variable Stars in the Newly Discovered Milky Way Satellite in Bootes}",
  journal = {\apjl},
   eprint = {astro-ph/0611285},
     year = 2006,
    month = dec,
   volume = 653,
    pages = {L109-L112},
      doi = {10.1086/510665},
   adsurl = {http://adsabs.harvard.edu/abs/2006ApJ...653L.109D}
}

@ARTICLE{Walsh2008ApJ...688..245W,
       author = {{Walsh}, S.~M. and {Willman}, B. and {Sand}, D. and {Harris}, J. and {Seth}, A. and {Zaritsky}, D. and {Jerjen}, H.},
        title = "{Bo{\"o}tes II ReBo{\"o}ted: An MMT/MegaCam Study of an Ultrafaint Milky Way Satellite}",
      journal = {\apj},
         year = 2008,
        month = nov,
       volume = {688},
       number = {1},
        pages = {245-253},
          doi = {10.1086/592076},
archivePrefix = {arXiv},
       eprint = {0712.3054},
 primaryClass = {astro-ph},
       adsurl = {https://ui.adsabs.harvard.edu/abs/2008ApJ...688..245W}
}

@ARTICLE{Carlin2018ApJ...865....7C,
       author = {{Carlin}, Jeffrey L. and {Sand}, D.~J.},
        title = "{Bo{\"o}tes III is a Disrupting Dwarf Galaxy Associated with the Styx Stellar Stream}",
      journal = {\apj},
         year = 2018,
        month = sep,
       volume = {865},
       number = {1},
          eid = {7},
        pages = {7},
          doi = {10.3847/1538-4357/aad8c1},
archivePrefix = {arXiv},
       eprint = {1805.11624},
 primaryClass = {astro-ph.GA},
       adsurl = {https://ui.adsabs.harvard.edu/abs/2018ApJ...865....7C}
}

@ARTICLE{Kuehn2008ApJ...674L..81K,
   author = {{Kuehn}, C. and {Kinemuchi}, K. and {Ripepi}, V. and {Clementini}, G. and 
	{Dall'Ora}, M. and {Di Fabrizio}, L. and {Rodgers}, C.~T. and 
	{Greco}, C. and {Marconi}, M. and {Musella}, I. and {Smith}, H.~A. and 
	{Catelan}, M. and {Beers}, T.~C. and {Pritzl}, B.~J.},
    title = "{Variable Stars in the Newly Discovered Milky Way Dwarf Spheroidal Satellite Canes Venatici I}",
  journal = {\apjl},
archivePrefix = "arXiv",
   eprint = {0709.3281},
     year = 2008,
    month = feb,
   volume = 674,
      eid = {L81},
    pages = {L81},
      doi = {10.1086/529137},
   adsurl = {http://adsabs.harvard.edu/abs/2008ApJ...674L..81K}
}

@ARTICLE{Greco2008ApJ...675L..73G,
       author = {{Greco}, Claudia and {Dall'Ora}, Massimo and {Clementini}, Gisella and {Ripepi}, Vincenzo and {Di Fabrizio}, Luca and {Kinemuchi}, Karen and {Marconi}, Marcella and {Musella}, Ilaria and {Smith}, Horace A. and {Rodgers}, Christopher T. and {Kuehn}, Charles and {Beers}, Timothy C. and {Catelan}, M{\'a}rcio and {Pritzl}, Barton J.},
        title = "{On the Newly Discovered Canes Venatici II dSph Galaxy}",
      journal = {\apjl},
         year = 2008,
        month = mar,
       volume = {675},
       number = {2},
        pages = {L73},
          doi = {10.1086/533585},
archivePrefix = {arXiv},
       eprint = {0712.2241},
 primaryClass = {astro-ph},
       adsurl = {https://ui.adsabs.harvard.edu/abs/2008ApJ...675L..73G}
}

@ARTICLE{Karczmarek2015AJ....150...90K,
   author = {{Karczmarek}, P. and {Pietrzy{\'n}ski}, G. and {Gieren}, W. and 
	{Suchomska}, K. and {Konorski}, P. and {G{\'o}rski}, M. and 
	{Pilecki}, B. and {Graczyk}, D. and {Wielg{\'o}rski}, P.},
    title = "{The Araucaria Project: The Distance to the Carina Dwarf Galaxy from Infrared Photometry of RR Lyrae Stars}",
  journal = {\aj},
archivePrefix = "arXiv",
   eprint = {1507.07713},
     year = 2015,
    month = sep,
   volume = 150,
      eid = {90},
    pages = {90},
      doi = {10.1088/0004-6256/150/3/90},
   adsurl = {http://adsabs.harvard.edu/abs/2015AJ....150...90K}
}

@ARTICLE{DrlicaWagner2015ApJ...813..109D,
   author = {{Drlica-Wagner}, A. and {Bechtol}, K. and {Rykoff}, E.~S. and 
	{Luque}, E. and {Queiroz}, A. and {Mao}, Y.-Y. and {Wechsler}, R.~H. and 
	{Simon}, J.~D. and {Santiago}, B. and {Yanny}, B. and {Balbinot}, E. and 
	{Dodelson}, S. and {Fausti Neto}, A. and {James}, D.~J. and 
	{Li}, T.~S. and {Maia}, M.~A.~G. and {Marshall}, J.~L. and {Pieres}, A. and 
	{Stringer}, K. and {Walker}, A.~R. and {Abbott}, T.~M.~C. and 
	{Abdalla}, F.~B. and {Allam}, S. and {Benoit-L{\'e}vy}, A. and 
	{Bernstein}, G.~M. and {Bertin}, E. and {Brooks}, D. and {Buckley-Geer}, E. and 
	{Burke}, D.~L. and {Carnero Rosell}, A. and {Carrasco Kind}, M. and 
	{Carretero}, J. and {Crocce}, M. and {da Costa}, L.~N. and {Desai}, S. and 
	{Diehl}, H.~T. and {Dietrich}, J.~P. and {Doel}, P. and {Eifler}, T.~F. and 
	{Evrard}, A.~E. and {Finley}, D.~A. and {Flaugher}, B. and {Fosalba}, P. and 
	{Frieman}, J. and {Gaztanaga}, E. and {Gerdes}, D.~W. and {Gruen}, D. and 
	{Gruendl}, R.~A. and {Gutierrez}, G. and {Honscheid}, K. and 
	{Kuehn}, K. and {Kuropatkin}, N. and {Lahav}, O. and {Martini}, P. and 
	{Miquel}, R. and {Nord}, B. and {Ogando}, R. and {Plazas}, A.~A. and 
	{Reil}, K. and {Roodman}, A. and {Sako}, M. and {Sanchez}, E. and 
	{Scarpine}, V. and {Schubnell}, M. and {Sevilla-Noarbe}, I. and 
	{Smith}, R.~C. and {Soares-Santos}, M. and {Sobreira}, F. and 
	{Suchyta}, E. and {Swanson}, M.~E.~C. and {Tarle}, G. and {Tucker}, D. and 
	{Vikram}, V. and {Wester}, W. and {Zhang}, Y. and {Zuntz}, J. and 
	{DES Collaboration}},
    title = "{Eight Ultra-faint Galaxy Candidates Discovered in Year Two of the Dark Energy Survey}",
  journal = {\apj},
archivePrefix = "arXiv",
   eprint = {1508.03622},
     year = 2015,
    month = nov,
   volume = 813,
      eid = {109},
    pages = {109},
      doi = {10.1088/0004-637X/813/2/109},
   adsurl = {http://adsabs.harvard.edu/abs/2015ApJ...813..109D}
}

@ARTICLE{Carlin2017AJ....154..267C,
   author = {{Carlin}, J.~L. and {Sand}, D.~J. and {Mu{\~n}oz}, R.~R. and 
	{Spekkens}, K. and {Willman}, B. and {Crnojevi{\'c}}, D. and 
	{Forbes}, D.~A. and {Hargis}, J. and {Kirby}, E. and {Peter}, A.~H.~G. and 
	{Romanowsky}, A.~J. and {Strader}, J.},
    title = "{Deep Subaru Hyper Suprime-Cam Observations of Milky Way Satellites Columba I and Triangulum II}",
  journal = {\aj},
archivePrefix = "arXiv",
   eprint = {1710.06444},
     year = 2017,
    month = dec,
   volume = 154,
      eid = {267},
    pages = {267},
      doi = {10.3847/1538-3881/aa94d0},
   adsurl = {http://adsabs.harvard.edu/abs/2017AJ....154..267C}
}

@ARTICLE{Torrealba2016MNRAS.459.2370T,
   author = {{Torrealba}, G. and {Koposov}, S.~E. and {Belokurov}, V. and 
	{Irwin}, M.},
    title = "{The feeble giant. Discovery of a large and diffuse Milky Way dwarf galaxy in the constellation of Crater}",
  journal = {\mnras},
archivePrefix = "arXiv",
   eprint = {1601.07178},
     year = 2016,
    month = jul,
   volume = 459,
    pages = {2370-2378},
      doi = {10.1093/mnras/stw733},
   adsurl = {http://adsabs.harvard.edu/abs/2016MNRAS.459.2370T}
}

@ARTICLE{Simon2020ApJ...892..137S,
       author = {{Simon}, J.~D. and {Li}, T.~S. and {Erkal}, D. and {Pace}, A.~B. and
         {Drlica-Wagner}, A. and {James}, D.~J. and {Marshall}, J.~L. and
         {Bechtol}, K. and {Hansen}, T. and {Kuehn}, K. and {Lidman}, C. and
         {Allam}, S. and {Annis}, J. and {Avila}, S. and {Bertin}, E. and
         {Brooks}, D. and {Burke}, D.~L. and {Rosell}, A. Carnero and
         {Carrasco Kind}, M. and {Carretero}, J. and {da Costa}, L.~N. and
         {De Vicente}, J. and {Desai}, S. and {Doel}, P. and {Eifler}, T.~F. and
         {Everett}, S. and {Fosalba}, P. and {Frieman}, J. and
         {Garc{\'\i}a-Bellido}, J. and {Gaztanaga}, E. and {Gerdes}, D.~W. and
         {Gruen}, D. and {Gruendl}, R.~A. and {Gschwend}, J. and
         {Gutierrez}, G. and {Hollowood}, D.~L. and {Honscheid}, K. and
         {Krause}, E. and {Kuropatkin}, N. and {MacCrann}, N. and
         {Maia}, M.~A.~G. and {March}, M. and {Miquel}, R. and {Palmese}, A. and
         {Paz-Chinch{\'o}n}, F. and {Plazas}, A.~A. and {Reil}, K. and
         {Roodman}, A. and {Sanchez}, E. and {Santiago}, B. and {Scarpine}, V. and
         {Schubnell}, M. and {Serrano}, S. and {Smith}, M. and {Suchyta}, E. and
         {Tarle}, G. and {Walker}, A.~R. and {DES Collaboration}},
        title = "{Birds of a Feather? Magellan/IMACS Spectroscopy of the Ultra-faint Satellites Grus II, Tucana IV, and Tucana V}",
      journal = {\apj},
         year = 2020,
        month = apr,
       volume = {892},
       number = {2},
          eid = {137},
        pages = {137},
          doi = {10.3847/1538-4357/ab7ccb},
archivePrefix = {arXiv},
       eprint = {1911.08493},
 primaryClass = {astro-ph.GA},
       adsurl = {https://ui.adsabs.harvard.edu/abs/2020ApJ...892..137S}
}

@ARTICLE{Cantu2021ApJ...916...81C,
       author = {{Cantu}, Sarah A. and {Pace}, Andrew B. and {Marshall}, Jennifer and {Strigari}, Louis E. and {Crnojevic}, Denija and {Simon}, Joshua D. and {Drlica-Wagner}, A. and {Bechtol}, K. and {Mart{\'\i}nez-V{\'a}zquez}, Clara E. and {Santiago}, B. and {Amara}, A. and {Stringer}, K.~M. and {Diehl}, H.~T. and {Aguena}, M. and {Allam}, S. and {Avila}, S. and {Brooks}, D. and {Carnero Rosell}, A. and {Carrasco Kind}, M. and {Carretero}, J. and {Costanzi}, M. and {Da Costa}, L.~N. and {De Vicente}, J. and {Desai}, S. and {Doel}, P. and {Eifler}, T.~F. and {Everett}, S. and {Frieman}, J. and {Garc{\'\i}a-Bellido}, J. and {Gaztanaga}, E. and {Gruen}, D. and {Gruendl}, R.~A. and {Gschwend}, J. and {Gutierrez}, G. and {Hinton}, S.~R. and {Hollowood}, D.~L. and {Honscheid}, K. and {James}, D.~J. and {Kuehn}, K. and {Maia}, M.~A.~G. and {Menanteau}, F. and {Miquel}, R. and {Palmese}, A. and {Paz-Chinch{\'o}n}, F. and {Plazas}, A.~A. and {Sanchez}, E. and {Scarpine}, V. and {Schubnell}, M. and {Serrano}, S. and {Sevilla-Noarbe}, I. and {Smith}, M. and {Soares-Santos}, M. and {Suchyta}, E. and {Swanson}, M.~E.~C. and {Tarle}, G. and {Walker}, A.~R. and {Wilkinson}, R.~D. and {DES Collaboration}},
        title = "{A Deeper Look at DES Dwarf Galaxy Candidates: Grus I and Indus II}",
      journal = {\apj},
         year = 2021,
        month = aug,
       volume = {916},
       number = {2},
          eid = {81},
        pages = {81},
          doi = {10.3847/1538-4357/ac0443},
archivePrefix = {arXiv},
       eprint = {2005.06478},
 primaryClass = {astro-ph.GA},
       adsurl = {https://ui.adsabs.harvard.edu/abs/2021ApJ...916...81C}
}

@ARTICLE{Richstein2022ApJ...933..217R,
       author = {{Richstein}, Hannah and {Patel}, Ekta and {Kallivayalil}, Nitya and {Simon}, Joshua D. and {Zivick}, Paul and {Tollerud}, Erik and {Fritz}, Tobias and {Warfield}, Jack T. and {Besla}, Gurtina and {van der Marel}, Roeland P. and {Wetzel}, Andrew and {Choi}, Yumi and {Deason}, Alis and {Geha}, Marla and {Guhathakurta}, Puragra and {Jeon}, Myoungwon and {Kirby}, Evan N. and {Libralato}, Mattia and {Sacchi}, Elena and {Sohn}, Sangmo Tony},
        title = "{Structural Parameters and Possible Association of the Ultra-faint Dwarfs Pegasus III and Pisces II from Deep Hubble Space Telescope Photometry}",
      journal = {\apj},
         year = 2022,
        month = jul,
       volume = {933},
       number = {2},
          eid = {217},
        pages = {217},
          doi = {10.3847/1538-4357/ac7226},
archivePrefix = {arXiv},
       eprint = {2204.01917},
 primaryClass = {astro-ph.GA},
       adsurl = {https://ui.adsabs.harvard.edu/abs/2022ApJ...933..217R}
}

@ARTICLE{MutluPakdil2018ApJ...863...25M,
   author = {{Mutlu-Pakdil}, B. and {Sand}, D.~J. and {Carlin}, J.~L. and 
	{Spekkens}, K. and {Caldwell}, N. and {Crnojevi{\'c}}, D. and 
	{Hughes}, A.~K. and {Willman}, B. and {Zaritsky}, D.},
    title = "{A Deeper Look at the New Milky Way Satellites: Sagittarius II, Reticulum II, Phoenix II, and Tucana III}",
  journal = {\apj},
archivePrefix = "arXiv",
   eprint = {1804.08627},
     year = 2018,
    month = aug,
   volume = 863,
      eid = {25},
    pages = {25},
      doi = {10.3847/1538-4357/aacd0e},
   adsurl = {http://adsabs.harvard.edu/abs/2018ApJ...863...25M}
}

@ARTICLE{Koposov2015ApJ...811...62K,
   author = {{Koposov}, S.~E. and {Casey}, A.~R. and {Belokurov}, V. and 
	{Lewis}, J.~R. and {Gilmore}, G. and {Worley}, C. and {Hourihane}, A. and 
	{Randich}, S. and {Bensby}, T. and {Bragaglia}, A. and {Bergemann}, M. and 
	{Carraro}, G. and {Costado}, M.~T. and {Flaccomio}, E. and {Francois}, P. and 
	{Heiter}, U. and {Hill}, V. and {Jofre}, P. and {Lando}, C. and 
	{Lanzafame}, A.~C. and {de Laverny}, P. and {Monaco}, L. and 
	{Morbidelli}, L. and {Sbordone}, L. and {Mikolaitis}, {\v S}. and 
	{Ryde}, N.},
    title = "{Kinematics and Chemistry of Recently Discovered Reticulum 2 and Horologium 1 Dwarf Galaxies}",
  journal = {\apj},
archivePrefix = "arXiv",
   eprint = {1504.07916},
     year = 2015,
    month = sep,
   volume = 811,
      eid = {62},
    pages = {62},
      doi = {10.1088/0004-637X/811/1/62},
   adsurl = {http://adsabs.harvard.edu/abs/2015ApJ...811...62K}
}

@ARTICLE{Kirby2015ApJ...810...56K,
   author = {{Kirby}, E.~N. and {Simon}, J.~D. and {Cohen}, J.~G.},
    title = "{Spectroscopic Confirmation of the Dwarf Galaxies Hydra II and Pisces II and the Globular Cluster Laevens 1}",
  journal = {\apj},
archivePrefix = "arXiv",
   eprint = {1506.01021},
     year = 2015,
    month = sep,
   volume = 810,
      eid = {56},
    pages = {56},
      doi = {10.1088/0004-637X/810/1/56},
   adsurl = {http://adsabs.harvard.edu/abs/2015ApJ...810...56K}
}

@ARTICLE{Musella2009ApJ...695L..83M,
   author = {{Musella}, I. and {Ripepi}, V. and {Clementini}, G. and {Dall'Ora}, M. and 
	{Kinemuchi}, K. and {di Fabrizio}, L. and {Greco}, C. and {Marconi}, M. and 
	{Smith}, H.~A. and {Radovich}, M. and {Beers}, T.~C.},
    title = "{Pulsating Variable Stars in the Coma Berenices Dwarf Spheroidal Galaxy}",
  journal = {\apjl},
archivePrefix = "arXiv",
   eprint = {0902.4230},
 primaryClass = "astro-ph.GA",
     year = 2009,
    month = apr,
   volume = 695,
    pages = {L83-L87},
      doi = {10.1088/0004-637X/695/1/L83},
   adsurl = {http://adsabs.harvard.edu/abs/2009ApJ...695L..83M}
}

@ARTICLE{Simon2011ApJ...733...46S,
   author = {{Simon}, J.~D. and {Geha}, M. and {Minor}, Q.~E. and {Martinez}, G.~D. and 
	{Kirby}, E.~N. and {Bullock}, J.~S. and {Kaplinghat}, M. and 
	{Strigari}, L.~E. and {Willman}, B. and {Choi}, P.~I. and {Tollerud}, E.~J. and 
	{Wolf}, J.},
    title = "{A Complete Spectroscopic Survey of the Milky Way Satellite Segue 1: The Darkest Galaxy}",
  journal = {\apj},
archivePrefix = "arXiv",
   eprint = {1007.4198},
 primaryClass = "astro-ph.GA",
     year = 2011,
    month = may,
   volume = 733,
    pages = {46-+},
      doi = {10.1088/0004-637X/733/1/46},
   adsurl = {http://adsabs.harvard.edu/abs/2011ApJ...733...46S}
}

@ARTICLE{Kirby2013ApJ...770...16K,
   author = {{Kirby}, E.~N. and {Boylan-Kolchin}, M. and {Cohen}, J.~G. and 
	{Geha}, M. and {Bullock}, J.~S. and {Kaplinghat}, M.},
    title = "{Segue 2: The Least Massive Galaxy}",
  journal = {\apj},
archivePrefix = "arXiv",
   eprint = {1304.6080},
 primaryClass = "astro-ph.CO",
     year = 2013,
    month = jun,
   volume = 770,
      eid = {16},
    pages = {16},
      doi = {10.1088/0004-637X/770/1/16},
   adsurl = {http://adsabs.harvard.edu/abs/2013ApJ...770...16K}
}

@ARTICLE{Carlin2009ApJ...702L...9C,
   author = {{Carlin}, J.~L. and {Grillmair}, C.~J. and {Mu{\~n}oz}, R.~R. and 
	{Nidever}, D.~L. and {Majewski}, S.~R.},
    title = "{Kinematics and Metallicities in the Bo{\"o}tes III Stellar Overdensity: A Disrupted Dwarf Galaxy?}",
  journal = {\apjl},
archivePrefix = "arXiv",
   eprint = {0907.3738},
 primaryClass = "astro-ph.GA",
     year = 2009,
    month = sep,
   volume = 702,
    pages = {L9-L13},
      doi = {10.1088/0004-637X/702/1/L9},
   adsurl = {http://adsabs.harvard.edu/abs/2009ApJ...702L...9C}
}

@ARTICLE{Li2018ApJ...857..145L,
   author = {{Li}, T.~S. and {Simon}, J.~D. and {Pace}, A.~B. and {Torrealba}, G. and 
	{Kuehn}, K. and {Drlica-Wagner}, A. and {Bechtol}, K. and {Vivas}, A.~K. and 
	{van der Marel}, R.~P. and {Wood}, M. and {Yanny}, B. and {Belokurov}, V. and 
	{Jethwa}, P. and {Zucker}, D.~B. and {Lewis}, G. and {Kron}, R. and 
	{Nidever}, D.~L. and {S{\'a}nchez-Conde}, M.~A. and {Ji}, A.~P. and 
	{Conn}, B.~C. and {James}, D.~J. and {Martin}, N.~F. and {Martinez-Delgado}, D. and 
	{No{\"e}l}, N.~E.~D. and {MagLiteS Collaboration}},
    title = "{Ships Passing in the Night: Spectroscopic Analysis of Two Ultra-faint Satellites in the Constellation Carina}",
  journal = {\apj},
archivePrefix = "arXiv",
   eprint = {1802.06810},
     year = 2018,
    month = apr,
   volume = 857,
      eid = {145},
    pages = {145},
      doi = {10.3847/1538-4357/aab666},
   adsurl = {http://adsabs.harvard.edu/abs/2018ApJ...857..145L}
}

@ARTICLE{Fritz2019AA...623A.129F,
       author = {{Fritz}, T.~K. and {Carrera}, R. and {Battaglia}, G. and {Taibi}, S.},
        title = "{Gaia DR 2 and VLT/FLAMES search for new satellites of the LMC}",
      journal = {\aap},
         year = "2019",
        month = "Mar",
       volume = {623},
          eid = {A129},
        pages = {A129},
          doi = {10.1051/0004-6361/201833458},
archivePrefix = {arXiv},
       eprint = {1805.07350},
 primaryClass = {astro-ph.GA},
       adsurl = {https://ui.adsabs.harvard.edu/abs/2019A&A...623A.129F}
}

@ARTICLE{Chiti2022ApJ...939...41C,
       author = {{Chiti}, Anirudh and {Simon}, Joshua D. and {Frebel}, Anna and {Pace}, Andrew B. and {Ji}, Alexander P. and {Li}, Ting S.},
        title = "{Magellan/IMACS Spectroscopy of Grus. I. A Low Metallicity Ultra-faint Dwarf Galaxy}",
      journal = {\apj},
         year = 2022,
        month = nov,
       volume = {939},
       number = {1},
          eid = {41},
        pages = {41},
          doi = {10.3847/1538-4357/ac96ed},
archivePrefix = {arXiv},
       eprint = {2206.04580},
 primaryClass = {astro-ph.GA},
       adsurl = {https://ui.adsabs.harvard.edu/abs/2022ApJ...939...41C}
}

@ARTICLE{Willman2011AJ....142..128W,
       author = {{Willman}, Beth and {Geha}, Marla and {Strader}, Jay and {Strigari}, Louis E. and {Simon}, Joshua D. and {Kirby}, Evan and {Ho}, Nhung and {Warres}, Alex},
        title = "{Willman 1{\textemdash}A Probable Dwarf Galaxy with an Irregular Kinematic Distribution}",
      journal = {\aj},
         year = 2011,
        month = oct,
       volume = {142},
       number = {4},
          eid = {128},
        pages = {128},
          doi = {10.1088/0004-6256/142/4/128},
archivePrefix = {arXiv},
       eprint = {1007.3499},
 primaryClass = {astro-ph.GA},
       adsurl = {https://ui.adsabs.harvard.edu/abs/2011AJ....142..128W}
}

@ARTICLE{Simon2007ApJ...670..313S,
       author = {{Simon}, Joshua D. and {Geha}, Marla},
        title = "{The Kinematics of the Ultra-faint Milky Way Satellites: Solving the Missing Satellite Problem}",
      journal = {\apj},
         year = 2007,
        month = nov,
       volume = {670},
       number = {1},
        pages = {313-331},
          doi = {10.1086/521816},
archivePrefix = {arXiv},
       eprint = {0706.0516},
 primaryClass = {astro-ph},
       adsurl = {https://ui.adsabs.harvard.edu/abs/2007ApJ...670..313S}
}

@ARTICLE{Walker2009AJ....137.3100W,
   author = {{Walker}, M.~G. and {Mateo}, M. and {Olszewski}, E.~W.},
    title = "{Stellar Velocities in the Carina, Fornax, Sculptor, and Sextans dSph Galaxies: Data From the Magellan/MMFS Survey}",
  journal = {\aj},
archivePrefix = "arXiv",
   eprint = {0811.0118},
     year = 2009,
    month = feb,
   volume = 137,
    pages = {3100-3108},
      doi = {10.1088/0004-6256/137/2/3100},
   adsurl = {http://adsabs.harvard.edu/abs/2009AJ....137.3100W}
}

@ARTICLE{Walker2015MNRAS.448.2717W,
   author = {{Walker}, M.~G. and {Olszewski}, E.~W. and {Mateo}, M.},
    title = "{Bayesian analysis of resolved stellar spectra: application to MMT/Hectochelle observations of the Draco dwarf spheroidal}",
  journal = {\mnras},
archivePrefix = "arXiv",
   eprint = {1503.02589},
     year = 2015,
    month = apr,
   volume = 448,
    pages = {2717-2732},
      doi = {10.1093/mnras/stv099},
   adsurl = {http://adsabs.harvard.edu/abs/2015MNRAS.448.2717W}
}

@ARTICLE{Mateo2008ApJ...675..201M,
   author = {{Mateo}, M. and {Olszewski}, E.~W. and {Walker}, M.~G.},
    title = "{The Velocity Dispersion Profile of the Remote Dwarf Spheroidal Galaxy Leo I: A Tidal Hit and Run?}",
  journal = {\apj},
archivePrefix = "arXiv",
   eprint = {0708.1327},
     year = 2008,
    month = mar,
   volume = 675,
      eid = {201-233},
    pages = {201-233},
      doi = {10.1086/522326},
   adsurl = {http://adsabs.harvard.edu/abs/2008ApJ...675..201M}
}

@ARTICLE{MartinezVazquez2019MNRAS.490.2183M,
       author = {{Mart{\'\i}nez-V{\'a}zquez}, C.~E. and {Vivas}, A.~K. and {Gurevich}, M. and {Walker}, A.~R. and {McCarthy}, M. and {Pace}, A.~B. and {Stringer}, K.~M. and {Santiago}, B. and {Hounsell}, R. and {Macri}, L. and {Li}, T.~S. and {Bechtol}, K. and {Riley}, A.~H. and {Kim}, A.~G. and {Simon}, J.~D. and {Drlica-Wagner}, A. and {Nadler}, E.~O. and {Marshall}, J.~L. and {Annis}, J. and {Avila}, S. and {Bertin}, E. and {Brooks}, D. and {Buckley-Geer}, E. and {Burke}, D.~L. and {Carnero Rosell}, A. and {Carrasco Kind}, M. and {da Costa}, L.~N. and {De Vicente}, J. and {Desai}, S. and {Diehl}, H.~T. and {Doel}, P. and {Everett}, S. and {Frieman}, J. and {Garc{\'\i}a-Bellido}, J. and {Gaztanaga}, E. and {Gruen}, D. and {Gruendl}, R.~A. and {Gschwend}, J. and {Gutierrez}, G. and {Hollowood}, D.~L. and {Honscheid}, K. and {James}, D.~J. and {Kuehn}, K. and {Kuropatkin}, N. and {Maia}, M.~A.~G. and {Menanteau}, F. and {Miller}, C.~J. and {Miquel}, R. and {Paz-Chinch{\'o}n}, F. and {Plazas}, A.~A. and {Sanchez}, E. and {Scarpine}, V. and {Serrano}, S. and {Sevilla-Noarbe}, I. and {Smith}, M. and {Soares-Santos}, M. and {Sobreira}, F. and {Swanson}, M.~E.~C. and {Tarle}, G. and {Vikram}, V. and {DES Collaboration}},
        title = "{Search for RR Lyrae stars in DES ultrafaint systems: Grus I, Kim 2, Phoenix II, and Grus II}",
      journal = {\mnras},
         year = 2019,
        month = dec,
       volume = {490},
       number = {2},
        pages = {2183-2199},
          doi = {10.1093/mnras/stz2609},
archivePrefix = {arXiv},
       eprint = {1909.06308},
 primaryClass = {astro-ph.GA},
       adsurl = {https://ui.adsabs.harvard.edu/abs/2019MNRAS.490.2183M}
}

@ARTICLE{MutluPakdil2020ApJ...902..106M,
       author = {{Mutlu-Pakdil}, Bur{\c{c}}in and {Sand}, David J. and {Crnojevi{\'c}}, Denija and {Olszewski}, Edward W. and {Zaritsky}, Dennis and {Strader}, Jay and {Collins}, Michelle L. and {Seth}, Anil C. and {Willman}, Beth},
        title = "{The Elusive Distance Gradient in the Ultrafaint Dwarf Galaxy Hercules: A Combined Hubble Space Telescope and Gaia View}",
      journal = {\apj},
         year = 2020,
        month = oct,
       volume = {902},
       number = {2},
          eid = {106},
        pages = {106},
          doi = {10.3847/1538-4357/abb40b},
archivePrefix = {arXiv},
       eprint = {2007.11589},
 primaryClass = {astro-ph.GA},
       adsurl = {https://ui.adsabs.harvard.edu/abs/2020ApJ...902..106M}
}

@ARTICLE{Vivas2016AJ....151..118V,
   author = {{Vivas}, A.~K. and {Olsen}, K. and {Blum}, R. and {Nidever}, D.~L. and 
	{Walker}, A.~R. and {Martin}, N.~F. and {Besla}, G. and {Gallart}, C. and 
	{van der Marel}, R.~P. and {Majewski}, S.~R. and {Kaleida}, C.~C. and 
	{Mu{\~n}oz}, R.~R. and {Saha}, A. and {Conn}, B.~C. and {Jin}, S.
	},
    title = "{Variable Stars in the Field of the Hydra II Ultra-faint Dwarf Galaxy}",
  journal = {\aj},
archivePrefix = "arXiv",
   eprint = {1510.05539},
     year = 2016,
    month = may,
   volume = 151,
      eid = {118},
    pages = {118},
      doi = {10.3847/0004-6256/151/5/118},
   adsurl = {http://adsabs.harvard.edu/abs/2016AJ....151..118V}
}

@ARTICLE{Stetson2014PASP..126..616S,
   author = {{Stetson}, P.~B. and {Fiorentino}, G. and {Bono}, G. and {Bernard}, E.~J. and 
	{Monelli}, M. and {Iannicola}, G. and {Gallart}, C. and {Ferraro}, I.
	},
    title = "{Homogeneous Photometry VI: Variable Stars in the Leo I Dwarf Spheroidal Galaxy}",
  journal = {\pasp},
archivePrefix = "arXiv",
   eprint = {1406.6704},
     year = 2014,
    month = jul,
   volume = 126,
    pages = {616},
      doi = {10.1086/677352},
   adsurl = {http://adsabs.harvard.edu/abs/2014PASP..126..616S}
}

@ARTICLE{Medina2018ApJ...855...43M,
       author = {{Medina}, Gustavo E. and {Mu{\~n}oz}, Ricardo R. and {Vivas}, A. Katherina and {Carlin}, Jeffrey L. and {F{\"o}rster}, Francisco and {Mart{\'\i}nez}, Jorge and {Galbany}, Llu{\'\i}s and {Gonz{\'a}lez-Gait{\'a}n}, Santiago and {Hamuy}, Mario and {de Jaeger}, Thomas and {Maureira}, Juan Carlos and {San Mart{\'\i}n}, Jaime},
        title = "{Discovery of Distant RR Lyrae Stars in the Milky Way Using DECam}",
      journal = {\apj},
         year = 2018,
        month = mar,
       volume = {855},
       number = {1},
          eid = {43},
        pages = {43},
          doi = {10.3847/1538-4357/aaad02},
archivePrefix = {arXiv},
       eprint = {1802.01581},
 primaryClass = {astro-ph.GA},
       adsurl = {https://ui.adsabs.harvard.edu/abs/2018ApJ...855...43M}
}

@ARTICLE{Kim2016ApJ...833...16K,
   author = {{Kim}, D. and {Jerjen}, H. and {Geha}, M. and {Chiti}, A. and 
	{Milone}, A.~P. and {Da Costa}, G. and {Mackey}, D. and {Frebel}, A. and 
	{Conn}, B.},
    title = "{Portrait of a Dark Horse: a Photometric and Spectroscopic Study of the Ultra-faint Milky Way Satellite Pegasus III}",
  journal = {\apj},
archivePrefix = "arXiv",
   eprint = {1608.04934},
     year = 2016,
    month = dec,
   volume = 833,
      eid = {16},
    pages = {16},
      doi = {10.3847/0004-637X/833/1/16},
   adsurl = {http://adsabs.harvard.edu/abs/2016ApJ...833...16K}
}

@ARTICLE{Sand2012ApJ...756...79S,
   author = {{Sand}, D.~J. and {Strader}, J. and {Willman}, B. and {Zaritsky}, D. and 
	{McLeod}, B. and {Caldwell}, N. and {Seth}, A. and {Olszewski}, E.
	},
    title = "{Tidal Signatures in the Faintest Milky Way Satellites: The Detailed Properties of Leo V, Pisces II, and Canes Venatici II}",
  journal = {\apj},
archivePrefix = "arXiv",
   eprint = {1111.6608},
     year = 2012,
    month = sep,
   volume = 756,
      eid = {79},
    pages = {79},
      doi = {10.1088/0004-637X/756/1/79},
   adsurl = {http://adsabs.harvard.edu/abs/2012ApJ...756...79S}
}

@ARTICLE{MartinezVazquez2015MNRAS.454.1509M,
   author = {{Mart{\'{\i}}nez-V{\'a}zquez}, C.~E. and {Monelli}, M. and {Bono}, G. and 
	{Stetson}, P.~B. and {Ferraro}, I. and {Bernard}, E.~J. and 
	{Gallart}, C. and {Fiorentino}, G. and {Iannicola}, G. and {Udalski}, A.
	},
    title = "{Variable stars in Local Group Galaxies - I. Tracing the early chemical enrichment and radial gradients in the Sculptor dSph with RR Lyrae stars}",
  journal = {\mnras},
archivePrefix = "arXiv",
   eprint = {1508.06942},
     year = 2015,
    month = dec,
   volume = 454,
    pages = {1509-1516},
      doi = {10.1093/mnras/stv2014},
   adsurl = {http://adsabs.harvard.edu/abs/2015MNRAS.454.1509M}
}

@ARTICLE{Boettcher2013AJ....146...94B,
   author = {{Boettcher}, E. and {Willman}, B. and {Fadely}, R. and {Strader}, J. and 
	{Baker}, M. and {Hopkins}, E. and {Tasnim Ananna}, T. and {Cunningham}, E.~C. and 
	{Douglas}, T. and {Gilbert}, J. and {Preston}, A. and {Sturner}, A.~P.
	},
    title = "{A Search for RR Lyrae Stars in Segue 2 and Segue 3}",
  journal = {\aj},
archivePrefix = "arXiv",
   eprint = {1308.2227},
 primaryClass = "astro-ph.GA",
     year = 2013,
    month = oct,
   volume = 146,
      eid = {94},
    pages = {94},
      doi = {10.1088/0004-6256/146/4/94},
   adsurl = {http://adsabs.harvard.edu/abs/2013AJ....146...94B}
}

@ARTICLE{Garofalo2013ApJ...767...62G,
   author = {{Garofalo}, A. and {Cusano}, F. and {Clementini}, G. and {Ripepi}, V. and 
	{Dall'Ora}, M. and {Moretti}, M.~I. and {Coppola}, G. and {Musella}, I. and 
	{Marconi}, M.},
    title = "{Variable Stars in the Ultra-faint Dwarf Spheroidal Galaxy Ursa Major I}",
  journal = {\apj},
archivePrefix = "arXiv",
   eprint = {1302.3230},
     year = 2013,
    month = apr,
   volume = 767,
      eid = {62},
    pages = {62},
      doi = {10.1088/0004-637X/767/1/62},
   adsurl = {http://adsabs.harvard.edu/abs/2013ApJ...767...62G}
}

@ARTICLE{DallOra2012ApJ...752...42D,
   author = {{Dall'Ora}, M. and {Kinemuchi}, K. and {Ripepi}, V. and {Rodgers}, C.~T. and 
	{Clementini}, G. and {Di Fabrizio}, L. and {Smith}, H.~A. and 
	{Marconi}, M. and {Musella}, I. and {Greco}, C. and {Kuehn}, C.~A. and 
	{Catelan}, M. and {Pritzl}, B.~J. and {Beers}, T.~C.},
    title = "{Stellar Archaeology in the Galactic Halo with the Ultra-faint Dwarfs. VI. Ursa Major II}",
  journal = {\apj},
archivePrefix = "arXiv",
   eprint = {1203.5321},
     year = 2012,
    month = jun,
   volume = 752,
      eid = {42},
    pages = {42},
      doi = {10.1088/0004-637X/752/1/42},
   adsurl = {http://adsabs.harvard.edu/abs/2012ApJ...752...42D}
}

@ARTICLE{Willman2006astro.ph..3486W,
       author = {{Willman}, Beth and {Masjedi}, Morad and {Hogg}, David W. and {Dalcanton}, Julianne J. and {Martinez-Delgado}, David and {Blanton}, Michael and {West}, Andrew A. and {Dotter}, Aaron and {Chaboyer}, Brian},
        title = "{Willman 1 - A Galactic Satellite at 40 kpc With Multiple Stellar Tails}",
      journal = {arXiv e-prints},
         year = 2006,
        month = mar,
          eid = {astro-ph/0603486},
        pages = {astro-ph/0603486},
archivePrefix = {arXiv},
       eprint = {astro-ph/0603486},
 primaryClass = {astro-ph},
       adsurl = {https://ui.adsabs.harvard.edu/abs/2006astro.ph..3486W}
}

@ARTICLE{Martin2016ApJ...833..167M,
   author = {{Martin}, N.~F. and {Ibata}, R.~A. and {Lewis}, G.~F. and {McConnachie}, A. and 
	{Babul}, A. and {Bate}, N.~F. and {Bernard}, E. and {Chapman}, S.~C. and 
	{Collins}, M.~M.~L. and {Conn}, A.~R. and {Crnojevi{\'c}}, D. and 
	{Fardal}, M.~A. and {Ferguson}, A.~M.~N. and {Irwin}, M. and 
	{Mackey}, A.~D. and {McMonigal}, B. and {Navarro}, J.~F. and 
	{Rich}, R.~M.},
    title = "{The PAndAS View of the Andromeda Satellite System. II. Detailed Properties of 23 M31 Dwarf Spheroidal Galaxies}",
  journal = {\apj},
archivePrefix = "arXiv",
   eprint = {1610.01158},
     year = 2016,
    month = dec,
   volume = 833,
      eid = {167},
    pages = {167},
      doi = {10.3847/1538-4357/833/2/167},
   adsurl = {http://adsabs.harvard.edu/abs/2016ApJ...833..167M}
}

@ARTICLE{McConnachie2006MNRAS.365.1263M,
   author = {{McConnachie}, A.~W. and {Irwin}, M.~J.},
    title = "{Structural properties of the M31 dwarf spheroidal galaxies}",
  journal = {\mnras},
   eprint = {astro-ph/0511004},
     year = 2006,
    month = feb,
   volume = 365,
    pages = {1263-1276},
      doi = {10.1111/j.1365-2966.2005.09806.x},
   adsurl = {http://adsabs.harvard.edu/abs/2006MNRAS.365.1263M}
}

@ARTICLE{Slater2015ApJ...806..230S,
       author = {{Slater}, Colin T. and {Bell}, Eric F. and {Martin}, Nicolas F. and {Tollerud}, Erik J. and {Ho}, Nhung},
        title = "{A Deep Study of the Dwarf Satellites Andromeda XXVIII and Andromeda XXIX}",
      journal = {\apj},
         year = 2015,
        month = jun,
       volume = {806},
       number = {2},
          eid = {230},
        pages = {230},
          doi = {10.1088/0004-637X/806/2/230},
archivePrefix = {arXiv},
       eprint = {1505.02161},
 primaryClass = {astro-ph.GA},
       adsurl = {https://ui.adsabs.harvard.edu/abs/2015ApJ...806..230S}
}

@ARTICLE{MartinezVazquez2017ApJ...850..137M,
       author = {{Mart{\'\i}nez-V{\'a}zquez}, Clara E. and {Monelli}, Matteo and {Bernard}, Edouard J. and {Gallart}, Carme and {Stetson}, Peter B. and {Skillman}, Evan D. and {Bono}, Giuseppe and {Cassisi}, Santi and {Fiorentino}, Giuliana and {McQuinn}, Kristen B.~W. and {Cole}, Andrew A. and {McConnachie}, Alan W. and {Martin}, Nicolas F. and {Dolphin}, Andrew E. and {Boylan-Kolchin}, Michael and {Aparicio}, Antonio and {Hidalgo}, Sebastian L. and {Weisz}, Daniel R.},
        title = "{The ISLAnds Project. III. Variable Stars in Six Andromeda Dwarf Spheroidal Galaxies}",
      journal = {\apj},
         year = 2017,
        month = dec,
       volume = {850},
       number = {2},
          eid = {137},
        pages = {137},
          doi = {10.3847/1538-4357/aa9381},
archivePrefix = {arXiv},
       eprint = {1710.09038},
 primaryClass = {astro-ph.GA},
       adsurl = {https://ui.adsabs.harvard.edu/abs/2017ApJ...850..137M}
}

@ARTICLE{Collins2020MNRAS.491.3496C,
       author = {{Collins}, Michelle L.~M. and {Tollerud}, Erik J. and {Rich}, R. Michael and {Ibata}, Rodrigo A. and {Martin}, Nicolas F. and {Chapman}, Scott C. and {Gilbert}, Karoline M. and {Preston}, Janet},
        title = "{A detailed study of Andromeda XIX, an extreme local analogue of ultradiffuse galaxies}",
      journal = {\mnras},
         year = 2020,
        month = jan,
       volume = {491},
       number = {3},
        pages = {3496-3514},
          doi = {10.1093/mnras/stz3252},
archivePrefix = {arXiv},
       eprint = {1910.12879},
 primaryClass = {astro-ph.GA},
       adsurl = {https://ui.adsabs.harvard.edu/abs/2020MNRAS.491.3496C}
}

@ARTICLE{Tollerud2012ApJ...752...45T,
   author = {{Tollerud}, E.~J. and {Beaton}, R.~L. and {Geha}, M.~C. and 
	{Bullock}, J.~S. and {Guhathakurta}, P. and {Kalirai}, J.~S. and 
	{Majewski}, S.~R. and {Kirby}, E.~N. and {Gilbert}, K.~M. and 
	{Yniguez}, B. and {Patterson}, R.~J. and {Ostheimer}, J.~C. and 
	{Cooke}, J. and {Dorman}, C.~E. and {Choudhury}, A. and {Cooper}, M.~C.
	},
    title = "{The SPLASH Survey: Spectroscopy of 15 M31 Dwarf Spheroidal Satellite Galaxies}",
  journal = {\apj},
archivePrefix = "arXiv",
   eprint = {1112.1067},
 primaryClass = "astro-ph.CO",
     year = 2012,
    month = jun,
   volume = 752,
      eid = {45},
    pages = {45},
      doi = {10.1088/0004-637X/752/1/45},
   adsurl = {http://adsabs.harvard.edu/abs/2012ApJ...752...45T}
}

@ARTICLE{Collins2021MNRAS.505.5686C,
       author = {{Collins}, Michelle L.~M. and {Read}, Justin I. and {Ibata}, Rodrigo A. and {Rich}, R. Michael and {Martin}, Nicolas F. and {Pe{\~n}arrubia}, Jorge and {Chapman}, Scott C. and {Tollerud}, Erik J. and {Weisz}, Daniel R.},
        title = "{Andromeda XXI - a dwarf galaxy in a low-density dark matter halo}",
      journal = {\mnras},
         year = 2021,
        month = aug,
       volume = {505},
       number = {4},
        pages = {5686-5701},
          doi = {10.1093/mnras/stab1624},
archivePrefix = {arXiv},
       eprint = {2102.11890},
 primaryClass = {astro-ph.GA},
       adsurl = {https://ui.adsabs.harvard.edu/abs/2021MNRAS.505.5686C}
}

@ARTICLE{Martin2014ApJ...793L..14M,
   author = {{Martin}, N.~F. and {Chambers}, K.~C. and {Collins}, M.~L.~M. and 
	{Ibata}, R.~A. and {Rich}, R.~M. and {Bell}, E.~F. and {Bernard}, E.~J. and 
	{Ferguson}, A.~M.~N. and {Flewelling}, H. and {Kaiser}, N. and 
	{Magnier}, E.~A. and {Tonry}, J.~L. and {Wainscoat}, R.~J.},
    title = "{Spectroscopy of the Three Distant Andromedan Satellites Cassiopeia III, Lacerta I, and Perseus I}",
  journal = {\apjl},
archivePrefix = "arXiv",
   eprint = {1408.5130},
     year = 2014,
    month = sep,
   volume = 793,
      eid = {L14},
    pages = {L14},
      doi = {10.1088/2041-8205/793/1/L14},
   adsurl = {http://adsabs.harvard.edu/abs/2014ApJ...793L..14M}
}

@ARTICLE{Ho2012ApJ...758..124H,
   author = {{Ho}, N. and {Geha}, M. and {Munoz}, R.~R. and {Guhathakurta}, P. and 
	{Kalirai}, J. and {Gilbert}, K.~M. and {Tollerud}, E. and {Bullock}, J. and 
	{Beaton}, R.~L. and {Majewski}, S.~R.},
    title = "{Stellar Kinematics of the Andromeda II Dwarf Spheroidal Galaxy}",
  journal = {\apj},
archivePrefix = "arXiv",
   eprint = {1209.2116},
     year = 2012,
    month = oct,
   volume = 758,
      eid = {124},
    pages = {124},
      doi = {10.1088/0004-637X/758/2/124},
   adsurl = {http://adsabs.harvard.edu/abs/2012ApJ...758..124H}
}

@ARTICLE{Collins2013ApJ...768..172C,
   author = {{Collins}, M.~L.~M. and {Chapman}, S.~C. and {Rich}, R.~M. and 
	{Ibata}, R.~A. and {Martin}, N.~F. and {Irwin}, M.~J. and {Bate}, N.~F. and 
	{Lewis}, G.~F. and {Pe{\~n}arrubia}, J. and {Arimoto}, N. and 
	{Casey}, C.~M. and {Ferguson}, A.~M.~N. and {Koch}, A. and {McConnachie}, A.~W. and 
	{Tanvir}, N.},
    title = "{A Kinematic Study of the Andromeda Dwarf Spheroidal System}",
  journal = {\apj},
archivePrefix = "arXiv",
   eprint = {1302.6590},
     year = 2013,
    month = may,
   volume = 768,
      eid = {172},
    pages = {172},
      doi = {10.1088/0004-637X/768/2/172},
   adsurl = {http://adsabs.harvard.edu/abs/2013ApJ...768..172C}
}

@ARTICLE{Tollerud2013ApJ...768...50T,
       author = {{Tollerud}, Erik J. and {Geha}, Marla C. and {Vargas}, Luis C. and {Bullock}, James S.},
        title = "{The Outer Limits of the M31 System: Kinematics of the Dwarf Galaxy Satellites And XXVIII \& And XXIX}",
      journal = {\apj},
         year = 2013,
        month = may,
       volume = {768},
       number = {1},
          eid = {50},
        pages = {50},
          doi = {10.1088/0004-637X/768/1/50},
archivePrefix = {arXiv},
       eprint = {1302.0848},
 primaryClass = {astro-ph.CO},
       adsurl = {https://ui.adsabs.harvard.edu/abs/2013ApJ...768...50T}
}

@ARTICLE{Bruce2023ApJ...950..167B,
       author = {{Bruce}, Jordan and {Li}, Ting S. and {Pace}, Andrew B. and {Heiger}, Mairead and {Song}, Ying-Yi and {Simon}, Joshua D.},
        title = "{Spectroscopic Analysis of Milky Way Outer Halo Satellites: Aquarius II and Bo{\"o}tes II}",
      journal = {\apj},
         year = 2023,
        month = jun,
       volume = {950},
       number = {2},
          eid = {167},
        pages = {167},
          doi = {10.3847/1538-4357/acc943},
archivePrefix = {arXiv},
       eprint = {2302.03708},
 primaryClass = {astro-ph.GA},
       adsurl = {https://ui.adsabs.harvard.edu/abs/2023ApJ...950..167B}
}

@ARTICLE{Correnti2009MNRAS.397L..26C,
       author = {{Correnti}, M. and {Bellazzini}, M. and {Ferraro}, F.~R.},
        title = "{Red Clump stars in the Bo{\"o}tes III stellar system}",
      journal = {\mnras},
         year = 2009,
        month = jul,
       volume = {397},
       number = {1},
        pages = {L26-L30},
          doi = {10.1111/j.1745-3933.2009.00677.x},
archivePrefix = {arXiv},
       eprint = {0904.3068},
 primaryClass = {astro-ph.GA},
       adsurl = {https://ui.adsabs.harvard.edu/abs/2009MNRAS.397L..26C}
}

@article{sandford_chemodynamics_2026,
	title = {Chemodynamics of {Boötes} {I} with {S5}: {Revised} {Velocity} {Gradient}, {Dark} {Matter} {Density}, and {Galactic} {Chemical} {Evolution} {Constraints}},
	volume = {998},
	issn = {0004-637X},
	shorttitle = {Chemodynamics of {Boötes} {I} with {S5}},
	url = {https://ui.adsabs.harvard.edu/abs/2026ApJ...998...47S},
	doi = {10.3847/1538-4357/ae2fe5},
	urldate = {2026-08-28},
	journal = {The Astrophysical Journal},
	publisher = {IOP},
	author = {Sandford, Nathan R. and Li, Ting S. and Koposov, Sergey E. and Hayashi, Kohei and Pace, Andrew B. and Erkal, Denis and Bovy, Jo and Da Costa, Gary S. and Cullinane, Lara R. and Ji, Alexander P. and Kuehn, Kyler and Lewis, Geraint F. and Zucker, Daniel B. and Limberg, Guilherme and Medina, Gustavo E. and Simon, Joshua D. and Yang, Yong and {(S}},
	month = feb,
	year = {2026},
	note = {ADS Bibcode: 2026ApJ...998...47S},
	pages = {47},
}

@ARTICLE{Charles2023MNRAS.521.3527C,
       author = {{Charles}, Emily J.~E. and {Collins}, Michelle L.~M. and {Rich}, R. Michael and {Read}, Justin I. and {Kim}, Stacy Y. and {Ibata}, Rodrigo A. and {Martin}, Nicolas F. and {Chapman}, Scott C. and {Balbinot}, Eduardo and {Weisz}, Daniel R.},
        title = "{Andromeda XXV - a dwarf galaxy with a low central dark matter density}",
      journal = {\mnras},
         year = 2023,
        month = may,
       volume = {521},
       number = {3},
        pages = {3527-3539},
          doi = {10.1093/mnras/stad752},
archivePrefix = {arXiv},
       eprint = {2209.15022},
 primaryClass = {astro-ph.GA},
       adsurl = {https://ui.adsabs.harvard.edu/abs/2023MNRAS.521.3527C}
}

@ARTICLE{Vivas2020ApJS..247...35V,
       author = {{Vivas}, A. Katherina and {Mart{\'\i}nez-V{\'a}zquez}, Clara and
         {Walker}, Alistair R.},
        title = "{Gaia RR Lyrae Stars in Nearby Ultra-faint Dwarf Satellite Galaxies}",
      journal = {\apjs},
         year = 2020,
        month = mar,
       volume = {247},
       number = {1},
          eid = {35},
        pages = {35},
          doi = {10.3847/1538-4365/ab67c0},
archivePrefix = {arXiv},
       eprint = {2001.01107},
 primaryClass = {astro-ph.GA},
       adsurl = {https://ui.adsabs.harvard.edu/abs/2020ApJS..247...35V}
}

@ARTICLE{Warfield2023MNRAS.519.1189W,
       author = {{Warfield}, Jack T. and {Kallivayalil}, Nitya and {Zivick}, Paul and {Fritz}, Tobias and {Richstein}, Hannah and {Sohn}, Sangmo Tony and {del Pino}, Andr{\'e}s and {Savino}, Alessandro and {Weisz}, Daniel R.},
        title = "{HUBPUG: proper motions for local group dwarfs observed with HST utilizing Gaia as a reference frame}",
      journal = {\mnras},
         year = 2023,
        month = feb,
       volume = {519},
       number = {1},
        pages = {1189-1200},
          doi = {10.1093/mnras/stac3647},
archivePrefix = {arXiv},
       eprint = {2209.02751},
 primaryClass = {astro-ph.GA},
       adsurl = {https://ui.adsabs.harvard.edu/abs/2023MNRAS.519.1189W}
}

@ARTICLE{Heiger2024ApJ...961..234H,
       author = {{Heiger}, M.~E. and {Li}, T.~S. and {Pace}, A.~B. and {Simon}, J.~D. and {Ji}, A.~P. and {Chiti}, A. and {Bom}, C.~R. and {Carballo-Bello}, J.~A. and {Carlin}, J.~L. and {Cerny}, W. and {Choi}, Y. and {Drlica-Wagner}, A. and {James}, D.~J. and {Mart{\'\i}nez-V{\'a}zquez}, C.~E. and {Medina}, G.~E. and {Mutlu-Pakdil}, B. and {Navabi}, M. and {No{\"e}l}, N.~E.~D. and {Sakowska}, J.~D. and {Stringfellow}, G.~S. and {DELVE Collaboration}},
        title = "{Reading between the (Spectral) Lines: Magellan/IMACS Spectroscopy of the Ultrafaint Dwarf Galaxies Eridanus IV and Centaurus I}",
      journal = {\apj},
         year = 2024,
        month = feb,
       volume = {961},
       number = {2},
          eid = {234},
        pages = {234},
          doi = {10.3847/1538-4357/ad0cf7},
archivePrefix = {arXiv},
       eprint = {2308.08602},
 primaryClass = {astro-ph.GA},
       adsurl = {https://ui.adsabs.harvard.edu/abs/2024ApJ...961..234H}
}

@ARTICLE{Collins2024MNRAS.528.2614C,
       author = {{Collins}, Michelle L.~M. and {Karim}, Noushin and {Martinez-Delgado}, David and {Monelli}, Matteo and {Tollerud}, Erik J. and {Donatiello}, Giuseppe and {Navabi}, Mahdieh and {Charles}, Emily and {Boschin}, Walter},
        title = "{Pisces VII/Triangulum III - M33's second dwarf satellite galaxy}",
      journal = {\mnras},
         year = 2024,
        month = feb,
       volume = {528},
       number = {2},
        pages = {2614-2620},
          doi = {10.1093/mnras/stae199},
archivePrefix = {arXiv},
       eprint = {2305.13966},
 primaryClass = {astro-ph.GA},
       adsurl = {https://ui.adsabs.harvard.edu/abs/2024MNRAS.528.2614C}
}

@ARTICLE{Pietrzynski2019Natur.567..200P,
       author = {{Pietrzy{\'n}ski}, G. and {Graczyk}, D. and {Gallenne}, A. and {Gieren}, W. and {Thompson}, I.~B. and {Pilecki}, B. and {Karczmarek}, P. and {G{\'o}rski}, M. and {Suchomska}, K. and {Taormina}, M. and {Zgirski}, B. and {Wielg{\'o}rski}, P. and {Ko{\l}aczkowski}, Z. and {Konorski}, P. and {Villanova}, S. and {Nardetto}, N. and {Kervella}, P. and {Bresolin}, F. and {Kudritzki}, R.~P. and {Storm}, J. and {Smolec}, R. and {Narloch}, W.},
        title = "{A distance to the Large Magellanic Cloud that is precise to one per cent}",
      journal = {\nat},
         year = 2019,
        month = mar,
       volume = {567},
       number = {7747},
        pages = {200-203},
          doi = {10.1038/s41586-019-0999-4},
archivePrefix = {arXiv},
       eprint = {1903.08096},
 primaryClass = {astro-ph.GA},
       adsurl = {https://ui.adsabs.harvard.edu/abs/2019Natur.567..200P}
}

@ARTICLE{Kallivayalil2013ApJ...764..161K,
       author = {{Kallivayalil}, Nitya and {van der Marel}, Roeland P. and {Besla}, Gurtina and {Anderson}, Jay and {Alcock}, Charles},
        title = "{Third-epoch Magellanic Cloud Proper Motions. I. Hubble Space Telescope/WFC3 Data and Orbit Implications}",
      journal = {\apj},
         year = 2013,
        month = feb,
       volume = {764},
       number = {2},
          eid = {161},
        pages = {161},
          doi = {10.1088/0004-637X/764/2/161},
archivePrefix = {arXiv},
       eprint = {1301.0832},
 primaryClass = {astro-ph.CO},
       adsurl = {https://ui.adsabs.harvard.edu/abs/2013ApJ...764..161K}
}

@ARTICLE{vanderMarel2002AJ....124.2639V,
       author = {{van der Marel}, Roeland P. and {Alves}, David R. and {Hardy}, Eduardo and {Suntzeff}, Nicholas B.},
        title = "{New Understanding of Large Magellanic Cloud Structure, Dynamics, and Orbit from Carbon Star Kinematics}",
      journal = {\aj},
         year = 2002,
        month = nov,
       volume = {124},
       number = {5},
        pages = {2639-2663},
          doi = {10.1086/343775},
archivePrefix = {arXiv},
       eprint = {astro-ph/0205161},
 primaryClass = {astro-ph},
       adsurl = {https://ui.adsabs.harvard.edu/abs/2002AJ....124.2639V}
}

@BOOK{deVaucouleurs1991rc3..book.....D,
       author = {{de Vaucouleurs}, Gerard and {de Vaucouleurs}, Antoinette and {Corwin}, Herold G., Jr. and {Buta}, Ronald J. and {Paturel}, Georges and {Fouque}, Pascal},
        title = "{Third Reference Catalogue of Bright Galaxies}",
         year = 1991,
       adsurl = {https://ui.adsabs.harvard.edu/abs/1991rc3..book.....D}
}

@ARTICLE{Cioni2000AA...359..601C,
       author = {{Cioni}, M. -R.~L. and {van der Marel}, R.~P. and {Loup}, C. and {Habing}, H.~J.},
        title = "{The tip of the red giant branch and distance of the Magellanic Clouds: results from the DENIS survey}",
      journal = {\aap},
         year = 2000,
        month = jul,
       volume = {359},
        pages = {601-614},
          doi = {10.48550/arXiv.astro-ph/0003223},
archivePrefix = {arXiv},
       eprint = {astro-ph/0003223},
 primaryClass = {astro-ph},
       adsurl = {https://ui.adsabs.harvard.edu/abs/2000A&A...359..601C}
}

@ARTICLE{Zivick2018ApJ...864...55Z,
       author = {{Zivick}, Paul and {Kallivayalil}, Nitya and {van der Marel}, Roeland P. and {Besla}, Gurtina and {Linden}, Sean T. and {Koz{\l}owski}, Szymon and {Fritz}, Tobias K. and {Kochanek}, C.~S. and {Anderson}, J. and {Sohn}, Sangmo Tony and {Geha}, Marla C. and {Alcock}, Charles R.},
        title = "{The Proper Motion Field of the Small Magellanic Cloud: Kinematic Evidence for Its Tidal Disruption}",
      journal = {\apj},
         year = 2018,
        month = sep,
       volume = {864},
       number = {1},
          eid = {55},
        pages = {55},
          doi = {10.3847/1538-4357/aad4b0},
archivePrefix = {arXiv},
       eprint = {1804.04110},
 primaryClass = {astro-ph.GA},
       adsurl = {https://ui.adsabs.harvard.edu/abs/2018ApJ...864...55Z}
}

@ARTICLE{Harris2006AJ....131.2514H,
       author = {{Harris}, Jason and {Zaritsky}, Dennis},
        title = "{Spectroscopic Survey of Red Giants in the Small Magellanic Cloud. I. Kinematics}",
      journal = {\aj},
         year = 2006,
        month = may,
       volume = {131},
       number = {5},
        pages = {2514-2524},
          doi = {10.1086/500974},
archivePrefix = {arXiv},
       eprint = {astro-ph/0601025},
 primaryClass = {astro-ph},
       adsurl = {https://ui.adsabs.harvard.edu/abs/2006AJ....131.2514H}
}

@ARTICLE{Rhode2023AJ....166..180R,
       author = {{Rhode}, Katherine L. and {Smith}, Nicholas J. and {Crnojevic}, Denija and {Sand}, David J. and {Lambert}, Ryan A. and {Vesperini}, Enrico and {Smith}, Madison V. and {Janowiecki}, Steven and {Salzer}, John J. and {Karunakaran}, Ananthan and {Spekkens}, Kristine},
        title = "{Exploring the Structures and Substructures of the Andromeda Satellite Dwarf Galaxies Cassiopeia III, Perseus I, and Lacerta I}",
      journal = {\aj},
         year = 2023,
        month = nov,
       volume = {166},
       number = {5},
          eid = {180},
        pages = {180},
          doi = {10.3847/1538-3881/acf859},
archivePrefix = {arXiv},
       eprint = {2309.01045},
 primaryClass = {astro-ph.GA},
       adsurl = {https://ui.adsabs.harvard.edu/abs/2023AJ....166..180R}
}

@ARTICLE{Brunthaler2007AA...462..101B,
       author = {{Brunthaler}, A. and {Reid}, M.~J. and {Falcke}, H. and {Henkel}, C. and {Menten}, K.~M.},
        title = "{The proper motion of the Local Group galaxy IC 10}",
      journal = {\aap},
         year = 2007,
        month = jan,
       volume = {462},
       number = {1},
        pages = {101-106},
          doi = {10.1051/0004-6361:20066430},
archivePrefix = {arXiv},
       eprint = {astro-ph/0610774},
 primaryClass = {astro-ph},
       adsurl = {https://ui.adsabs.harvard.edu/abs/2007A&A...462..101B}
}

@ARTICLE{McQuinn2017ApJ...834...78M,
       author = {{McQuinn}, Kristen B.~W. and {Boyer}, Martha L. and {Mitchell}, Mallory B. and {Skillman}, Evan D. and {Gehrz}, R.~D. and {Groenewegen}, Martin A.~T. and {McDonald}, Iain and {Sloan}, G.~C. and {van Loon}, Jacco Th. and {Whitelock}, Patricia A. and {Zijlstra}, Albert A.},
        title = "{DUSTiNGS. III. Distribution of Intermediate-age and Old Stellar Populations in Disks and Outer Extremities of Dwarf Galaxies}",
      journal = {\apj},
         year = 2017,
        month = jan,
       volume = {834},
       number = {1},
          eid = {78},
        pages = {78},
          doi = {10.3847/1538-4357/834/1/78},
archivePrefix = {arXiv},
       eprint = {1611.05044},
 primaryClass = {astro-ph.GA},
       adsurl = {https://ui.adsabs.harvard.edu/abs/2017ApJ...834...78M}
}

@ARTICLE{Bennet2024ApJ...971...98B,
       author = {{Bennet}, Paul and {Patel}, Ekta and {Sohn}, Sangmo Tony and {del Pino Molina}, Andr{\'e}s and {van der Marel}, Roeland P. and {Libralato}, Mattia and {Watkins}, Laura L. and {Aparicio}, Antonio and {Besla}, Gurtina and {Gallart}, Carme and {Fardal}, Mark A. and {Monelli}, Matteo and {Sacchi}, Elena and {Tollerud}, Erik and {Weisz}, Daniel R.},
        title = "{Proper Motions and Orbits of Distant Local Group Dwarf Galaxies from a Combination of Gaia and Hubble Data}",
      journal = {\apj},
         year = 2024,
        month = aug,
       volume = {971},
       number = {1},
          eid = {98},
        pages = {98},
          doi = {10.3847/1538-4357/ad5349},
archivePrefix = {arXiv},
       eprint = {2312.09276},
 primaryClass = {astro-ph.GA},
       adsurl = {https://ui.adsabs.harvard.edu/abs/2024ApJ...971...98B}
}

@ARTICLE{MartinezVazquez2021AJ....162..253M,
       author = {{Mart{\'\i}nez-V{\'a}zquez}, C.~E. and {Cerny}, W. and {Vivas}, A.~K. and {Drlica-Wagner}, A. and {Pace}, A.~B. and {Simon}, J.~D. and {Munoz}, R.~R. and {Walker}, A.~R. and {Allam}, S. and {Tucker}, D.~L. and {Adam{\'o}w}, M. and {Carlin}, J.~L. and {Choi}, Y. and {Ferguson}, P.~S. and {Ji}, A.~P. and {Kuropatkin}, N. and {Li}, T.~S. and {Mart{\'\i}nez-Delgado}, D. and {Mau}, S. and {Mutlu-Pakdil}, B. and {Nidever}, D.~L. and {Riley}, A.~H. and {Sakowska}, J.~D. and {Sand}, D.~J. and {Stringfellow}, G.~S. and {Stringfellow}, G.~S.},
        title = "{RR Lyrae Stars in the Newly Discovered Ultra-faint Dwarf Galaxy Centaurus I}",
      journal = {\aj},
         year = 2021,
        month = dec,
       volume = {162},
       number = {6},
          eid = {253},
        pages = {253},
          doi = {10.3847/1538-3881/ac2368},
archivePrefix = {arXiv},
       eprint = {2107.05688},
 primaryClass = {astro-ph.GA},
       adsurl = {https://ui.adsabs.harvard.edu/abs/2021AJ....162..253M}
}

@ARTICLE{Richstein2024ApJ...967...72R,
       author = {{Richstein}, Hannah and {Kallivayalil}, Nitya and {Simon}, Joshua D. and {Garling}, Christopher T. and {Wetzel}, Andrew and {Warfield}, Jack T. and {van der Marel}, Roeland P. and {Jeon}, Myoungwon and {Rose}, Jonah C. and {Torrey}, Paul and {Engelhardt}, Anna Claire and {Besla}, Gurtina and {Choi}, Yumi and {Geha}, Marla and {Guhathakurta}, Puragra and {Kirby}, Evan N. and {Patel}, Ekta and {Sacchi}, Elena and {Sohn}, Sangmo Tony},
        title = "{Deep Hubble Space Telescope Photometry of Large Magellanic Cloud and Milky Way Ultrafaint Dwarfs: A Careful Look into the Magnitude{\textendash}Size Relation}",
      journal = {\apj},
         year = 2024,
        month = may,
       volume = {967},
       number = {1},
          eid = {72},
        pages = {72},
          doi = {10.3847/1538-4357/ad393c},
archivePrefix = {arXiv},
       eprint = {2402.08731},
 primaryClass = {astro-ph.GA},
       adsurl = {https://ui.adsabs.harvard.edu/abs/2024ApJ...967...72R}
}

@ARTICLE{Bhardwaj2024AJ....167..247B,
       author = {{Bhardwaj}, Anupam and {Rejkuba}, Marina and {Ngeow}, Chow-Choong and {Marconi}, Marcella and {Ripepi}, Vincenzo and {Samantaray}, Abhinna Sundar and {Singh}, Harinder P.},
        title = "{Stellar Variability and Distance Indicators in the Near-infrared in Nearby Galaxies. I. RR Lyrae and Anomalous Cepheids in Draco Dwarf Spheroidal}",
      journal = {\aj},
         year = 2024,
        month = may,
       volume = {167},
       number = {5},
          eid = {247},
        pages = {247},
          doi = {10.3847/1538-3881/ad38b6},
archivePrefix = {arXiv},
       eprint = {2404.01394},
 primaryClass = {astro-ph.GA},
       adsurl = {https://ui.adsabs.harvard.edu/abs/2024AJ....167..247B}
}

@ARTICLE{Hansen2024ApJ...968...21H,
       author = {{Hansen}, Terese T. and {Simon}, Joshua D. and {Li}, Ting S. and {Sharkey}, Domani and {Ji}, Alexander P. and {Thompson}, Ian B. and {Reggiani}, Henrique M. and {Galarza}, Jhon Yana},
        title = "{Chemical Diversity on Small Scales: Abundance Analysis of the Tucana V Ultrafaint Dwarf Galaxy}",
      journal = {\apj},
         year = 2024,
        month = jun,
       volume = {968},
       number = {1},
          eid = {21},
        pages = {21},
          doi = {10.3847/1538-4357/ad3a52},
archivePrefix = {arXiv},
       eprint = {2403.13060},
 primaryClass = {astro-ph.GA},
       adsurl = {https://ui.adsabs.harvard.edu/abs/2024ApJ...968...21H}
}

@ARTICLE{An2024MNRAS.532.3713A,
       author = {{An}, Zhaozhou and {Walker}, Matthew G. and {Pace}, Andrew B.},
        title = "{Offset of M54 from the Sagittarius dwarf spheroidal galaxy}",
      journal = {\mnras},
         year = 2024,
        month = aug,
       volume = {532},
       number = {4},
        pages = {3713-3728},
          doi = {10.1093/mnras/stae1680},
archivePrefix = {arXiv},
       eprint = {2404.16184},
 primaryClass = {astro-ph.GA},
       adsurl = {https://ui.adsabs.harvard.edu/abs/2024MNRAS.532.3713A}
}

@ARTICLE{Cerny2025ApJ...979..164C,
       author = {{Cerny}, W. and {Chiti}, A. and {Geha}, M. and {Mutlu-Pakdil}, B. and {Drlica-Wagner}, A. and {Tan}, C.~Y. and {Adam{\'o}w}, M. and {Pace}, A.~B. and {Simon}, J.~D. and {Sand}, D.~J. and {Ji}, A.~P. and {Li}, T.~S. and {Vivas}, A.~K. and {Bell}, E.~F. and {Carlin}, J.~L. and {Carballo-Bello}, J.~A. and {Chaturvedi}, A. and {Choi}, Y. and {Doliva-Dolinsky}, A. and {Gnedin}, O.~Y. and {Limberg}, G. and {Mart{\'\i}nez-V{\'a}zquez}, C.~E. and {Mau}, S. and {Medina}, G.~E. and {Navabi}, M. and {No{\"e}l}, N.~E.~D. and {Placco}, V.~M. and {Riley}, A.~H. and {Roederer}, I.~U. and {Stringfellow}, G.~S. and {Bom}, C.~R. and {Ferguson}, P.~S. and {James}, D.~J. and {Mart{\'\i}nez-Delgado}, D. and {Massana}, P. and {Nidever}, D.~L. and {Sakowska}, J.~D. and {Santana-Silva}, L. and {Sherman}, N.~F. and {Tollerud}, E.~J.},
        title = "{Discovery and Spectroscopic Confirmation of Aquarius III: A Low-mass Milky Way Satellite Galaxy}",
      journal = {\apj},
         year = 2025,
        month = feb,
       volume = {979},
       number = {2},
          eid = {164},
        pages = {164},
          doi = {10.3847/1538-4357/ad8eba},
archivePrefix = {arXiv},
       eprint = {2410.00981},
 primaryClass = {astro-ph.GA},
       adsurl = {https://ui.adsabs.harvard.edu/abs/2025ApJ...979..164C}
}

@ARTICLE{Garofalo2025AA...695A..88G,
       author = {{Garofalo}, A. and {Clementini}, G. and {Cusano}, F. and {Muraveva}, T. and {Monti}, L.},
        title = "{Give to Ursa Minor what is Ursa Minor's: An updated census of the RR Lyrae population in the Ursa Minor dwarf galaxy based on Gaia DR3}",
      journal = {\aap},
         year = 2025,
        month = mar,
       volume = {695},
          eid = {A88},
        pages = {A88},
          doi = {10.1051/0004-6361/202452619},
archivePrefix = {arXiv},
       eprint = {2410.12433},
 primaryClass = {astro-ph.GA},
       adsurl = {https://ui.adsabs.harvard.edu/abs/2025A&A...695A..88G}
}

@ARTICLE{Casey2025ApJ...984..148C,
       author = {{Casey}, Quinn O. and {Mutlu-Pakdil}, Bur{\c{c}}in and {Sand}, David J. and {Pace}, Andrew B. and {Crnojevi{\'c}}, Denija and {Doliva-Dolinsky}, Amandine and {Cerny}, William and {Heiger}, Mairead E. and {Riley}, Alex H. and {Ji}, Alexander P. and {Limberg}, Guilherme and {Marin}, Laurella and {Mart{\'\i}nez-V{\'a}zquez}, Clara E. and {Medina}, Gustavo E. and {Li}, Ting S. and {Campana}, Sasha N. and {Chaturvedi}, Astha and {Sakowska}, Joanna D. and {Zenteno}, Alfredo and {Carballo-Bello}, Julio A. and {Navabi}, Mahdieh and {Bom}, Clecio R. and {DELVE Collaboration}},
        title = "{Deep Photometric Observations of Ultrafaint Milky Way Satellites Centaurus I and Eridanus IV}",
      journal = {\apj},
         year = 2025,
        month = may,
       volume = {984},
       number = {2},
          eid = {148},
        pages = {148},
          doi = {10.3847/1538-4357/adc67e},
archivePrefix = {arXiv},
       eprint = {2501.04772},
 primaryClass = {astro-ph.GA},
       adsurl = {https://ui.adsabs.harvard.edu/abs/2025ApJ...984..148C}
}

@ARTICLE{Tan2025ApJ...979..176T,
       author = {{Tan}, C.~Y. and {Cerny}, W. and {Drlica-Wagner}, A. and {Pace}, A.~B. and {Geha}, M. and {Ji}, A.~P. and {Li}, T.~S. and {Adam{\'o}w}, M. and {Anbajagane}, D. and {Bom}, C.~R. and {Carballo-Bello}, J.~A. and {Carlin}, J.~L. and {Chang}, C. and {Chaturvedi}, A. and {Chiti}, A. and {Choi}, Y. and {Collins}, M.~L.~M. and {Doliva-Dolinsky}, A. and {Ferguson}, P.~S. and {Gruendl}, R.~A. and {James}, D.~J. and {Limberg}, G. and {Navabi}, M. and {Mart{\'\i}nez-Delgado}, D. and {Mart{\'\i}nez-V{\'a}zquez}, C.~E. and {Medina}, G.~E. and {Mutlu-Pakdil}, B. and {Nidever}, D.~L. and {No{\"e}l}, N.~E.~D. and {Riley}, A.~H. and {Sakowska}, J.~D. and {Sand}, D.~J. and {Sharp}, J. and {Stringfellow}, G.~S. and {Tolley}, C. and {Tucker}, D.~L. and {Vivas}, A.~K. and {Delve Collaboration}},
        title = "{A Pride of Satellites in the Constellation Leo? Discovery of the Leo VI Milky Way Satellite Ultra-faint Dwarf Galaxy with DELVE Early Data Release 3}",
      journal = {\apj},
         year = 2025,
        month = feb,
       volume = {979},
       number = {2},
          eid = {176},
        pages = {176},
          doi = {10.3847/1538-4357/ad9b0c},
archivePrefix = {arXiv},
       eprint = {2408.00865},
 primaryClass = {astro-ph.GA},
       adsurl = {https://ui.adsabs.harvard.edu/abs/2025ApJ...979..176T}
}

@ARTICLE{Monelli2025AA...700A.201M,
       author = {{Monelli}, M. and {Piersimoni}, A.~M. and {Mart{\'\i}nez-V{\'a}zquez}, C.~E. and {Cassisi}, S. and {Stetson}, P.~B. and {Marconi}, M. and {P{\'e}rez-Garc{\'\i}a}, P.~J. and {Bono}, G. and {Dall'Ora}, M. and {Fiorentino}, G. and {Gallart}, C. and {Marengo}, M.},
        title = "{Variable stars in Local Group galaxies: VII. New insights into the Leo II dSph}",
      journal = {\aap},
         year = 2025,
        month = aug,
       volume = {700},
          eid = {A201},
        pages = {A201},
          doi = {10.1051/0004-6361/202453567},
       adsurl = {https://ui.adsabs.harvard.edu/abs/2025A&A...700A.201M}
}
\bibliographystyle{aasjournalv7}



\appendix

\section{References for the observational data}\label{app:references}
We list the literature references for the observational quantities used in our analysis in Tables~\ref{tab:references_mw} and~\ref{tab:references_m31} for the MW and M31 satellites, respectively. Stellar masses are taken from the Local Volume Database, calculated from $M_V$ assuming $M/L_V=2$ \citep{pace_local_2025}.

\begin{sidewaystable*}[ht]
\caption{Summary of confirmed MW satellites ($d_{\text{host}} \le 350$ kpc). References correspond to the distance, systemic line-of-sight velocity, proper motions, and $M_V$, respectively.}
\label{tab:references_mw}
\centering
\scriptsize
\begin{tabular}{lllll}
\toprule
Object & Distance & $v_{\rm los}$ & Proper motion & $M_V$ \\
\midrule
Antlia II & \cite{Vivas2022ApJ...926...78V} & \cite{Ji2021ApJ...921...32J} & \cite{pace_proper_2022} & \cite{Ji2021ApJ...921...32J} \\
Aquarius II & \cite{torrealba_feeble_2016} & \cite{Bruce2023ApJ...950..167B} & \cite{pace_proper_2022} & \cite{torrealba_feeble_2016} \\
Aquarius III & \cite{Cerny2025ApJ...979..164C} & \cite{Cerny2025ApJ...979..164C} & \cite{Cerny2025ApJ...979..164C} & \cite{Cerny2025ApJ...979..164C} \\
Bootes I & \cite{DallOra2006ApJ...653L.109D} & \cite{sandford_chemodynamics_2026} & \cite{pace_proper_2022} & \cite{Munoz2018ApJ...860...66M} \\
Bootes II & \cite{Walsh2008ApJ...688..245W} & \cite{Bruce2023ApJ...950..167B} & \cite{pace_proper_2022} & \cite{Munoz2018ApJ...860...66M} \\
Bootes III & \cite{Carlin2018ApJ...865....7C} & \cite{Carlin2009ApJ...702L...9C} & \cite{pace_proper_2022} & \cite{Correnti2009MNRAS.397L..26C} \\
Canes Venatici I & \cite{Kuehn2008ApJ...674L..81K} & \cite{Simon2007ApJ...670..313S} & \cite{pace_proper_2022} & \cite{Munoz2018ApJ...860...66M} \\
Canes Venatici II & \cite{Greco2008ApJ...675L..73G} & \cite{Simon2007ApJ...670..313S} & \cite{pace_proper_2022} & \cite{Munoz2018ApJ...860...66M} \\
Carina & \cite{Karczmarek2015AJ....150...90K} & \cite{Walker2009AJ....137.3100W} & \cite{pace_proper_2022} & \cite{Munoz2018ApJ...860...66M} \\
Carina II & \cite{Torrealba2018MNRAS.475.5085T} & \cite{Li2018ApJ...857..145L} & \cite{pace_proper_2022} & \cite{Torrealba2018MNRAS.475.5085T} \\
Carina III & \cite{Torrealba2018MNRAS.475.5085T} & \cite{Li2018ApJ...857..145L} & \cite{pace_proper_2022} & \cite{Torrealba2018MNRAS.475.5085T} \\
Centaurus I & \cite{MartinezVazquez2021AJ....162..253M} & \cite{Heiger2024ApJ...961..234H} & \cite{Heiger2024ApJ...961..234H} & \cite{Casey2025ApJ...984..148C} \\
Columba I & \cite{Carlin2017AJ....154..267C} & \cite{Fritz2019AA...623A.129F} & \cite{pace_proper_2022} & \cite{Carlin2017AJ....154..267C} \\
Coma Berenices & \cite{Musella2009ApJ...695L..83M} & \cite{Simon2007ApJ...670..313S} & \cite{pace_proper_2022} & \cite{Munoz2018ApJ...860...66M} \\
Crater II & \cite{Torrealba2016MNRAS.459.2370T} & \cite{Ji2021ApJ...921...32J} & \cite{pace_proper_2022} & \cite{Torrealba2016MNRAS.459.2370T} \\
Draco & \cite{Bhardwaj2024AJ....167..247B} & \cite{Walker2015MNRAS.448.2717W} & \cite{pace_proper_2022} & \cite{Munoz2018ApJ...860...66M} \\
Eridanus IV & \cite{Cerny2021ApJ...920L..44C} & \cite{Heiger2024ApJ...961..234H} & \cite{Cerny2021ApJ...920L..44C} & \cite{Casey2025ApJ...984..148C} \\
Fornax & \cite{Oakes2022ApJ...929..116O} & \cite{Walker2009AJ....137.3100W} & \cite{pace_proper_2022} & \cite{Munoz2018ApJ...860...66M} \\
Grus I & \cite{MartinezVazquez2019MNRAS.490.2183M} & \cite{Chiti2022ApJ...939...41C} & \cite{pace_proper_2022} & \cite{Cantu2021ApJ...916...81C} \\
Grus II & \cite{MartinezVazquez2019MNRAS.490.2183M} & \cite{Simon2020ApJ...892..137S} & \cite{pace_proper_2022} & \cite{Simon2020ApJ...892..137S} \\
Hercules & \cite{MutluPakdil2020ApJ...902..106M} & \cite{Simon2007ApJ...670..313S} & \cite{pace_proper_2022} & \cite{Munoz2018ApJ...860...66M} \\
Horologium I & \cite{Koposov2015ApJ...805..130K} & \cite{Koposov2015ApJ...811...62K} & \cite{pace_proper_2022} & \cite{Richstein2024ApJ...967...72R} \\
Hydra II & \cite{Vivas2016AJ....151..118V} & \cite{Kirby2015ApJ...810...56K} & \cite{pace_proper_2022} & \cite{Richstein2024ApJ...967...72R} \\
Hydrus I & \cite{Koposov2018MNRAS.479.5343K} & \cite{Koposov2018MNRAS.479.5343K} & \cite{pace_proper_2022} & \cite{Koposov2018MNRAS.479.5343K} \\
Leo I & \cite{Stetson2014PASP..126..616S} & \cite{Mateo2008ApJ...675..201M} & \cite{Bennet2024ApJ...971...98B} & \cite{Munoz2018ApJ...860...66M} \\
Leo II & \cite{Monelli2025AA...700A.201M} & \cite{Spencer2017ApJ...836..202S} & \cite{Bennet2024ApJ...971...98B} & \cite{Munoz2018ApJ...860...66M} \\
Leo IV & \cite{Medina2018ApJ...855...43M} & \cite{Jenkins2021ApJ...920...92J} & \cite{julio_satellite_2024} & \cite{Munoz2018ApJ...860...66M} \\
Leo V & \cite{Medina2018ApJ...855...43M} & \cite{Jenkins2021ApJ...920...92J} & \cite{julio_satellite_2024} & \cite{Munoz2018ApJ...860...66M} \\
Leo VI & \cite{Tan2025ApJ...979..176T} & \cite{Tan2025ApJ...979..176T} & \cite{Tan2025ApJ...979..176T} & \cite{Tan2025ApJ...979..176T} \\
LMC & \cite{Pietrzynski2019Natur.567..200P} & \cite{vanderMarel2002AJ....124.2639V} & \cite{Kallivayalil2013ApJ...764..161K} & \cite{deVaucouleurs1991rc3..book.....D} \\
Pegasus III & \cite{Kim2016ApJ...833...16K} & \cite{Kim2016ApJ...833...16K} & \cite{pace_proper_2022} & \cite{richstein_structural_2022} \\
Pegasus IV & \cite{Cerny2023ApJ...942..111C} & \cite{Cerny2023ApJ...942..111C} & \cite{Cerny2023ApJ...942..111C} & \cite{Cerny2023ApJ...942..111C} \\
Phoenix II & \cite{MutluPakdil2018ApJ...863...25M} & \cite{Fritz2019AA...623A.129F} & \cite{pace_proper_2022} & \cite{Richstein2024ApJ...967...72R} \\
Pictor II & \cite{pace_local_2025} & \cite{pace_local_2025} & \cite{pace_local_2025} & \cite{pace_local_2025} \\
Pisces II & \cite{Sand2012ApJ...756...79S} & \cite{Kirby2015ApJ...810...56K} & \cite{pace_proper_2022} & \cite{Richstein2022ApJ...933..217R} \\
Reticulum II & \cite{MutluPakdil2018ApJ...863...25M} & \cite{Walker2015ApJ...808..108W} & \cite{pace_proper_2022} & \cite{MutluPakdil2018ApJ...863...25M} \\
Reticulum III & \cite{DrlicaWagner2015ApJ...813..109D} & \cite{Fritz2019AA...623A.129F} & \cite{pace_proper_2022} & \cite{DrlicaWagner2015ApJ...813..109D} \\
Sagittarius & \cite{mcconnachie_observed_2012} & \cite{mcconnachie_observed_2012} & \cite{An2024MNRAS.532.3713A} & \cite{mcconnachie_observed_2012} \\
Sculptor & \cite{MartinezVazquez2015MNRAS.454.1509M} & \cite{Walker2009AJ....137.3100W} & \cite{pace_proper_2022} & \cite{Munoz2018ApJ...860...66M} \\
Segue 1 & \cite{belokurov_cats_2007} & \cite{Simon2011ApJ...733...46S} & \cite{pace_proper_2022} & \cite{Munoz2018ApJ...860...66M} \\
Segue 2 & \cite{Boettcher2013AJ....146...94B} & \cite{Kirby2013ApJ...770...16K} & \cite{pace_proper_2022} & \cite{Munoz2018ApJ...860...66M} \\
Sextans & \cite{Lee2009ApJ...703..692L} & \cite{Walker2009AJ....137.3100W} & \cite{pace_proper_2022} & \cite{Munoz2018ApJ...860...66M} \\
SMC & \cite{Cioni2000AA...359..601C} & \cite{Harris2006AJ....131.2514H} & \cite{Zivick2018ApJ...864...55Z} & \cite{deVaucouleurs1991rc3..book.....D} \\
Triangulum II & \cite{Carlin2017AJ....154..267C} & \cite{Buttry2022MNRAS.514.1706B} & \cite{pace_proper_2022} & \cite{Richstein2024ApJ...967...72R} \\
Tucana II & \cite{Vivas2020ApJS..247...35V} & \cite{Chiti2023AJ....165...55C} & \cite{pace_proper_2022} & \cite{Koposov2015ApJ...805..130K} \\
Tucana IV & \cite{Simon2020ApJ...892..137S} & \cite{Simon2020ApJ...892..137S} & \cite{pace_proper_2022} & \cite{Simon2020ApJ...892..137S} \\
Tucana V & \cite{Simon2020ApJ...892..137S} & \cite{Hansen2024ApJ...968...21H} & \cite{battaglia_gaia_2022} & \cite{Simon2020ApJ...892..137S} \\
Ursa Major I & \cite{Garofalo2013ApJ...767...62G} & \cite{Simon2007ApJ...670..313S} & \cite{pace_proper_2022} & \cite{Munoz2018ApJ...860...66M} \\
Ursa Major II & \cite{DallOra2012ApJ...752...42D} & \cite{Simon2007ApJ...670..313S} & \cite{pace_proper_2022} & \cite{Munoz2018ApJ...860...66M} \\
Ursa Minor & \cite{Garofalo2025AA...695A..88G} & \cite{Spencer2018AJ....156..257S} & \cite{pace_proper_2022} & \cite{Munoz2018ApJ...860...66M} \\
Willman 1 & \cite{Willman2006astro.ph..3486W} & \cite{Willman2011AJ....142..128W} & \cite{pace_proper_2022} & \cite{Munoz2018ApJ...860...66M} \\
\bottomrule
\end{tabular}
\end{sidewaystable*}

\begin{sidewaystable*}[ht]
\caption{Summary of confirmed M31 satellites ($d_{\text{host}} \le 350$ kpc). References correspond to the distance, systemic line-of-sight velocity, proper motions (when available), and $M_V$, respectively.}
\label{tab:references_m31}
\centering
\scriptsize
\begin{tabular}{lllll}
\toprule
Object & Distance & $v_{\rm los}$ & Proper motion & $M_V$ \\
\midrule
Andromeda I & \cite{savino_hubble_2022} & \cite{Tollerud2012ApJ...752...45T} & --- & \cite{Martin2016ApJ...833..167M} \\
Andromeda II & \cite{savino_hubble_2022} & \cite{Ho2012ApJ...758..124H} & --- & \cite{Martin2016ApJ...833..167M} \\
Andromeda III & \cite{savino_hubble_2022} & \cite{Tollerud2012ApJ...752...45T} & \cite{casetti-dinescu_hst_2024} & \cite{Martin2016ApJ...833..167M} \\
Andromeda V & \cite{savino_hubble_2022} & \cite{Tollerud2012ApJ...752...45T} & \cite{casetti-dinescu_hst_2025} & \cite{Martin2016ApJ...833..167M} \\
Andromeda VI & \cite{savino_hubble_2022} & \cite{Pickett2025MNRAS.540.1701P} & \cite{casetti-dinescu_hst_2025} & \cite{McConnachie2006MNRAS.365.1263M} \\
Andromeda VII & \cite{savino_hubble_2022} & \cite{Tollerud2012ApJ...752...45T} & \cite{Warfield2023MNRAS.519.1189W} & \cite{McConnachie2006MNRAS.365.1263M} \\
Andromeda IX & \cite{savino_hubble_2022} & \cite{Tollerud2012ApJ...752...45T} & --- & \cite{Martin2016ApJ...833..167M} \\
Andromeda X & \cite{savino_hubble_2022} & \cite{Tollerud2012ApJ...752...45T} & --- & \cite{Martin2016ApJ...833..167M} \\
Andromeda XI & \cite{savino_hubble_2022} & \cite{Collins2013ApJ...768..172C} & --- & \cite{Martin2016ApJ...833..167M} \\
Andromeda XII & \cite{savino_hubble_2022} & \cite{Collins2013ApJ...768..172C} & --- & \cite{Martin2016ApJ...833..167M} \\
Andromeda XIII & \cite{savino_hubble_2022} & \cite{Tollerud2012ApJ...752...45T} & --- & \cite{Martin2016ApJ...833..167M} \\
Andromeda XIV & \cite{savino_hubble_2022} & \cite{Collins2013ApJ...768..172C} & --- & \cite{Martin2016ApJ...833..167M} \\
Andromeda XV & \cite{savino_hubble_2022} & \cite{Tollerud2012ApJ...752...45T} & --- & \cite{Martin2016ApJ...833..167M} \\
Andromeda XVI & \cite{MartinezVazquez2017ApJ...850..137M} & \cite{Tollerud2012ApJ...752...45T} & --- & \cite{Martin2016ApJ...833..167M} \\
Andromeda XVII & \cite{savino_hubble_2022} & \cite{Collins2013ApJ...768..172C} & --- & \cite{Martin2016ApJ...833..167M} \\
Andromeda XIX & \cite{savino_hubble_2022} & \cite{Collins2020MNRAS.491.3496C} & --- & \cite{Martin2016ApJ...833..167M} \\
Andromeda XX & \cite{savino_hubble_2022} & \cite{Collins2013ApJ...768..172C} & --- & \cite{Martin2016ApJ...833..167M} \\
Andromeda XXI & \cite{savino_hubble_2022} & \cite{Collins2021MNRAS.505.5686C} & --- & \cite{Martin2016ApJ...833..167M} \\
Andromeda XXII & \cite{savino_hubble_2022} & \cite{Collins2013ApJ...768..172C} & --- & \cite{Martin2016ApJ...833..167M} \\
Andromeda XXIII & \cite{savino_hubble_2022} & \cite{Pickett2025MNRAS.540.1701P} & --- & \cite{Martin2016ApJ...833..167M} \\
Andromeda XXIV & \cite{savino_hubble_2022} & \cite{Collins2013ApJ...768..172C} & --- & \cite{Martin2016ApJ...833..167M} \\
Andromeda XXV & \cite{savino_hubble_2022} & \cite{Charles2023MNRAS.521.3527C} & --- & \cite{Martin2016ApJ...833..167M} \\
Andromeda XXVI & \cite{savino_hubble_2022} & \cite{Collins2013ApJ...768..172C} & --- & \cite{Martin2016ApJ...833..167M} \\
Andromeda XXVII & \cite{Richardson2011ApJ...732...76R} & \cite{Collins2013ApJ...768..172C} & --- & \cite{Richardson2011ApJ...732...76R} \\
Andromeda XXIX & \cite{savino_hubble_2022} & \cite{Tollerud2013ApJ...768...50T} & --- & \cite{Slater2015ApJ...806..230S} \\
Andromeda XXXV & \cite{Arias2025ApJ...982L...3A} & --- & --- & \cite{Arias2025ApJ...982L...3A} \\
Cassiopeia II & \cite{savino_hubble_2022} & \cite{Collins2013ApJ...768..172C} & --- & \cite{Martin2016ApJ...833..167M} \\
Cassiopeia III & \cite{savino_hubble_2022} & \cite{Martin2014ApJ...793L..14M} & --- & \cite{Rhode2023AJ....166..180R} \\
IC 10 & \cite{McQuinn2017ApJ...834...78M} & \cite{mcconnachie_observed_2012} & \cite{Brunthaler2007AA...462..101B} & \cite{mcconnachie_observed_2012} \\
Lacerta I & \cite{savino_hubble_2022} & \cite{Martin2014ApJ...793L..14M} & --- & \cite{Rhode2023AJ....166..180R} \\
LGS 3 & \cite{savino_hubble_2022} & \cite{mcconnachie_observed_2012} & \cite{Bennet2025ApJ...993..228B} & \cite{mcconnachie_observed_2012} \\
M 32 & \cite{savino_hubble_2022} & \cite{mcconnachie_observed_2012} & --- & \cite{mcconnachie_observed_2012} \\
NGC 147 & \cite{savino_hubble_2022} & \cite{mcconnachie_observed_2012} & \cite{sohn_hst_2020} & \cite{mcconnachie_observed_2012} \\
NGC 185 & \cite{savino_hubble_2022} & \cite{mcconnachie_observed_2012} & \cite{sohn_hst_2020} & \cite{mcconnachie_observed_2012} \\
NGC 205 & \cite{savino_hubble_2022} & \cite{mcconnachie_observed_2012} & --- & \cite{mcconnachie_observed_2012} \\
Pegasus V & \cite{collins_observational_2022} & --- & --- & \cite{collins_observational_2022} \\
Pegasus VII & \cite{Smith2025ApJ...983...59S} & --- & --- & \cite{Smith2025ApJ...983...59S} \\
Perseus I & \cite{savino_hubble_2022} & \cite{Martin2014ApJ...793L..14M} & --- & \cite{Rhode2023AJ....166..180R} \\
Pisces VII & \cite{Collins2024MNRAS.528.2614C} & --- & --- & \cite{Collins2024MNRAS.528.2614C} \\
\bottomrule
\end{tabular}
\end{sidewaystable*}

\section{Cumulative radial distribution of simulations}\label{app:cdfs_sims}
\begin{figure}[ht]
    \includegraphics[width=0.47\columnwidth]{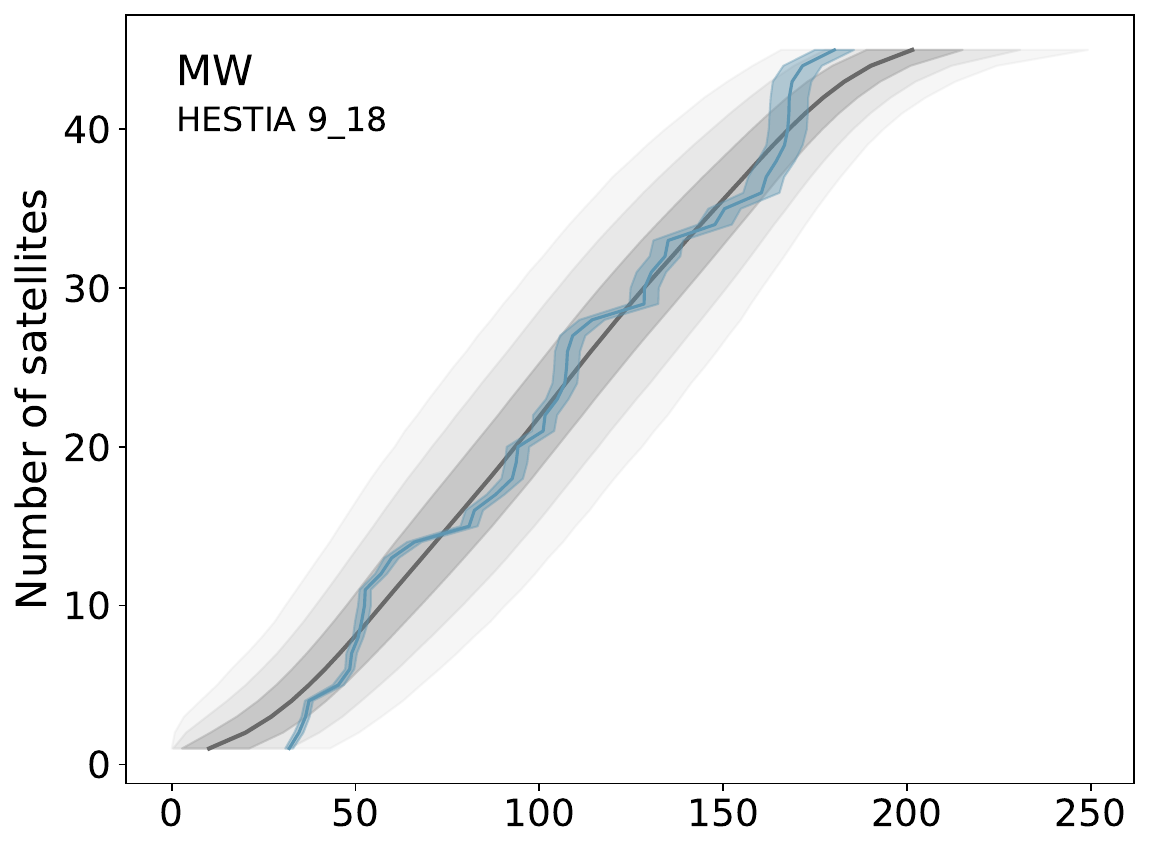}
    \includegraphics[width=0.47\columnwidth]{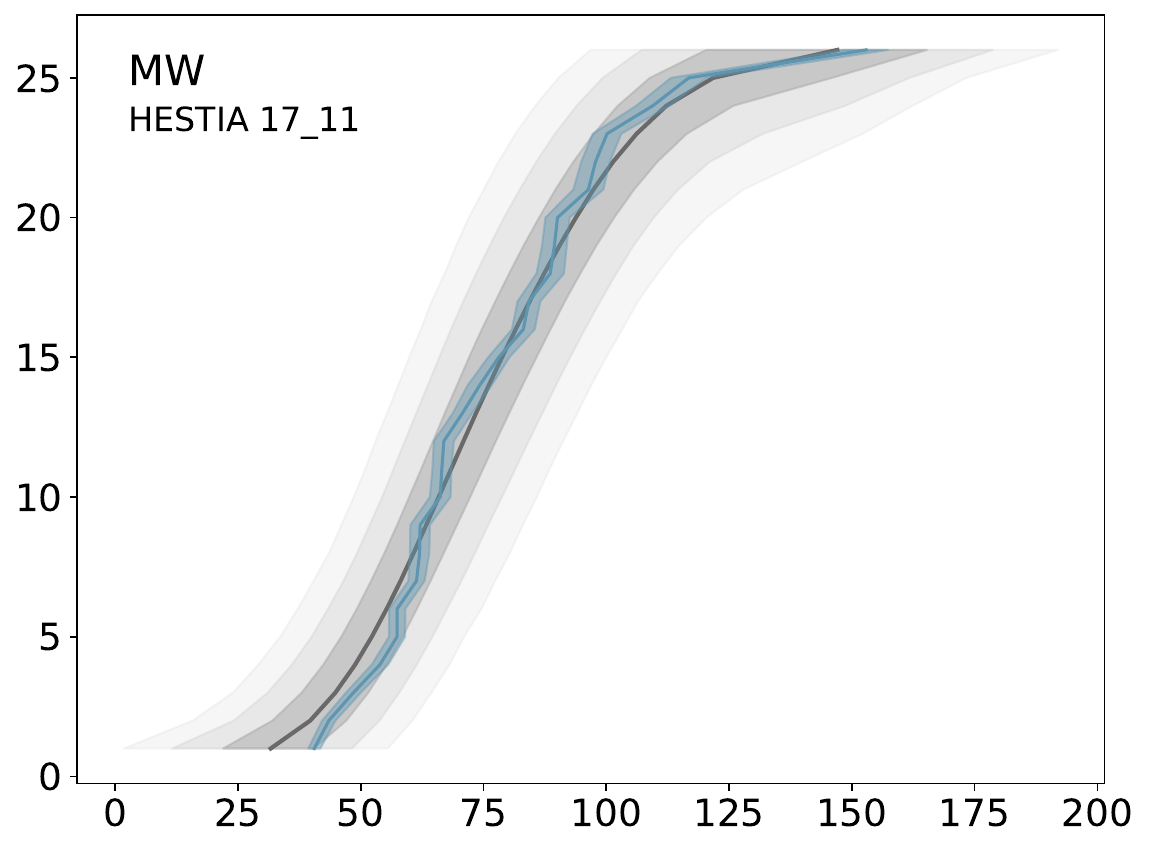}
    \includegraphics[width=0.47\columnwidth]{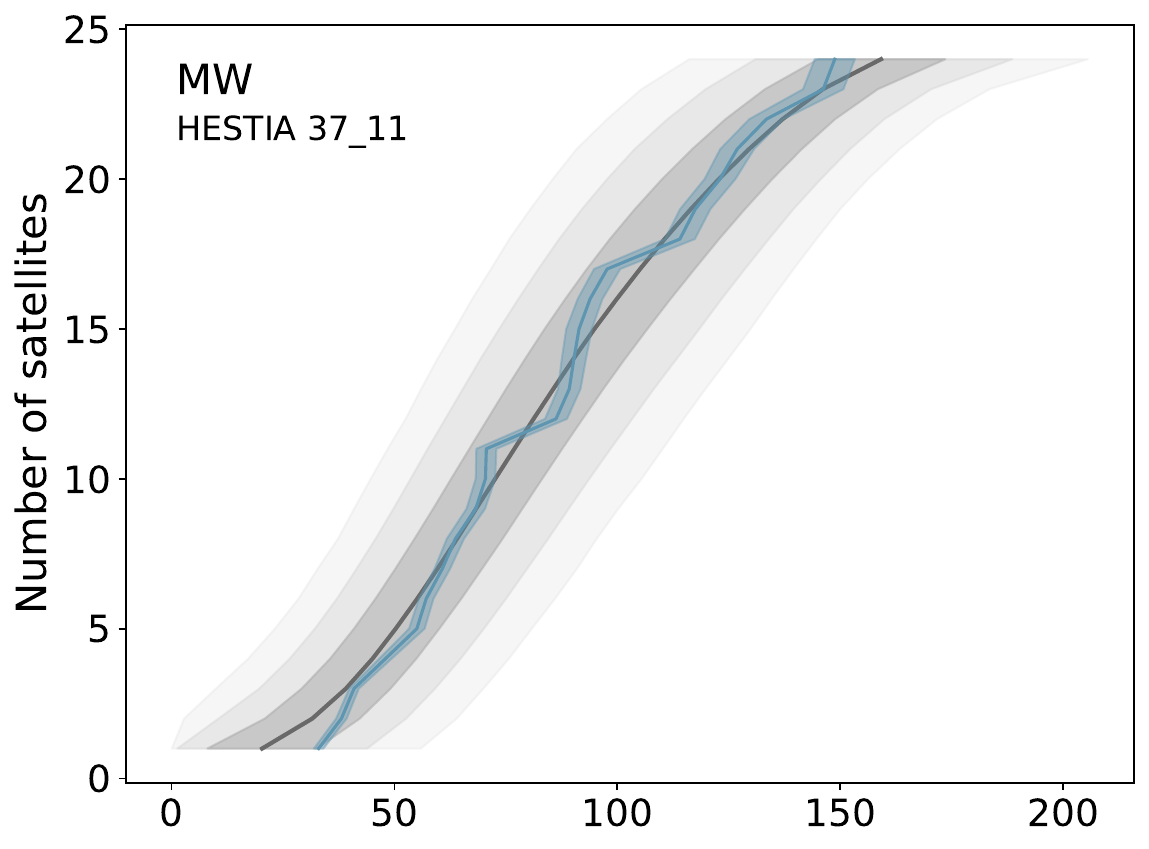}
    \includegraphics[width=0.47\columnwidth]{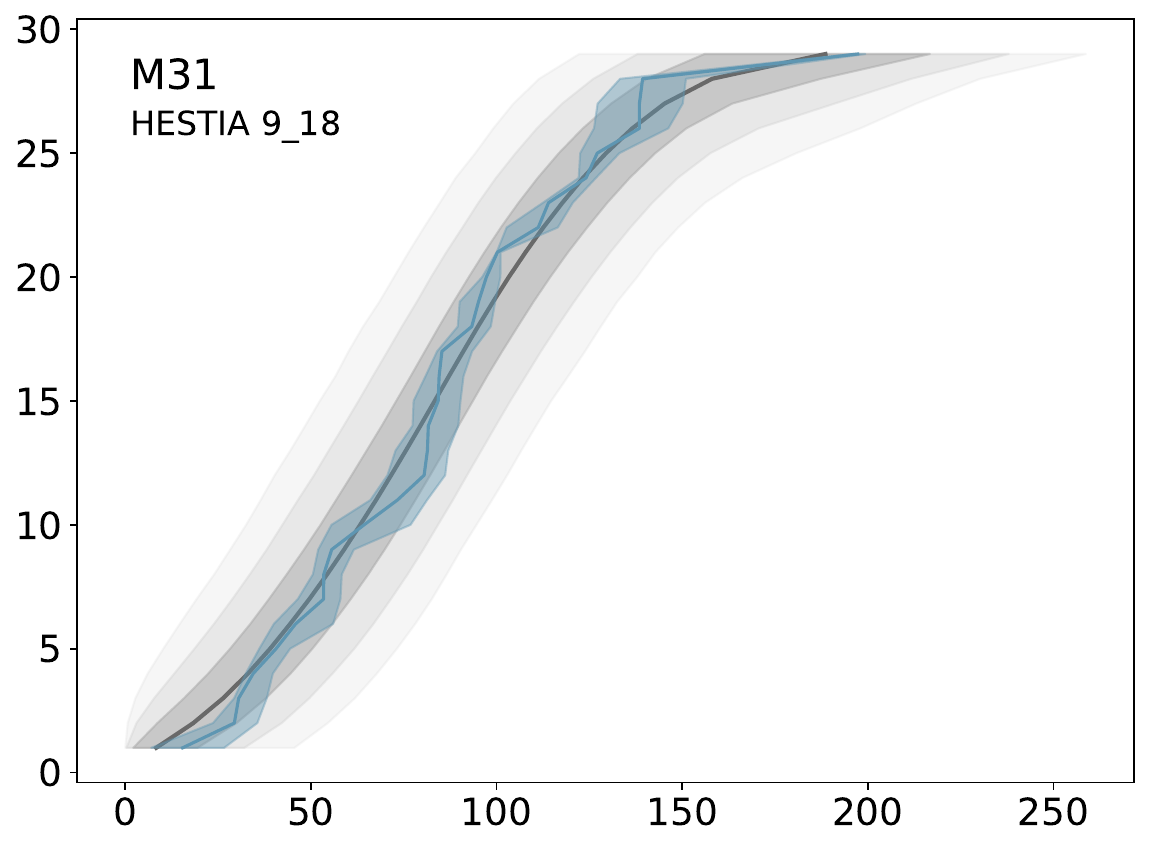}
    \includegraphics[width=0.47\columnwidth]{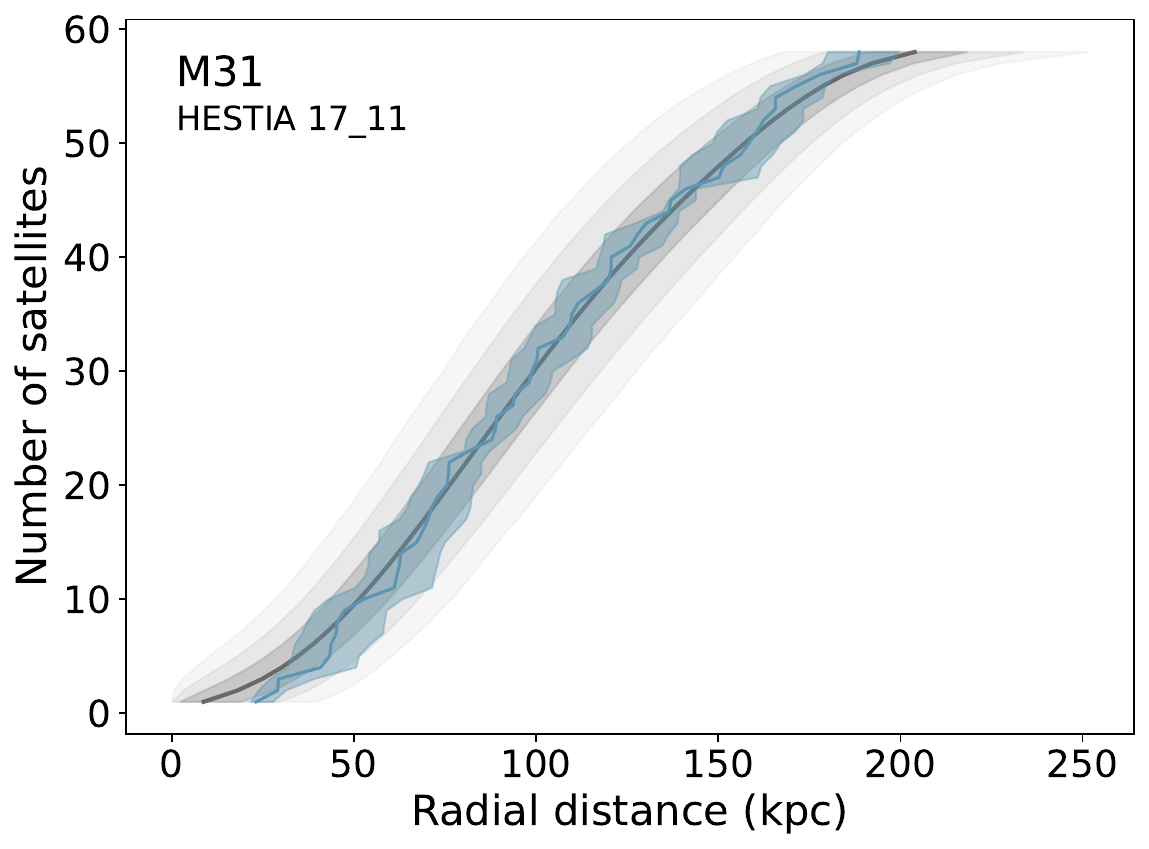}
    \hspace{2.5mm}
    \includegraphics[width=0.47\columnwidth]{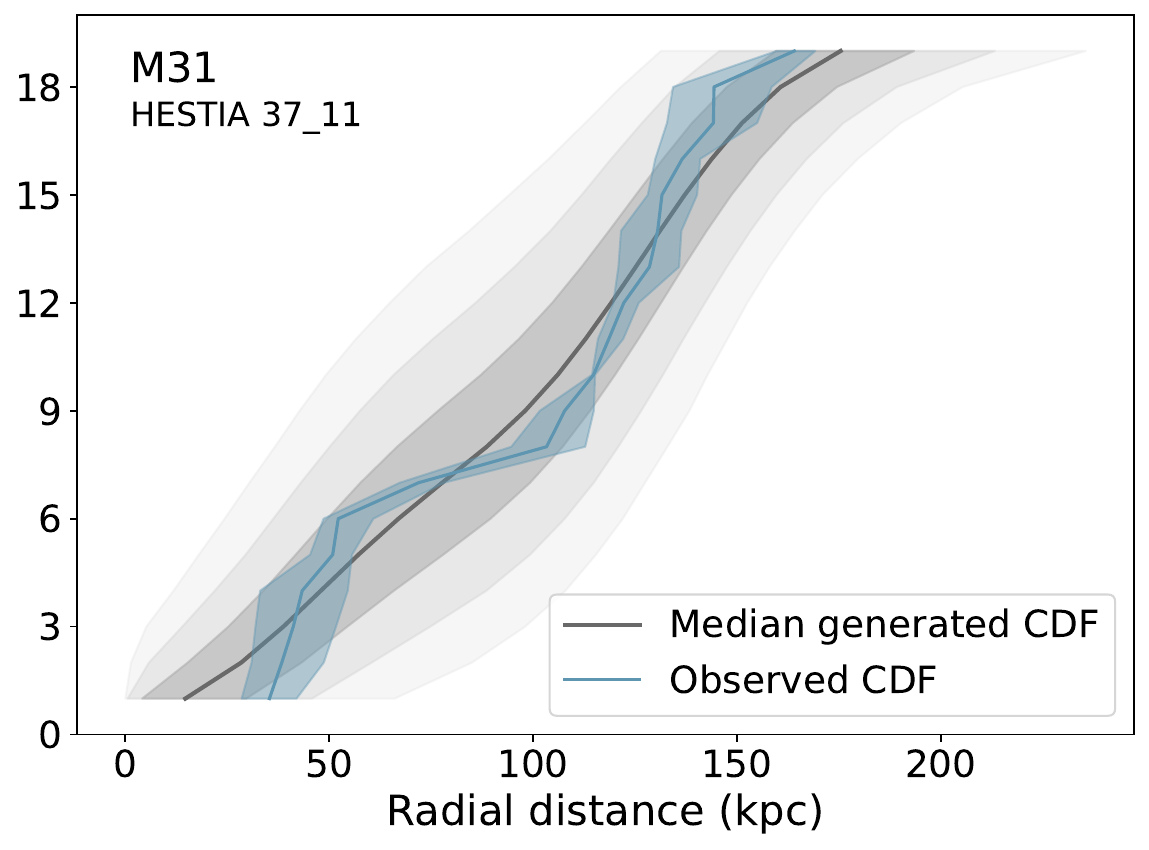}
    \caption{Same as Figure~\ref{fig:cdf_obs}, but for the \textsc{hestia} MW and M31 analogues. Each panel corresponds to one simulated host.}
    \label{fig:cdf_sims}
\end{figure}

Following the methodology applied to the observed satellite systems described in Section~\ref{sec:methods}, we show here the radial distributions of the \textsc{hestia} simulated analogues. Figure~\ref{fig:cdf_sims} presents the cumulative radial distributions of satellites around each MW- and M31-like host in the three simulation runs, compared against $10^5$ isotropic realisations constructed from the same radial profiles.

In all cases, the simulated satellite distributions fall within the $1\sigma$ to $3\sigma$ confidence intervals of the isotropic populations, confirming that the radial profiles of the simulated analogues are statistically consistent with the isotropic realisations used to generate the NN reference distributions in our analysis.

\section{$P$-value distributions of opposite host analogues}\label{app:cdfs_opposite_hosts}
\begin{figure}[ht]
    \centering
    \includegraphics[width=0.9\columnwidth]{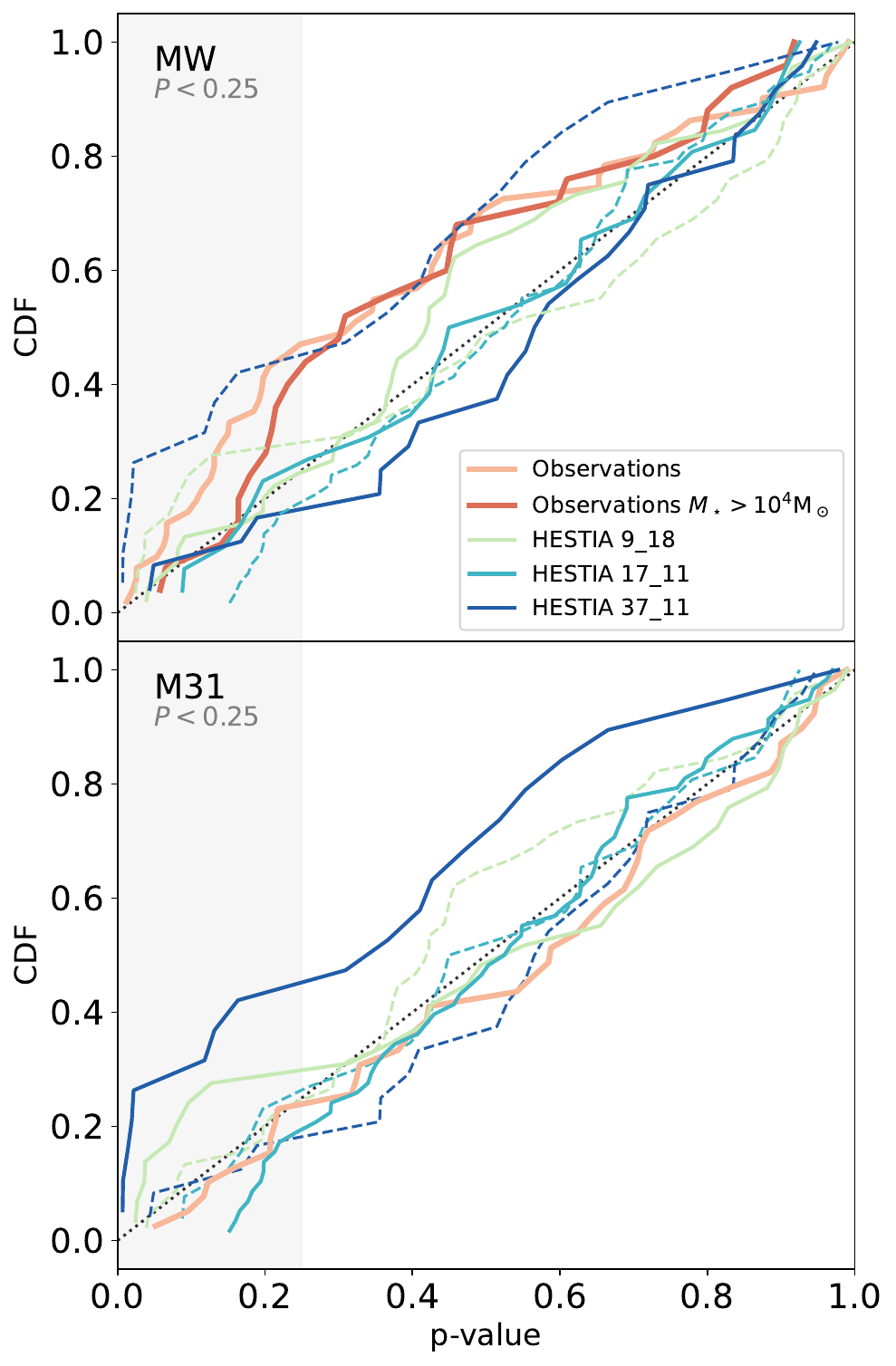}
    \caption{Same as Figure~\ref{fig:pvalues}, but with the opposite host analogue from each \textsc{hestia} realisation shown as dashed lines (M31 analogues in the top panel and MW analogues in the bottom panel).}
    \label{fig:pvalues_v2}
\end{figure}

In Section~\ref{sec:statistical} we noted that the level of apparent clustering depends partly on the host halo mass and the resulting satellite population. Figure~\ref{fig:pvalues_v2} therefore reproduces Figure~\ref{fig:pvalues}, but additionally shows the $P$-value distribution of the opposite host analogue from each \textsc{hestia} realisation.

In some cases, the $P$-value distribution of the opposite host analogue reproduces the observed system more closely than the nominal host counterpart. This is most evident for the 37\_11 realisation, where the M31 analogue matches the observed MW $P$-value distribution more closely than the corresponding MW analogue.

\clearpage

\section{Kinematic analysis of pair candidates}\label{app:kinematic_coherence}
The NN analysis presented in Section~\ref{sec:paircandidates} identifies satellites that are unusually close in space compared to expectations from isotropy. However, spatial proximity alone does not guarantee a physical association. Satellites that were accreted together or belong to the same substructure are expected to exhibit additional signatures of coherence, such as similar line-of-sight velocities, aligned orbital motions, or the possibility of being gravitationally bound.

To qualitatively explore the likelihood that the identified candidate pairs correspond to real physical associations, we therefore examine several additional diagnostics: (i) kinematic coherence based on the difference in line-of-sight velocities, (ii) consistency of the orbital motions when PMs are available, and (iii) whether the minimum dynamical mass required for the pair to be gravitationally bound is compatible with the expected halo masses of the satellites.

Tables~\ref{tab:pairs_observations_possibilities} and \ref{tab:pairs_simulations_possibilities} summarise these diagnostics for the observational and simulated candidate pairs, respectively. These indicators provide a qualitative assessment of whether the spatially identified pairs are consistent with a common origin or represent chance alignments. For two of the observed M31 pair candidates, there are no kinematic measurements, so none of the criteria can be qualified.

\begin{table}[ht]
\caption{Qualitative assessment of the physical coherence of the pair candidates identified in observations.}
\label{tab:pairs_observations_possibilities}
\begin{center}
\begin{tabular}{lccc} 
\toprule
Pair & $\Delta v_\mathrm{los}$ & Motions & Bound \\ 
\midrule
\textbf{MW} & & & \\
Bootes II -- Bootes III & no & no & no \\
Can.Ven. I -- Can.Ven. II* & no & yes & no \\
Carina II -- Pictor II & no & no & no \\ 
Carina II -- Carina III & no & yes & no \\
Grus II -- Tucana II & yes & no & yes\\
Horo. I -- Reti. III & yes & yes & no \\
Cent. I  -- Hydra II* & no & no & no \\
Leo I -- Leo II* & no & yes & no \\
Leo IV -- Leo V & yes & yes & yes \\
LMC -- Pictor II & yes & yes & yes \\
Pegasus III -- Pisces II & yes & yes & yes \\ 
Segue II -- Trian. II & no & no & no \\
SMC -- Tucana V & no & no & yes \\ 
Tucana II -- Tucana V & yes & no & yes \\
Tucana IV -- Tucana V & yes & no & yes \\ 
U. Major II -- Willman I & yes & no & yes \\
\addlinespace[0.5em] 
\textbf{MW ($M_\star > 10^4\, \mathrm{M_\odot}$)} & & & \\ 
Bootes I -- Bootes III & yes & no & no \\
Carina II -- LMC & no & yes & yes \\
Draco -- Ursa Minor & yes & yes & yes \\
LMC -- SMC & yes & yes & yes \\ 
\addlinespace[0.5em] 
\textbf{M31} & & & \\
And. X -- And. XXIV & yes & & yes \\
And. XI -- And. XII & no & & no \\
And. XV -- And. XXIII & yes & & no \\
And. XXV -- NGC 147 & yes & & no \\
\bottomrule
\end{tabular}
\end{center}
\small{\textbf{Columns from left to right:} (1) Names of the satellites part of a pair candidate; (2) kinematic coherence based on the difference in line-of-sight velocity ($\Delta v_\mathrm{los} < 117\,\mathrm{km\,s^{-1}}$); (3) possible alignment of the orbital motions based on PMs and great-circle trajectories when available; (4) whether the minimum dynamical mass required for the pair to be gravitationally bound is compatible with expected satellite halo masses. The pairs marked with an asterisk are identified in both the full observational sample and the stellar mass limited sample.}
\end{table}

\begin{table}[ht]
\caption{Qualitative assessment of the physical coherence of the pair candidates identified in the \textsc{hestia} simulations.}
\label{tab:pairs_simulations_possibilities}
\begin{center}
\begin{tabular}{lcccc} 
\toprule
Pair & Sim. & $\Delta v_\mathrm{los}$ & Motions & Bound \\ 
\midrule
\textbf{MW analogues} & & & & \\
10--2963 & 9\_18 & no & yes & yes \\
774--26549 & 9\_18 & yes & no & no \\
931--2973 & 9\_18 & yes & no & no \\
4719--11410 & 9\_18 & no & yes & no \\
5715--7479 & 9\_18 & no & yes & no \\
2963--11736 & 9\_18 & no & yes & no \\
22879--26549 & 9\_18 & yes & no & no \\
\addlinespace[0.5em] 
3292--25953 & 17\_11 & yes & no & no \\
4602--31053 & 17\_11 & no & no & no \\
9356--18567 & 17\_11 & no & yes & no \\
\addlinespace[0.5em] 
1408--11290 & 37\_11 & no & no & no \\
25745--27628 & 37\_11 & yes & no & no \\
1408--27628 & 37\_11 & yes & yes & no \\
\textbf{M31 analogues} & & & &\\
884-8995 & 9\_18 & no & no & no \\
1479-67802 & 9\_18 & yes & no & no \\  
1584-2344 & 9\_18 & yes & no & no \\
1887-6560 & 9\_18 & no & no & no \\
2344-6723 & 9\_18  & yes & yes & no \\
\addlinespace[0.5em] 
41--4743 & 17\_11 & yes & no & no \\
993--1822 & 17\_11  & yes & no & no \\
2038--5836 & 17\_11 & yes & no & no \\
3262--7688 & 17\_11 & no & no & no  \\
3311--25757 & 17\_11 & yes & no & no \\
4743--12533 & 17\_11 & yes & yes & no \\
41--5926 & 17\_11 & no & yes & no \\
30--9254 & 17\_11 & yes & yes & yes \\
\addlinespace[0.5em] 
41--9146 & 37\_11 & yes & yes & yes \\
976--2481 & 37\_11 & yes & yes & no \\
2481--4294 & 37\_11  & yes & no & no \\
4294--11673 & 37\_11 & yes & yes & no \\
8263--17744 & 37\_11 & yes & yes & no \\
\bottomrule
\end{tabular}
\end{center}
\small{\textbf{Columns from left to right:} (1) Pair of satellites identified as a candidate association; (2) Simulation realisation; (3) kinematic coherence based on the difference in line-of-sight velocity ($\Delta v_\mathrm{los} < 117\,\mathrm{km\,s^{-1}}$); (4) possible alignment of the orbital motions; (5) whether the pair is gravitationally bound.}
\end{table}



{\end{document}
